\documentclass[oldversion]{aa}
\usepackage{graphicx}
\usepackage{txfonts}
\usepackage{natbib}

\def\gtap{\mathrel{\hbox{\rlap{\lower.55ex \hbox {$\sim$}}
                   \kern-.3em \raise.4ex \hbox{$>$}}}}
\def\ltap{\mathrel{\hbox{\rlap{\lower.55ex \hbox {$\sim$}}
                 \kern-.3em \raise.4ex \hbox{$<$}}}}

\begin{document}
  \title{X-ray and optical observations of M55 and NGC\,6366:\\ evidence
    for primordial binaries}

  \author{C.\ G.\ Bassa\inst{1,2}
    \and D.\ Pooley\inst{3}
    \and F.\ Verbunt\inst{1}
    \and L.\ Homer\inst{4}\thanks{Current address: Abbey College
      Cambridge, 7 Station Rd, Cambridge, CB1 2JB, United Kingdom}
    \and S.\ F.\ Anderson\inst{4}
    \and W.\ H.\ G.\ Lewin\inst{5}}

  \institute{Astronomical Institute, Utrecht University, PO Box 80\,000,
    3508 TA Utrecht, The Netherlands
    \and Physics Department, McGill University, Montreal, QC H3A 2T8,
    Canada; \email{bassa@physics.mcgill.ca}
    \and Astronomy Department, UC Berkeley, 601 Campbell Hall,
    Berkeley, CA 94720-3411, USA
    \and Department of Astronomy, University of Washington, Box
    351580, Seattle, WA 98195, USA
    \and Kavli Institute for Astrophysics and Space Research, 
    MIT, 77 Massachusetts Avenue, Cambridge, MA 02139, USA
  }

  \date{Received / Accepted}

  \abstract{ We present \emph{Chandra X-ray Observatory} ACIS-S3 X-ray
    imaging observations and VLT/FORS2 and \emph{Hubble Space
      Telescope} optical observations of two low-density Galactic
    globular clusters; NGC\,6366 and M55.  We detect 16 X-ray sources
    with 0.5--6.0\,keV luminosities above
    $L_\mathrm{X}=4\times10^{30}$\,erg\,s$^{-1}$ within the half-mass
    radius of M55, of which 8 or 9 are expected to be background
    sources, and 5 within the half-mass radius of NGC\,6366, of which
    4 are expected to be background sources. Optical counterparts are
    identified for several X-ray sources in both clusters and from
    these we conclude that 3 of the X-ray sources in M55 and 2 or 3 of
    the X-ray sources in NGC\,6366 are probably related to the
    cluster. Combining these results with those for other clusters, we
    find the best fit for a predicted number of X-ray sources in a
    globular cluster $\mu_\mathrm{c}=1.2\Gamma+1.1M_\mathrm{h}$, where
    $\Gamma$ is the collision number and $M_\mathrm{h}$ is (half of)
    the cluster mass, both normalized to the values for the globular
    cluster M4. Some sources tentatively classified as magnetically
    active binaries are more luminous in X-rays than the upper limit
    of $L_\mathrm{X}\simeq0.001L_\mathrm{bol}$ of such binaries in the
    solar neighbourhood. Comparison with \textit{XMM} and
    \textit{ROSAT} observations lead us to conclude that the brightest
    X-ray source in M55, a dwarf nova, becomes fainter in X-rays
    during the optical outburst, in accordance with other dwarf
    novae. The brightest X-ray source in NGC\,6366 is a point source
    surrounded by a slightly offset extended source. The absence of
    galaxies and H$\alpha$ emission in our optical observations argues
    against a cluster of galaxies and against a planetary nebula, and
    we suggest that the source may be an old nova.

  }

  \maketitle

  \section{Introduction}
  All stars emit X-rays, but some emit more than others.  Thanks to
  the \emph{Chandra} X-ray Observatory the study of X-ray sources in
  globular clusters now includes sources down to luminosities of
  $L_\mathrm{X}\sim10^{29-30}$\,erg\,s$^{-1}$ in the 0.5--2.5\,keV
  band.  At these low luminosities, most sources are low-mass
  main-sequence stars that rotate rapidly, which in old stellar
  clusters is the case only when they have been spun up by tidal
  forces or by accretion in close binaries.  At the high luminosity
  end, $L_\mathrm{X}\gtap10^{32}$\,erg\,s$^{-1}$, most sources are
  low-mass X-ray binaries in which a neutron star accretes matter from
  a companion.  At intermediate luminosities, most sources are
  cataclysmic variables (CVs).  The clusters best studied at low X-ray
  luminosities are 47\,Tuc (\citealt{hge+05} and references therein)
  and M4 \citep{bph+04}.

  Whereas the binaries of main-sequence stars are primordial, i.e.\
  formed as binaries when the component stars formed, the low-mass
  X-ray binaries in globular clusters are thought to be formed in
  close stellar encounters that bring a previously single neutron star
  into a binary (see the review by \citealt{vl06} and references
  therein).  For cataclysmic variables both formation mechanisms,
  evolution from a primordial binary and capture of a previously
  single white dwarf in a close encounter, are viable, depending on
  the circumstances. The progenitor binary of a cataclysmic variable
  must be wide enough to allow the more massive star to evolve into a
  fairly big giant, before it reaches its Roche lobe. Such wide
  binaries are destroyed (`ionized') in cores of globular clusters
  with densities $\rho_0\gtap10^3\,M_\odot$\,pc$^{-3}$
  \citep{dav97}.

  One would naively expect that all cataclysmic variables in these
  cores were formed in stellar encounters.  However, it has been
  pointed out that cataclysmic variables that evolved in the
  low-density outskirts of globular clusters can sink towards the core
  at late times.  Thus, the population of cataclysmic variables in
  dense cores can be a mixture of locally produced products of stellar
  encounters and recently arrived products of the evolution of
  primordial binaries (\citealt{ihr+06}; see also \citealt{has07}).

  The number of stellar encounters in a globular cluster scales
  roughly with the collision number
  $\Gamma\equiv{\rho_0}^2{r_\mathrm{c}}^3/v$, where $\rho_0$ is the
  central density of the cluster, $r_\mathrm{c}$ the core radius, and
  $v$ the central velocity dispersion.  Through the virial theorem
  $v\propto r_\mathrm{c}\sqrt{\rho_0}$ and thus $\Gamma\propto
  {\rho_0}^{1.5}{r_\mathrm{c}}^2$ \citep{vh87,ver03}. The mass $M$ of
  a cluster can be estimated from the total luminosity of a cluster,
  with use of a mass-to-light ratio appropriate for a cluster star
  population, and if necessary a bolometric correction.  The number of
  X-ray sources in a globular cluster then may be expected to depend
  on the collision number (dynamical origin) and on the cluster mass
  (primordial origin); the form of the dependence depends on the type
  of binary, and therefore on the luminosity range that is studied.

  The brightest low-mass X-ray binaries, with
  $L_\mathrm{X}>10^{35}$\,erg\,s$^{-1}$, are sufficiently rare that a
  cluster contains either none or one; in one case (M15) two. The
  probability that a cluster contains such a source scales with
  $\Gamma$ \citep{vh87}.  The fainter low-mass X-ray binaries, with
  $10^{32}\ltap L_\mathrm{X}\ltap10^{35}$\,erg\,s$^{-1}$, are more
  numerous; the number of them in a cluster also scales with $\Gamma$
  \citep{hgl+03,gbw03b,pla+03}.  These results appear to confirm the
  theoretical idea that neutron stars enter binaries in globular
  clusters via stellar encounters. Studies of globular clusters within
  other galaxies complicate the issue, because they indicate that the
  probability that a cluster contains a bright source depends on its
  metallicity, being higher for metal-rich clusters
  (e.g.\ \citealt{kmz02}), and furthermore scales with
  ${\rho_0}^{1.5}$, a shallower dependence on density than implied by
  a proportionality with $\Gamma$ \citep{jfb+04}.

  The numbers of less luminous sources, mostly cataclysmic variables
  -- hitherto only observed in globular clusters of our Milky Way --
  appear to have a more shallow dependence on central density
  $N\propto{\rho_0}^\gamma$ with $\gamma=0.6-0.7$
  \citep{hgl+03,pla+03}. \citet{hwc+06} find different values of
  $\Gamma$ for bright and faint cataclysmic variables, with the
  boundary near $10^{32}$\,erg\,s$^{-1}$.  Finally, \citet{kbp+06}
  show that the number of sources (2-5 only!)  in the low-density
  globular cluster NGC\,288, a mixture of cataclysmic variables and
  binaries of main-sequence stars, scales with the mass of this
  cluster. \citet{ph06} successfully describe the number $N$ of X-ray
  sources in globular clusters with an equation of the form $N=a M + b
  \Gamma$.

  There are many uncertainties in the theoretical description of the
  formation of X-ray sources in globular clusters. We mention only the
  retained fraction of neutron stars; and the details of the spiral-in
  process which changes the wide orbit of the progenitor binary into
  the close orbit of a cataclysmic variable.  To help untangle these
  uncertainties empirically we have observed a variety of globular
  clusters both in X-rays and in the optical.  We have selected these
  clusters to span a wide range in central density, core radius, and
  mass. In this paper we discuss our observations of two clusters with
  relatively large core radii and low central densities: M55 and
  NGC\,6366. Some parameters of both clusters are given in
  Tables\,\ref{t1} and \ref{t6}.

  \begin{table}
    \begin{minipage}[t]{\columnwidth}
      \centering
      \caption{Parameters of M55 and NGC\,6366 used in this paper;
        core radius $r_\mathrm{c}$, half-mass radius $r_\mathrm{h}$,
        distance $d$ and reddening $E_{B-V}=A_V/3.1$ (taken from
        \citealt{har96}, version Feb 2003). The interstellar column
        $N_\mathrm{H}$ is computed from
        $N_\mathrm{H}=1.79\times10^{21}\,\mathrm{cm}^{-2}\,A_V$, see
        \citet{ps95}.
	\label{t1}}
      \begin{tabular}{lccccc}
	\hline \hline
	Cluster & $r_\mathrm{c}$ (\arcmin) & $r_\mathrm{h}$ (\arcmin)
	& $d$ (kpc) & $E_{B-V}$ & $N_\mathrm{H}$ (cm$^{-2}$) \\
	\hline
	M55 & 2.83 & 2.89 & 5.3 & 0.08 & $4.44\times10^{20}$ \\ 
	NGC\,6366 & 1.83 & 2.63 & 3.6 & 0.71 & $3.94\times10^{21}$ \\
	\hline
      \end{tabular}
    \end{minipage}
  \end{table}

  Both clusters were observed with the \emph{ROSAT} PSPC, which
  detected one source in the core of each cluster \citep{jvh96}. The
  position of the source in M55 was more accurately determined with a
  \emph{ROSAT} HRI observation \citet{ver01}, and this source was
  later optically identified with a dwarf nova
  \citep{kpt+05}. \emph{XMM} detected five sources in the core of M55,
  one of them coincident with the \emph{ROSAT} source \citep{wwb06}.

  \section{X-ray observations}
  M55 was observed for 33.7\,ks on 2004 May 11, and NGC\,6366 was
  observed for 22.0\,ks on 2002 July 5, both with the Advanced CCD
  Imaging Spectrometer (ACIS) on the \emph{Chandra} X-Ray Observatory with
  the telescope aim point on the back-side illuminated S3 chip. The
  data were taken in timed-exposure mode with the standard integration
  time of 3.24\,s per frame and telemetered to the ground in faint
  mode.

  Data reduction was performed using the CIAO 3.3 software provided by
  the \emph{Chandra} X-ray Center\footnote{{http://asc.harvard.edu}}.  The
  data were reprocessed using the CALDB 3.2.2 set of calibration files
  (gain maps, quantum efficiency, quantum efficiency uniformity,
  effective area) including a new bad pixel list made with the
  \texttt{acis\_run\_hotpix} tool.  The reprocessing was done without
  including the pixel randomization that is added during standard
  processing.  This omission slightly improves the point spread
  function.  The data were filtered using the standard ASCA grades (0,
  2, 3, 4, and 6) and excluding both bad pixels and software-flagged
  cosmic ray events. Intervals of strong background flaring were
  searched for, but none were found. The extraction of counts and
  spectra and the generation of response files was accomplished with
  ACIS
  Extract\footnote{http://www.astro.psu.edu/xray/docs/TARA/ae\_users\_guide.html}
  (\citealt{btgb02}, version 3.107), which calls many standard CIAO
  routines.

  \subsection{Source detection}\label{s2s1}
\begin{figure*}
    \centering
    \includegraphics[width=8.5cm]{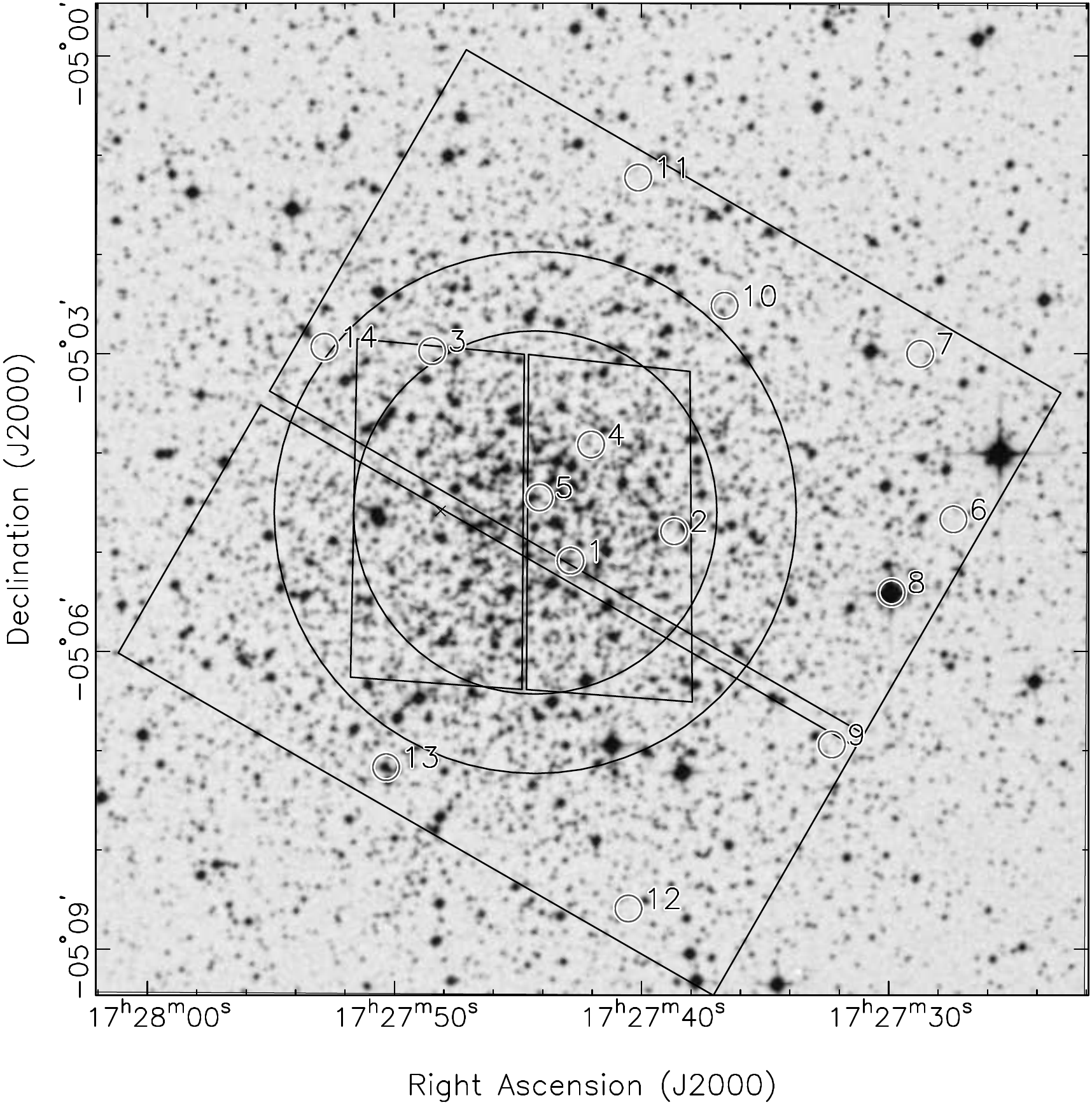}
    \includegraphics[width=8.5cm]{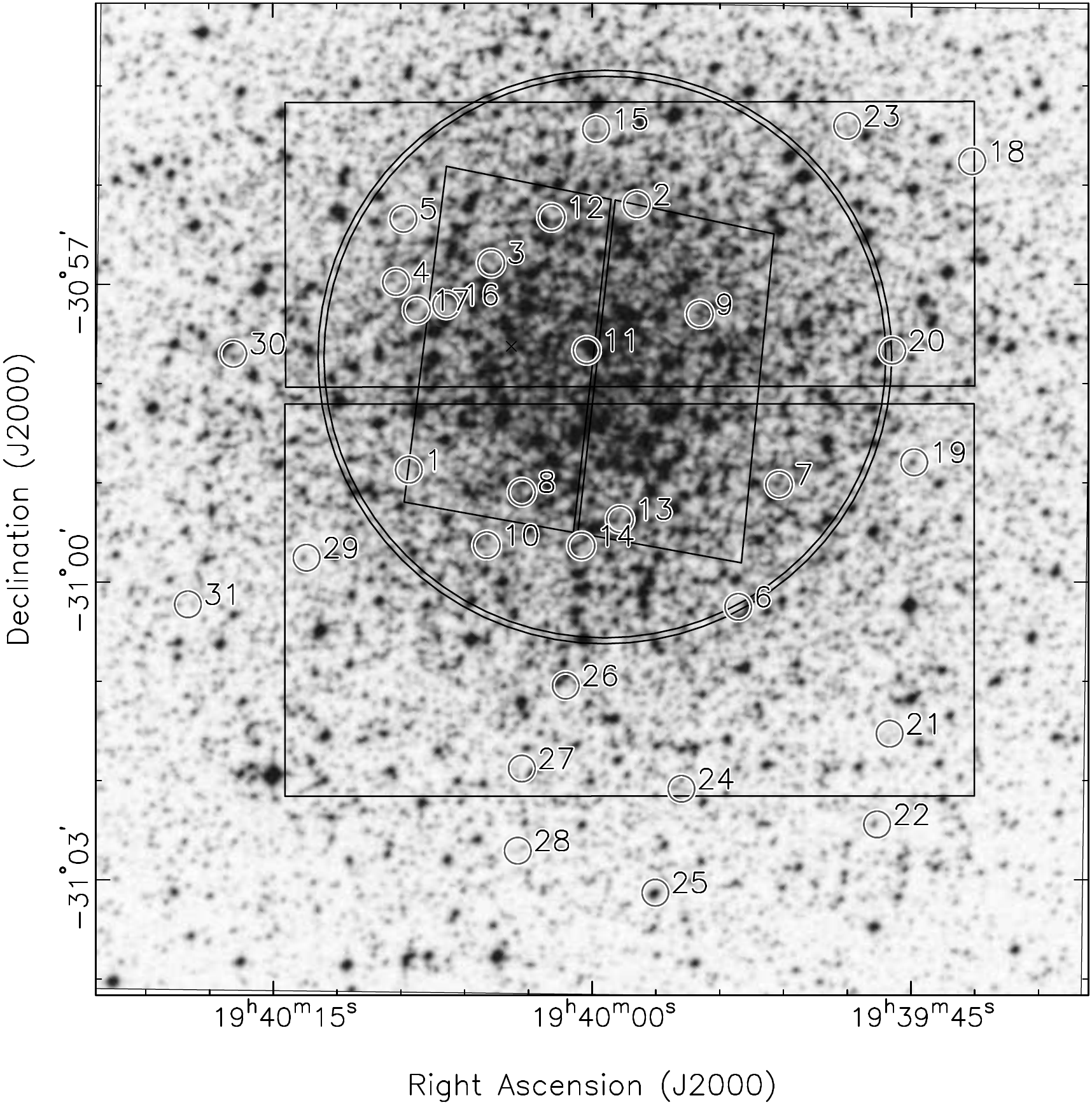}
    \caption{$10\arcmin\times10\arcmin$ Digitized Sky Survey $R$-band
    images of M55 (\textit{left}) and NGC\,6366 (\textit{right}). The
    positions of the \textit{Chandra} X-ray sources are depicted with
    circles and numbered according to their CX designation (see
    Table\,\ref{t2} and \ref{t3}). The two large rectangles on each image
    denotes the field-of-view of the two FORS2 chips, while the
    smaller slanted rectangles denote the field-of-view of the
    \emph{Hubble Space Telescope} ACS/WFC chips.}
    \label{f1}
  \end{figure*}

  The CIAO wavelet-based \texttt{wavdetect} tool was employed for
  source detection in both the 0.5--6.0 and 0.3--10.0\,keV bands. We
  detected 29 sources on the entire S3 chip in the M55 observation and
  12 in NGC\,6366 observation. Of these sources, 15 (14) lie within
  the 2\farcm89 half-mass radius (2\farcm83 core radius) of M55, and 3
  (3) lie within the 2\farcm63 half-mass radius (1\farcm83 core
  radius) of NGC\,6366.  We also searched part of the adjacent S4 CCD
  in the M55 observation since part of the half-mass region fell on
  this chip, but no sources were detected in this area.  Furthermore,
  we examined adaptively-smoothed images (made from the CIAO
  \texttt{csmooth} tool) for significant point sources.  In the M55
  observation, we found two additional possible sources within the
  core radius.  In the NGC\,6366 observation, we also found two
  additional possible sources within the half-mass radius, one of
  which was within the core radius. Fig.~\ref{f1} shows the location
  of the detected sources with respect to the cluster center and the
  core and half-mass radius.

\begin{figure*}
    \centering
    \includegraphics[width=5.8cm]{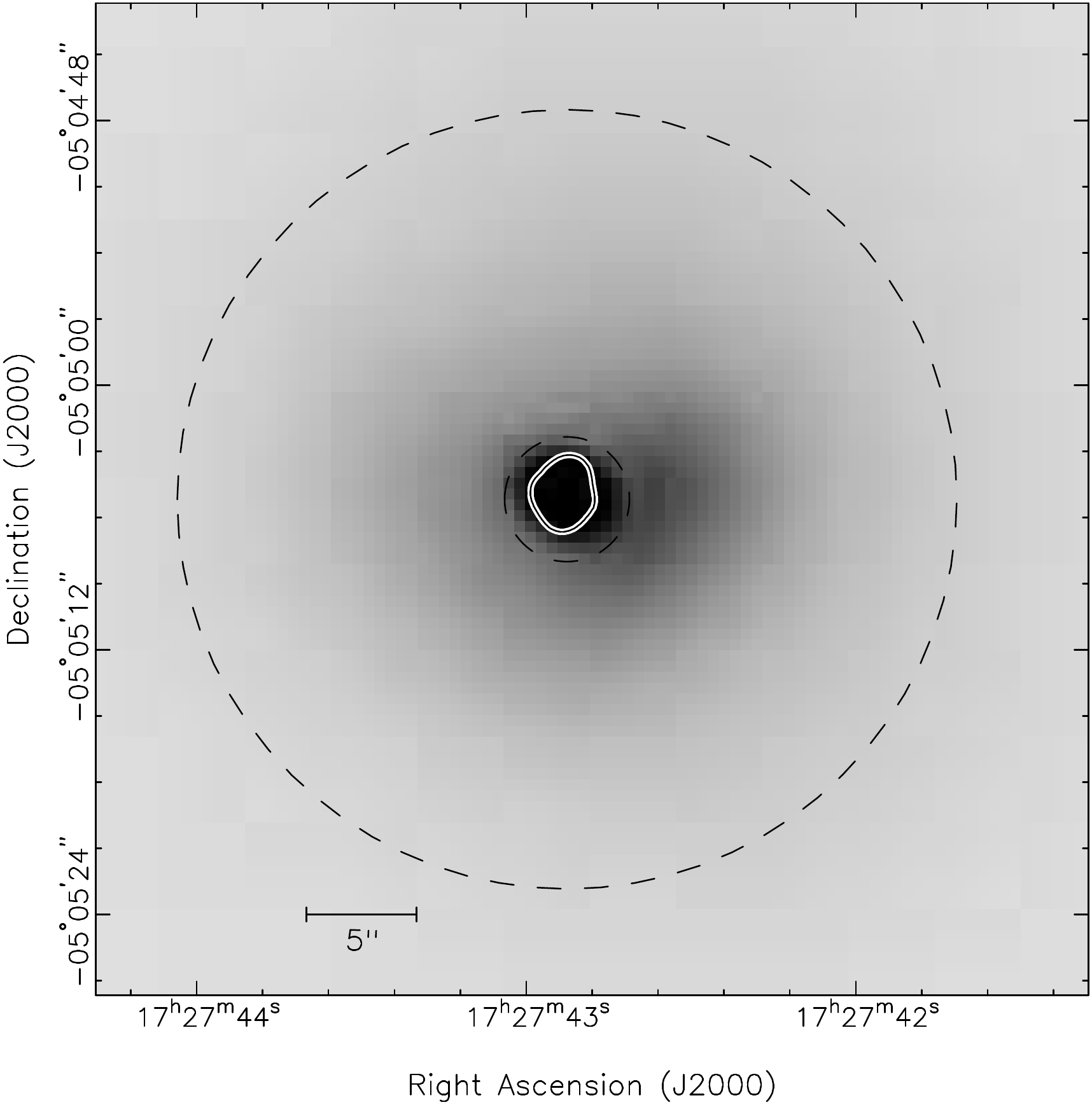}
    \includegraphics[width=5.8cm]{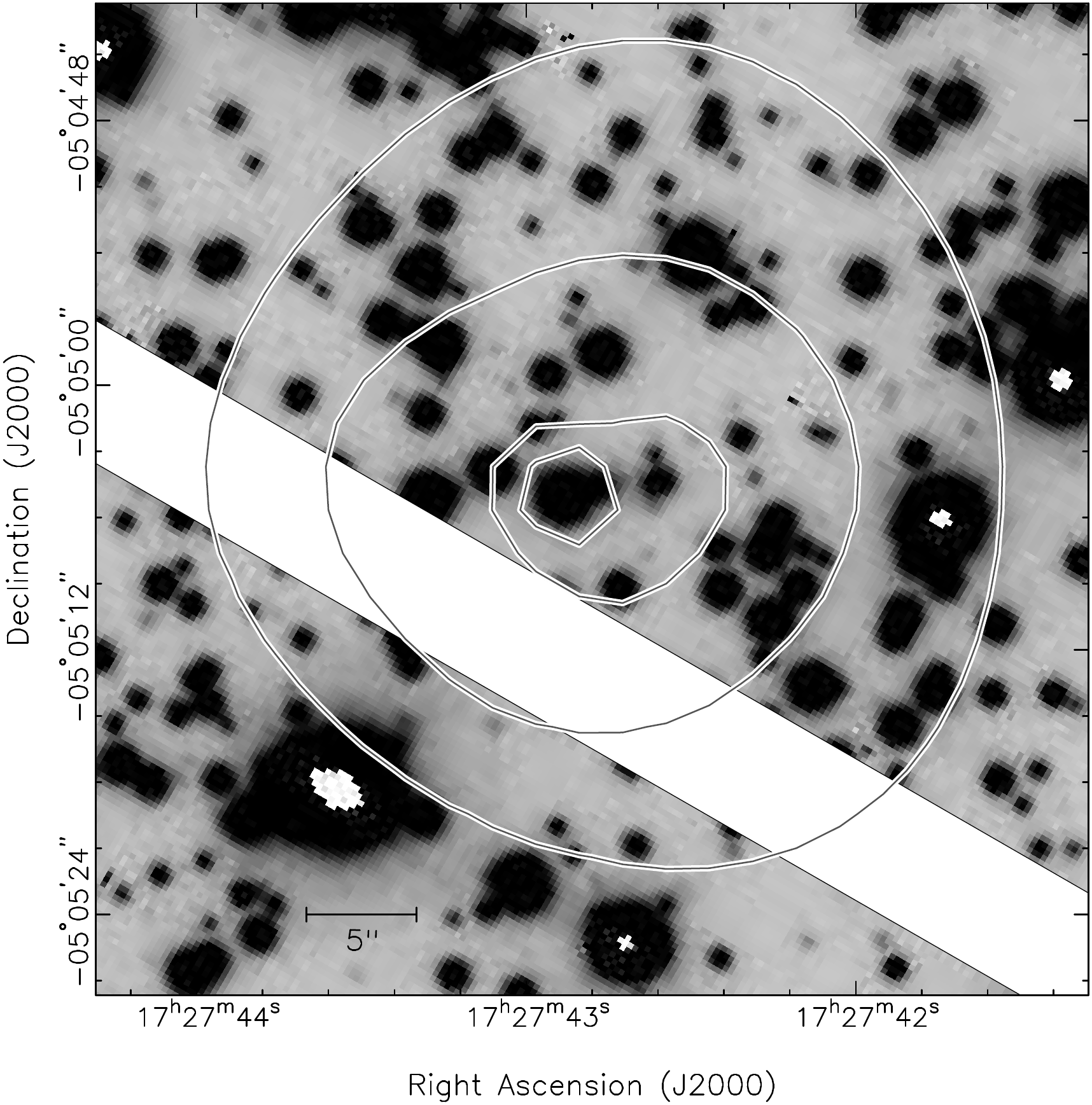}
    \includegraphics[width=5.8cm]{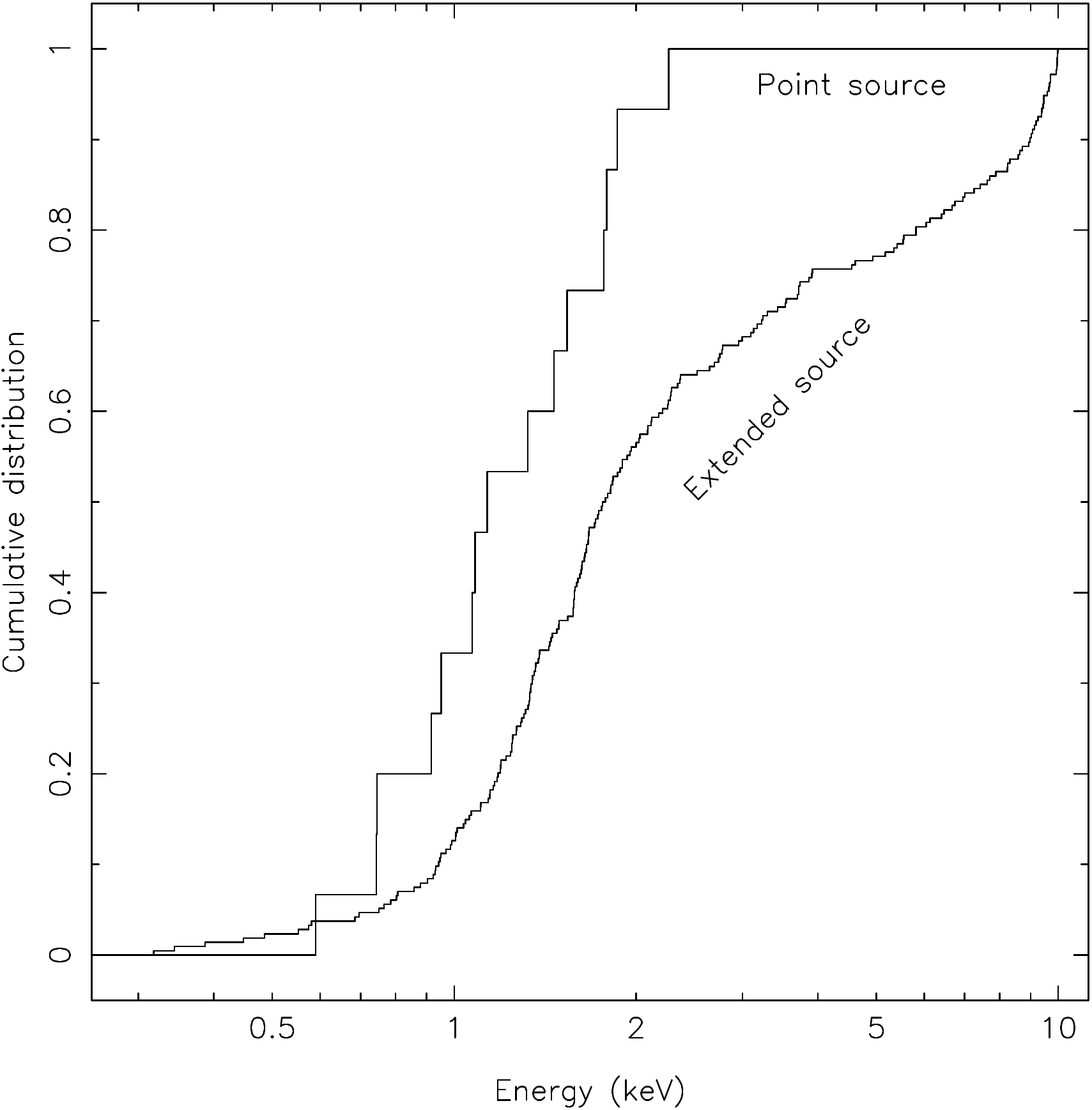}
    \caption{\emph{(left panel)} The adaptively smoothed brightness
      distribution of NGC\,6366 CX1 shows a point source and an
      extended source. The solid contour contains 90\% of the flux in
      the Chandra point spread function (at the detector location of
      the source); the dashed contours indicate the extraction region
      for the counts of the extended source. \emph{(middle panel)}
      Contours from the adaptively smoothed brightness distribution of
      CX1 overlaid on the 15\,minute H$\alpha$ images. The white
      diagonal strip is due to the gap between the FORS2 chips, see
      Fig.\,~\ref{f1}. There is no evidence for any extended H$\alpha$
      emission. \emph{(right panel)} The cumulative energy
      distributions of the point source (large-stepped distribution)
      and the extended source. A Kolmogorov Smirnov test gives a
      probability of 1\% that both energy distributions are the same,
      and we conclude that the extended source is harder.}
    \label{f2}
  \end{figure*}

  All sources are consistent with being point sources, except for CX1
  in NGC\,6366, which appears to extend to a radius of about
  $20\arcsec$ (Fig.\,\ref{f2}). Visual inspection suggests the
  superposition of a point source and extended emission.  To assess
  their spatial coincidence, we fit two gaussians plus a constant
  background to an image of the region.  One gaussian approximates the
  point spread function, and the other is broader and describes the
  extended emission.  Our best fit model gives source counts in rough
  agreement with the methods described in \S\,\ref{s2.2} and given in
  Table\,\ref{t3}; according to the model, the point source has 17.4
  net counts, and the extended source has 186.1 net counts.  The two
  gaussians are offset from each other by $2\farcs4\pm1\farcs1$.

  We will give two estimates of how many sources are related to each
  cluster. For the first estimate, we assume that all sources outside
  the half-mass radius are fore- or background sources,
  $N_\mathrm{o}=14$ for M55 and 9 for NGC\,6366.  Taking the size of
  the S3 detector as $8\farcm4\times8\farcm4=70.6$\,sq.arcmin, the
  expected number of fore- or background sources within the half-mass
  radius $r_\mathrm{h}$, expressed in arcmin, follows as
  $N_\mathrm{e}=
  N_\mathrm{o}\pi{r_\mathrm{h}}^2/(70.6-\pi{r_\mathrm{h}}^2)$. This
  gives $N_\mathrm{e}=8.3$ for M55 and 4.0 for NGC\,6366.  

  The second estimate may be found from the $\log N$-$\log S$
  relationships of \citet{grt+01}; at our detection limits, this leads
  to an expected number of background sources within the half-mass
  radius of 6--9 for M55 and 3--4 for NGC\,6366.  These numbers are
  wholly compatible with our first estimate.  In M55 we detect 17
  sources within the core, of which we estimate 8-11 cluster
  members. The probability to find 14 or more sources for expected 8.3
  is about 4\%; so it is likely that at least 4 sources are members.
  In the case of NGC\,6366 we estimate that 1 of the 4 sources within
  the half-mass radius is a cluster member, but there is a sizable
  probability (37\%) that all sources are fore- or background.

  We repeat our first estimate for sources with
  $L_\mathrm{X}>4\times10^{30}$\,erg\,s$^{-1}$ [0.5--6.0\,keV], the
  limit used in the \citet{pla+03} study.  Above the corresponding
  flux limit of $1.2\times10^{-15}$\,erg\,s$^{-1}$\,cm$^{-2}$ for M55,
  we detected 14 sources outside the half-mass radius of M55, and
  expect 8.3 sources unrelated to the cluster within the half-mass
  radius. 16 sources within the half-mass radius are observed above the
  flux limit.  Above the flux limit for NGC\,6366 of
  $2.6\times10^{-15}$\,erg\,s$^{-1}$\,cm$^{-2}$, we observe 9 sources
  outside the half-mass radius, which predicts 4.0 unrelated sources
  within the core, compared to the observed number of 5.  The
  probability to observe 14 or more sources when 8.3 are expected is
  about 4\%.  Thus from the source numbers alone we have marginal
  evidence for sources above the luminosity limit used by
  \citet{pla+03} related to M55; and no evidence for such sources
  related to NGC\,6366.

  In Tables\,\ref{t2} and \ref{t3} we list the
  properties of the X-ray sources detected by \textit{Chandra} in M55
  and NGC\,6366, respectively.

  \subsection{Count Rates}\label{s2.2}
  We extracted source counts in the following bands: 0.5--1.5\,keV
  ($X_\mathrm{soft}$), 0.5--4.5\,keV ($X_\mathrm{med}$), and
  1.5--6.0\,keV ($X_\mathrm{hard}$). The detected count rate was
  corrected for background, exposure variations, and foreground
  photoelectric absorption. We make these corrections in order to
  produce a X-ray color-magnitude diagram (CMD) that can be compared
  to the X-ray CMDs that have resulted from \emph{Chandra} observations of
  other globular clusters. In addition, however, attention must be
  paid to differences in detector responses and, of course, exposure
  times and distances.

  The background count rates in each band were estimated from
  source-free regions on the S3 chip outside the half-mass radii. The
  density of background counts in each band for the M55 observation is
  0.0050\,counts\,pixel$^{-1}$ in $X_\mathrm{soft}$,
  0.0103\,counts\,pixel$^{-1}$ in $X_\mathrm{med}$, and
  0.0072\,counts\,pixel$^{-1}$ in $X_\mathrm{hard}$. For the NGC\,6366
  observation, the background densities are
  0.0030\,counts\,pixel$^{-1}$ in $X_\mathrm{soft}$,
  0.0063\,counts\,pixel$^{-1}$ in $X_\mathrm{med}$, and
  0.0056\,counts\,pixel$^{-1}$ in $X_\mathrm{hard}$. The background
  count rates in the cores may be somewhat higher, but even factors of
  a few greater than this estimate have negligible effects on our
  analysis.

  \begin{table*}
    \begin{minipage}[t]{\textwidth}
     \centering
      \caption[]{\textit{Chandra} X-ray sources detected in our
        observation of M55. The positions of the X-ray
        sources have been corrected for the bore-sight correction of
        $-0\farcs270$ in right ascension and $+0\farcs080$ in
        declination. Positional uncertainties are given in parentheses
        and refer to the last quoted digit and are the centroiding
        uncertainties given by ACIS Extract. They do not include the
        uncertainties in the bore-sight correction ($0\farcs044$ in
        right ascension and $0\farcs045$ in declination). The X-ray
        bands are 0.5--1.5\,keV ($X_\mathrm{soft}$), 0.5--4.5\,keV
        ($X_\mathrm{med}$) and 1.5--6.0\,keV ($X_\mathrm{hard}$). The
        first 17 sources are located within the half-mass radius of
        this cluster and are ordered on 0.5--6.0\,keV countrate
        ($X_\mathrm{soft}+X_\mathrm{hard}$). The remaining sources are
        located outside the half-mass radius and are ordered on right
        ascension.  Some of the X-ray sources are also
        detected by \emph{ROSAT} \citep{jvh96,ver01} and \emph{XMM}
        \citep{wwb06}. These sources are denoted by R and X. CV1 is
        cataclysmic variable found by \citet{kpt+05}.
      }\label{t2} \renewcommand{\footnoterule}{}
      \begin{tabular}{l@{\hspace{0.2cm}}l@{\hspace{0.2cm}}l@{\hspace{0.2cm}}ccccc}
	\hline\hline
	ID & \multicolumn{1}{c}{R.A.} & \multicolumn{1}{c}{Decl.} &
	\multicolumn{3}{c}{Counts (Detected/Corrected)} &
	$f_\mathrm{X}$ (0.5--2.5\,keV) & $f_\mathrm{X}$ (2.5--6.0\,keV) \\
	 & \multicolumn{1}{c}{(J2000)} & \multicolumn{1}{c}{(J2000)} &
	$X_\mathrm{soft}$ & $X_\mathrm{med}$ & $X_\mathrm{hard}$ & (erg\,s$^{-1}$\,cm$^{-2}$) & (erg\,s$^{-1}$\,cm$^{-2}$) \\
	\hline 
CX1 (R9/X30/CV1) & $19^\mathrm{h}40^\mathrm{m}08\fs593(3)$  & $-30\degr58\arcmin52\farcs08(4)$  &   $60/74.3$ & $102/123.5$ &   $45/47.9$ & $1.2\times10^{-14}$ & $1.4\times10^{-14}$\\
CX2 (X42)        & $19^\mathrm{h}39^\mathrm{m}57\fs877(3)$  & $-30\degr56\arcmin11\farcs67(4)$  &   $27/32.8$ &   $44/51.4$ &   $18/18.5$ & $5.0\times10^{-15}$ & $5.8\times10^{-15}$\\
CX3 (X12)        & $19^\mathrm{h}40^\mathrm{m}04\fs720(3)$  & $-30\degr56\arcmin47\farcs09(4)$  &   $14/17.0$ &   $33/38.6$ &   $22/22.8$ & $3.4\times10^{-15}$ & $7.9\times10^{-15}$\\
CX4 (X13)        & $19^\mathrm{h}40^\mathrm{m}09\fs192(5)$  & $-30\degr56\arcmin58\farcs54(7)$  &     $4/4.9$ &   $21/24.7$ &   $19/19.8$ & $1.9\times10^{-15}$ & $7.2\times10^{-15}$\\
CX5              & $19^\mathrm{h}40^\mathrm{m}08\fs846(6)$  & $-30\degr56\arcmin20\farcs34(8)$  &   $10/11.8$ &   $15/17.5$ &     $7/7.5$ & $1.9\times10^{-15}$ & $2.1\times10^{-15}$\\
CX6              & $19^\mathrm{h}39^\mathrm{m}53\fs108(11)$ & $-31\degr00\arcmin14\farcs99(15)$ &     $5/6.0$ &   $12/13.8$ &     $8/8.2$ & $1.3\times10^{-15}$ & $2.3\times10^{-15}$\\
CX7              & $19^\mathrm{h}39^\mathrm{m}51\fs190(9)$  & $-30\degr59\arcmin01\farcs01(11)$ &   $10/11.8$ &   $13/14.7$ &     $3/3.1$ & $1.7\times10^{-15}$ & $1.1\times10^{-15}$\\
CX8              & $19^\mathrm{h}40^\mathrm{m}03\fs279(8)$  & $-30\degr59\arcmin05\farcs61(11)$ &     $5/6.2$ &   $10/12.1$ &     $6/6.5$ & $1.2\times10^{-15}$ & $1.6\times10^{-15}$\\
CX9              & $19^\mathrm{h}39^\mathrm{m}54\fs920(7)$  & $-30\degr57\arcmin18\farcs01(8)$  &     $0/0.0$ &     $8/9.3$ &   $10/10.2$ & $6.1\times10^{-16}$ & $3.6\times10^{-15}$\\
CX10             & $19^\mathrm{h}40^\mathrm{m}04\fs932(11)$ & $-30\degr59\arcmin37\farcs68(15)$ &     $3/3.7$ &    $9/10.9$ &     $6/6.5$ & $9.2\times10^{-16}$ & $1.7\times10^{-15}$\\
CX11             & $19^\mathrm{h}40^\mathrm{m}00\fs235(8)$  & $-30\degr57\arcmin39\farcs85(10)$ &     $3/3.3$ &     $7/7.3$ &     $4/3.6$ & $8.1\times10^{-16}$ & $6.4\times10^{-16}$\\
CX12             & $19^\mathrm{h}40^\mathrm{m}01\fs887(8)$  & $-30\degr56\arcmin19\farcs48(10)$ &     $2/2.4$ &     $5/5.9$ &     $4/4.2$ & $5.8\times10^{-16}$ & $1.2\times10^{-15}$\\
CX13             & $19^\mathrm{h}39^\mathrm{m}58\fs668(12)$ & $-30\degr59\arcmin22\farcs01(16)$ &     $0/0.0$ &     $5/6.3$ &     $6/7.0$ & $4.8\times10^{-16}$ & $2.4\times10^{-15}$\\
CX14             & $19^\mathrm{h}40^\mathrm{m}00\fs479(14)$ & $-30\degr59\arcmin38\farcs08(18)$ &     $0/0.0$ &     $5/5.4$ &     $5/5.1$ & $4.3\times10^{-16}$ & $1.4\times10^{-15}$\\
CX15             & $19^\mathrm{h}39^\mathrm{m}59\fs793(10)$ & $-30\degr55\arcmin26\farcs12(13)$ &     $4/5.4$ &     $5/6.3$ &     $1/0.8$ & $6.4\times10^{-16}$ & $5.8\times10^{-16}$\\
CX16             & $19^\mathrm{h}40^\mathrm{m}06\fs829(10)$ & $-30\degr57\arcmin12\farcs29(13)$ &     $5/6.4$ &     $5/5.9$ &     $0/0.0$ & $7.6\times10^{-16}$ & $2.0\times10^{-17}$\\
CX17             & $19^\mathrm{h}40^\mathrm{m}08\fs214(12)$ & $-30\degr57\arcmin15\farcs47(15)$ &     $3/3.7$ &     $4/4.6$ &     $1/0.8$ & $5.7\times10^{-16}$ & $6.9\times10^{-16}$\\[0.5em]
CX18 (R6/X9)     & $19^\mathrm{h}39^\mathrm{m}42\fs130(4)$  & $-30\degr55\arcmin45\farcs48(5)$  &  $71/101.4$ & $137/170.8$ &   $72/75.8$ & $1.8\times10^{-14}$ & $2.0\times10^{-14}$\\
CX19             & $19^\mathrm{h}39^\mathrm{m}44\fs842(13)$ & $-30\degr58\arcmin47\farcs54(17)$ &     $5/6.4$ &   $11/12.8$ &     $6/6.0$ & $1.2\times10^{-15}$ & $2.0\times10^{-15}$\\
CX20 (X14)       & $19^\mathrm{h}39^\mathrm{m}45\fs876(4)$  & $-30\degr57\arcmin40\farcs03(4)$  &   $29/38.5$ &  $92/112.6$ &   $78/82.9$ & $9.7\times10^{-15}$ & $2.9\times10^{-14}$\\
CX21             & $19^\mathrm{h}39^\mathrm{m}45\fs989(19)$ & $-31\degr01\arcmin31\farcs65(25)$ &     $4/5.2$ &   $14/15.2$ &   $13/12.0$ & $1.5\times10^{-15}$ & $4.7\times10^{-15}$\\
CX22 (X19)       & $19^\mathrm{h}39^\mathrm{m}46\fs573(12)$ & $-31\degr02\arcmin26\farcs47(15)$ &   $33/44.2$ &   $65/77.3$ &   $34/33.9$ & $8.1\times10^{-15}$ & $1.2\times10^{-14}$\\
CX23             & $19^\mathrm{h}39^\mathrm{m}47\fs994(11)$ & $-30\degr55\arcmin24\farcs12(14)$ &    $8/23.8$ &   $11/28.6$ &     $3/6.5$ & $3.5\times10^{-15}$ & $1.2\times10^{-15}$\\
CX24 (R13/X17)   & $19^\mathrm{h}39^\mathrm{m}55\fs778(4)$  & $-31\degr02\arcmin04\farcs98(5)$  & $207/264.9$ & $304/368.1$ & $106/112.5$ & $3.9\times10^{-14}$ & $2.6\times10^{-14}$\\
CX25 (X21)       & $19^\mathrm{h}39^\mathrm{m}57\fs005(15)$ & $-31\degr03\arcmin07\farcs82(19)$ &   $45/56.2$ &   $48/56.0$ &     $3/2.0$ & $7.1\times10^{-15}$ & $4.8\times10^{-16}$\\
CX26             & $19^\mathrm{h}40^\mathrm{m}01\fs219(18)$ & $-31\degr01\arcmin02\farcs60(24)$ &     $0/0.0$ &     $3/3.0$ &     $7/7.1$ & $6.9\times10^{-16}$ & $3.9\times10^{-15}$\\
CX27             & $19^\mathrm{h}40^\mathrm{m}03\fs284(24)$ & $-31\degr01\arcmin52\farcs73(31)$ &     $0/0.0$ &     $8/9.4$ &     $9/8.8$ & $6.4\times10^{-16}$ & $3.1\times10^{-15}$\\
CX28 (X20)       & $19^\mathrm{h}40^\mathrm{m}03\fs475(14)$ & $-31\degr02\arcmin42\farcs40(18)$ &   $25/30.0$ &   $37/42.7$ &   $13/12.6$ & $4.7\times10^{-15}$ & $4.2\times10^{-15}$\\
CX29 (X45)       & $19^\mathrm{h}40^\mathrm{m}13\fs402(15)$ & $-30\degr59\arcmin45\farcs22(19)$ &    $9/11.0$ &   $11/13.1$ &     $2/2.2$ & $1.6\times10^{-15}$ & $7.7\times10^{-16}$\\
CX30             & $19^\mathrm{h}40^\mathrm{m}16\fs842(13)$ & $-30\degr57\arcmin41\farcs89(17)$ &     $4/5.5$ &   $10/12.3$ &     $6/6.0$ & $1.0\times10^{-15}$ & $3.0\times10^{-15}$\\
CX31             & $19^\mathrm{h}40^\mathrm{m}18\fs981(20)$ & $-31\degr00\arcmin13\farcs35(26)$ &     $2/2.2$ &   $12/14.5$ &   $12/12.1$ & $1.3\times10^{-15}$ & $5.0\times10^{-15}$\\
        \hline
    \end{tabular}
    \end{minipage}
  \end{table*}

  \begin{table*}
    \begin{minipage}[t]{\textwidth}
     \centering
      \caption[]{\textit{Chandra} X-ray sources detected in our
        observation of NGC\,6366. The celestial positions of the X-ray
        sources have been corrected for the bore-sight correction of
        $+0\farcs030$ in right ascension and $-0\farcs256$ in
        declination. Positional uncertainties are given in parenthesis
        and refer to the last quoted digit and are the centroiding
        uncertainties given by ACIS Extract. They do not include the
        uncertainties in the bore-sight correction ($0\farcs091$ in
        right ascension and $0\farcs082$ in declination). The X-ray
        bands are as defined in Table\,\ref{t2}. The first 5 sources
        are located within the half-mass radius of this cluster and
        are ordered on 0.5--6.0\,keV countrate
        ($X_\mathrm{soft}+X_\mathrm{hard}$). The remaining sources are
        located outside the half-mass radius and are ordered on right
        ascension. CX1 was also detected by \emph{ROSAT}
        \citep{jvh96,ver01}.}\label{t3} \renewcommand{\footnoterule}{}
      \begin{tabular}{l@{\hspace{0.2cm}}l@{\hspace{0.2cm}}l@{\hspace{0.2cm}}ccccc}
	\hline\hline
	ID & \multicolumn{1}{c}{R.A.} & \multicolumn{1}{c}{Decl.} &
	\multicolumn{3}{c}{Counts (Detected/Corrected)} &
	$f_\mathrm{X}$ (0.5--2.5\,keV) & $f_\mathrm{X}$ (2.5--6.0\,keV) \\
	 & \multicolumn{1}{c}{(J2000)} & \multicolumn{1}{c}{(J2000)} &
	$X_\mathrm{soft}$ & $X_\mathrm{med}$ & $X_\mathrm{hard}$ & (erg\,s$^{-1}$\,cm$^{-2}$) & (erg\,s$^{-1}$\,cm$^{-2}$) \\
	\hline 
CX1a      & $17^\mathrm{h}27^\mathrm{m}42\fs892(7)$  & $-05\degr05\arcmin05\farcs28(11)$ &  $10/33.2$ &   $15/32.1$ &    $5/4.8$ & $8.5\times10^{-15}$ & $7.0\times10^{-16}$ \\
CX1b (R4) & $17^\mathrm{h}27^\mathrm{m}42\fs78(7)$   & $-05\degr05\arcmin04\farcs7(1.1)$ & $87/190.0$ & $185/280.9$ & $111/79.2$ & $4.6\times10^{-14}$ & $4.9\times10^{-14}$ \\
CX2       & $17^\mathrm{h}27^\mathrm{m}38\fs682(7)$  & $-05\degr04\arcmin47\farcs63(11)$ &    $2/7.5$ &    $6/15.0$ &    $5/5.9$ & $1.4\times10^{-15}$ & $2.8\times10^{-15}$ \\
CX3       & $17^\mathrm{h}27^\mathrm{m}48\fs470(19)$ & $-05\degr02\arcmin58\farcs69(28)$ &    $2/7.7$ &    $5/11.5$ &    $3/2.8$ & $1.3\times10^{-15}$ & $2.4\times10^{-15}$ \\
CX4       & $17^\mathrm{h}27^\mathrm{m}42\fs044(12)$ & $-05\degr03\arcmin55\farcs12(18)$ &    $0/0.0$ &    $4/10.8$ &    $4/5.2$ & $5.5\times10^{-16}$ & $3.1\times10^{-15}$ \\
CX5       & $17^\mathrm{h}27^\mathrm{m}44\fs137(12)$ & $-05\degr04\arcmin27\farcs20(17)$ &   $3/12.1$ &    $5/13.4$ &    $2/2.6$ & $2.7\times10^{-15}$ & $3.6\times10^{-16}$ \\[0.5em]
CX6       & $17^\mathrm{h}27^\mathrm{m}27\fs384(8)$  & $-05\degr04\arcmin40\farcs42(12)$ &   $5/20.1$ &   $15/38.6$ &  $10/12.0$ & $4.6\times10^{-15}$ & $3.7\times10^{-15}$ \\
CX7       & $17^\mathrm{h}27^\mathrm{m}28\fs726(16)$ & $-05\degr03\arcmin00\farcs34(24)$ &    $2/7.7$ &    $6/15.0$ &    $6/7.2$ & $1.7\times10^{-15}$ & $3.6\times10^{-15}$ \\
CX8       & $17^\mathrm{h}27^\mathrm{m}29\fs883(6)$  & $-05\degr05\arcmin24\farcs56(8)$  &  $15/54.7$ &   $17/40.9$ &    $2/2.1$ & $1.5\times10^{-14}$ & $1.7\times10^{-16}$ \\
CX9       & $17^\mathrm{h}27^\mathrm{m}32\fs307(7)$  & $-05\degr06\arcmin56\farcs56(10)$ &    $2/6.7$ &    $8/19.9$ &    $6/7.1$ & $2.2\times10^{-15}$ & $2.2\times10^{-15}$ \\
CX10      & $17^\mathrm{h}27^\mathrm{m}36\fs643(17)$ & $-05\degr02\arcmin31\farcs24(26)$ &    $1/4.0$ &     $2/5.4$ &   $ 2/2.6$ & $1.8\times10^{-15}$ & $5.1\times10^{-15}$ \\
CX11      & $17^\mathrm{h}27^\mathrm{m}40\fs138(7)$  & $-05\degr01\arcmin13\farcs76(10)$ & $36/141.8$ &  $80/207.4$ &  $45/55.2$ & $2.9\times10^{-14}$ & $1.6\times10^{-14}$ \\
CX12      & $17^\mathrm{h}27^\mathrm{m}40\fs533(9)$  & $-05\degr08\arcmin35\farcs70(14)$ &    $0/0.0$ &    $5/12.7$ &    $6/6.9$ & $7.0\times10^{-16}$ & $5.2\times10^{-15}$ \\
CX13      & $17^\mathrm{h}27^\mathrm{m}50\fs335(14)$ & $-05\degr07\arcmin10\farcs27(21)$ &   $4/15.1$ &    $6/14.7$ &    $2/2.1$ & $5.3\times10^{-15}$ & $8.3\times10^{-17}$ \\
CX14      & $17^\mathrm{h}27^\mathrm{m}52\fs830(11)$ & $-05\degr02\arcmin56\farcs05(16)$ &  $16/62.0$ &   $29/71.4$ &  $13/14.6$ & $1.0\times10^{-14}$ & $5.5\times10^{-15}$ \\
\hline
    \end{tabular}
    \end{minipage}
  \end{table*}

  The exposure variations among sources were at the 15\% level or less
  in both observations. To account for these variations in exposure,
  we applied multiplicative corrections based on the ratio of the
  average effective area of the detector at the location of a source
  in each of the three bands to that in the same band of the source
  which had the highest average exposure (CX11 in M55 and CX2 in
  NGC\,6366). The individual effective area curves for the sources
  were made using the CIAO tool \texttt{mkarf}. The average effective
  area of the detector at the location of CX11 in M55 in each of the
  bands was 494\,cm$^2$ ($X_\mathrm{soft}$), 449\,cm$^2$
  ($X_\mathrm{med}$), and 381\,cm$^2$ ($X_\mathrm{hard}$); for CX2 in
  NGC\,6366, the areas were 509\,cm$^2$ ($X_\mathrm{soft}$),
  447\,cm$^2$ ($X_\mathrm{med}$), and 373\,cm$^2$ ($X_\mathrm{hard}$).

  While the previous corrections were relatively minor (at the few
  percent level or less), the correction for photoelectric absorption
  is appreciable for NGC\,6366 (less so for M55). We investigated the
  effects of absorption by the column densities given in
  Table\,\ref{t1} on three characteristic spectra: a
  3\,keV thermal bremsstrahlung, a 0.3\,keV blackbody plus power law
  with photon index of 2, and a power law with a photon index of
  2. The effects were most prominent in the $X_\mathrm{soft}$
  band. Averaging the results of each spectrum in each band, we use
  the following correction factors for NGC\,6366: 3.72
  ($X_\mathrm{soft}$), 2.47 ($X_\mathrm{med}$), and 1.17
  ($X_\mathrm{hard}$). For M55, the factors are 1.21
  ($X_\mathrm{soft}$), 1.16 ($X_\mathrm{med}$), and 1.02
  ($X_\mathrm{hard}$).  Table\,\ref{t2} and \ref{t3} list
  both the observed and fully corrected counts in each band. The
  effect of the absorption correction on the X-ray CMD (Fig.~\ref{f3})
  is a uniform shift of the NGC\,6366 sources by 0.39\,units on the
  left axis and 0.50\,units on the bottom axis and a uniform shift of
  the M55 sources by 0.06\,units on the left axis and 0.07\,units on
  the bottom axis. The bottom and left axes give the X-ray color and
  magnitude without this shift (they do, however, include the small
  corrections for background subtraction and exposure variations).

  \subsection{Spectral Fitting}\label{s2.3}
  We fit all sources with absorbed power-law spectral models in
  \texttt{Sherpa} \citep{fds01} using \citet{cas79} statistics.  We
  fixed the column density to the value given in
  Table\,\ref{t1}, with only the power law photon index and
  normalization allowed to vary.  From the best fit spectra, we
  calculated the unabsorbed fluxes, given in Tables\,\ref{t2},
  \ref{t3}.

  For the point source NGC\,6366 CX1a, the fluxes quoted in
  Table\,\ref{t3} are from a fit of an absorbed power law to the
  unbinned spectrum, for which we use Cash statistics, and fix the
  absorption to that of NGC\,6366. This gave a power-law with photon
  index $3.5\pm0.6$. For the extended source NGC\,6366 CX1b we
  extracted the spectrum from within a $40\arcsec$ radius centered on
  CX1a. We subtracted the spectrum of the point source CX1a, and used
  a nearby, source-free, region with $1\arcmin$ radius to estimate the
  background. This gave a bremsstrahlung temperature of the extended
  component of $kT=2.0^{+1.3}_{-0.7}$\,keV, and unabsorbed fluxes as
  listed in Table\,\ref{t3}.

  \section{Optical observations}\label{s3}
  Optical observations of M55 and NGC\,6366 were obtained with the
  FORS2 instrument at the Unit Telescope 1 (UT1) of the ESO VLT in
  April and May 2005. Both globular clusters were observed in three
  filters, $B$, $R$ and H$\alpha$, with exposure times chosen to
  maximize the dynamic range. Table\,\ref{t4} provides a
  condensed log of the observations. FORS2 is a mosaic of two
  2k$\times$4k chips with a pixel scale of
  $0\farcs126$\,pix$^{-1}$. For the majority of the observations, we
  used $2\times2$ on-chip binning, providing a pixel scale of
  $0\farcs252$\,pix$^{-1}$, except when the seeing was below
  $0\farcs6$, when no binning was applied.
  
  The images were reduced using the Munich Image Data Analysis System
  (MIDAS). All images were bias-subtracted and flatfielded using
  twilight flats. Next, we grouped the images sharing the same
  combination of chip, filter and exposure time. The images in each
  group were aligned using integer pixel offsets and co-added to
  remove cosmic rays and increase the signal-to-noise. Hence, for each
  cluster we obtained 18 separate stacked images, 9 for each chip.

  Both M55 and NGC\,6366 are part of a survey of globular clusters
  with ACS/WFC on board the \emph{Hubble Space Telescope}. M55 was
  observed for 284\,s in a $V$-band filter (\texttt{F606W}) and 384\,s
  in an $I$-band filter (\texttt{F814W}). For NGC\,6366 the exposure
  times with 570\,s in both \texttt{F606W} and
  \texttt{F814W}. Compared to FORS2, \emph{ACS/WFC} has a smaller
  field-of-view (about $3\farcm4\times3\farcm4$) and contains only a
  few of the X-ray sources.  For M55, CX2, CX3, CX8, CX9, CX11, CX12,
  CX13 and CX16 are coincident with the $202\arcsec\times202\arcsec$
  field of view, while for NGC\,6366 CX1 through CX5 are coincident.

  \begin{figure}
    \resizebox{0.95\hsize}{!}{\includegraphics{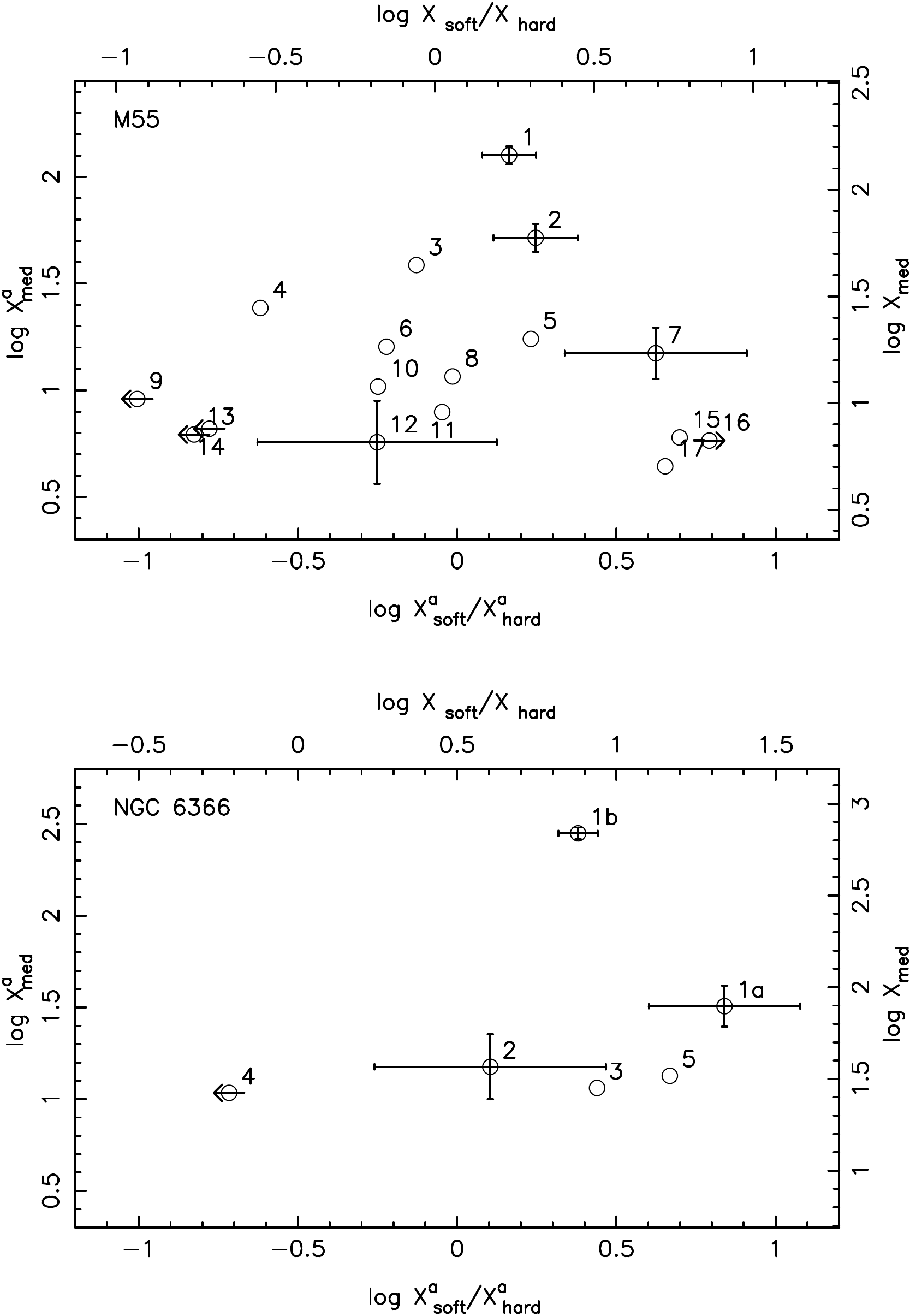}}
    \caption{X-ray color-magnitude diagram for the X-ray sources
    detected in the half-mass radius of M55 \textit{(top panel)}
    and NGC\,6366 \textit{(bottom panel}). The X-ray color is defined
    as the logarithm of the ratio of $X_\mathrm{soft}$ (0.5--1.5\,keV)
    counts to $X_\mathrm{hard}$ (1.5--6.0\,keV) counts. The X-ray
    magnitude is the logarithm of $X_\mathrm{med}$ (0.5--4.5\,keV)
    counts. The bottom and left axes provide the observed absorbed color and
    magnitude scales ($^\mathrm{a}$).}
    \label{f3}
  \end{figure}

  \subsection{Photometry}
  We used DAOPHOT II, running inside MIDAS, for the photometry of the
  stacked FORS2 images. We followed the recommendations of
  \citet{ste87} and obtained instrumental magnitudes using
  point-spread-function (PSF) fitting. The PSF for each stacked image
  was determined in an iterative manner. First, we used an analytic
  PSF to remove stars in the proximity of those stars used to
  determine the PSF. This subsequently improved the PSF for the next
  iteration. We found that allowing the final PSF to quadratically
  depend on the position on the chip decreased the fitting residuals.
  We used the final PSF to determine instrumental magnitudes of all
  stars above a $5\sigma$ detection threshold. The resulting
  star-subtracted image was then searched for objects that were missed
  in the first iteration. These stars were added to the star list
  after which the PSF-fitting process was repeated.

  In order to match the stars found on the separate stacked images, we
  employed the following method. First, we determined approximate
  coordinate transformations between the separate images taken with
  the same chip of each cluster to create a single list containing all
  stars. Next, we removed multiple entries of the same star from the
  list by removing stars located with 3\,pixels of another star. The
  resulting master star list was then matched against the star lists
  for each of the separate images, again using the approximate
  coordinate transformations. We determined improved, 6 parameter,
  coordinate transformations from the matched master list and repeated
  the process of creating a master list and matching stars. 

\begin{figure*}
    \centering
    \includegraphics[width=17cm]{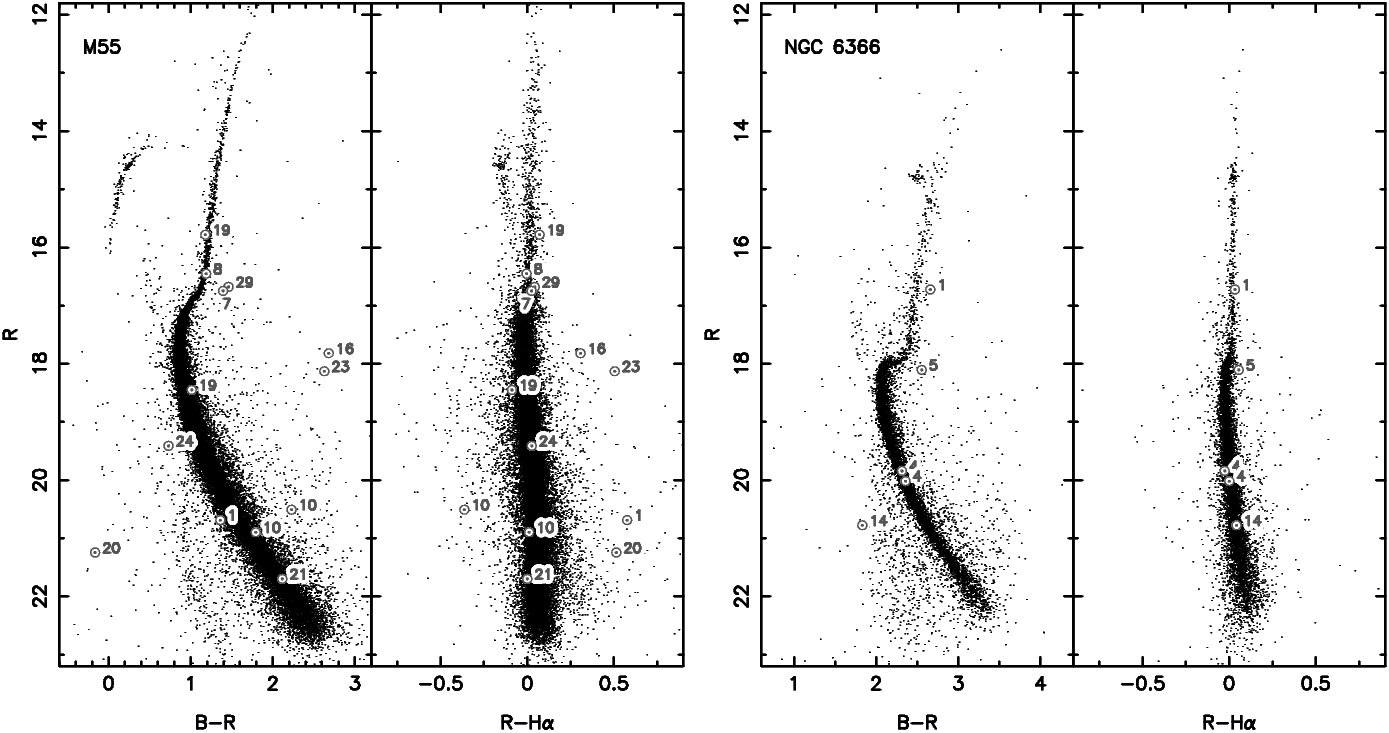}
    \caption{Optical color-magnitude diagrams for M55 and NGC\,6366
    constructed from the FORS2 data. Candidate optical counterparts to
    the \textit{Chandra} X-ray sources are indicated and numbered.}
    \label{f4}
  \end{figure*}

  Finally, the instrumental magnitudes for the separate chips were
  calibrated against photometric standards in the Mark\,A (for M55)
  and the Rubin\,152 fields (for NGC\,6366), using calibrated
  magnitudes from \citet{ste00} and fitting for zero-point and colour
  coefficients. Extinction coefficients in $B$ and $R$ were taken from
  the FORS2 webpage. The final rms residuals of the calibrations were
  0.03\,mag in $B$ and 0.04\,mag in $R$ for the calibration of the
  NGC\,6366 observations and 0.06\,mag in $B$ and 0.04\,mag in $R$ for
  the M55 observations. The H$\alpha$ magnitudes were calibrated by
  adopting $R-$H$\alpha=0$ for the bulk of the stars. Fig.~\ref{f4}
  shows colour-magnitude diagrams (CMDs) for both globular clusters.

  For both clusters, we estimate $5\sigma$ limiting magnitudes of
  $B=26.2$ and H$\alpha=24.8$. The $R$-band magnitudes are limited
  through the calibration, which relies on the instrumental $B-R$
  colors and which is not available if a star is not detected in the
  $B$-band images. Hence, the $R$-band limit depends on the
  $B-R$ color through $R=26.2-(B-R)$.

  For the ACS/WFC observations we used the photometry presented in
  \citet{asb+08}, who use highly specialized methods to determine
  accurate photometry of ACS/WFC images. For our purposes, we have
  used the photometry that was transferred from the ACS/WFC filters
  into ground-based $V$ and $I$-band magnitudes. Fig.\,\ref{f5} shows
  the CMDs for both globular clusters.

  \begin{table}
    \begin{minipage}[t]{\columnwidth}
      \centering
      \caption[]{A log of the optical observations. Here, $\sigma$
      denotes the seeing and $\sec z$ the airmass.}\label{t4}
      \renewcommand{\footnoterule}{}
      \begin{tabular}{l@{\hspace{0.2cm}}
	  l@{\hspace{0.3cm}}
	  l@{\hspace{0.2cm}}
	  c@{\hspace{0.15cm}}
	  c@{\hspace{0.15cm}}
	}
	\hline \hline
	\multicolumn{2}{l}{Date \& Time (UT)} & \multicolumn{1}{l}{Exp. ($B$, $R$, H$\alpha$)} & $\sigma$ ($\arcsec$) & $\sec z$ \\
	\hline
	\multicolumn{5}{l}{NGC\,6366} \\
        April 8 & 07:40--09:20 & $4\times$  (5\,m,  2\,m, 15\,m) & 0.6--0.7 & 1.06--1.13\\
        April 9 & 07:22--08:10 & $2\times$  (5\,m,  2\,m, 15\,m) & 0.7--1.0 & 1.09--1.16\\
                & 08:13--08:24 & $6\times$  (3\,s,  1\,s,  9\,s) & 0.7--1.1 & 1.07--1.08\\
                & 08:25--08:47 & $6\times$ (30\,s, 12\,s, 90\,s) & 0.6--0.8 & 1.06--1.07\\[0.5em]
	\multicolumn{5}{l}{M55} \\
        April 10 & 08:01--09:51 & $4\times$  (5\,m,  2\,m, 15\,m)$^\mathrm{a}$ & 0.4--0.6 & 1.06--1.33\\
           May 6 & 06:37--06:49 & $5\times$  (3\,s,  1\,s,  9\,s)              & 0.6--0.8 & 1.22--1.26\\
                 & 06:50--07:05 & $4\times$ (30\,s, 12\,s, 90\,s)              & 0.7--0.9 & 1.18--1.22\\
          May 12 & 07:07--08:00 & $2\times$  (5\,m,  2\,m, 15\,m)$^\mathrm{a}$ & 0.4--0.5 & 1.04--1.11\\
	\hline
	\multicolumn{5}{l}{$^\mathrm{a}$ These observations were obtained with $1\times1$ binning.}
      \end{tabular}
    \end{minipage}
  \end{table}

  \subsection{Astrometry}
  We aim to tie the stacked $R$-band images for each globular cluster
  and each CCD to the International Celestial Reference System (ICRS),
  to astrometrically calibrate our photometric catalog and to allow
  for a comparison between the X-ray and optical frames. For each
  globular cluster and each CCD, we calibrated the stacked 1\,s
  $R$-band images against UCAC2 astrometric standards
  \citep{zuz+04}. Even though the stellar density in the cores of
  these globular clusters is higher than a typical stellar field, our
  astrometry did not suffer from severe blending of stars, and for
  each image only 3 to 6 stars of the 40 to 80 that overlapped with
  each image were rejected. Fitting for zero-point position, scale and
  position angle, we obtained astrometric solutions for each image
  that had root-mean-square (rms) residuals of $0\farcs04$ to
  $0\farcs07$ in right ascension and declination. These solutions were
  transferred to the deeper $R$-band images using the calibrated
  positions of stars on the 1\,s images and used to calibrate the
  photometric catalog for each globular cluster. Typically, each
  solution used some 1300 to 1900 stars, resulting in rms residuals of
  $0\farcs01$ to $0\farcs03$.

  The drizzled ACS/WFC \texttt{F606W} images were calibrated against
  the FORS2 observations. Due to the large number of stars (500 to
  700) coincident between the FORS2 and ACS/WFC images, the
  astrometric calibration has rms uncertainties of less than
  $0\farcs02$ in right ascension and declination.

  To place the \emph{Chandra} frame onto the ICRS we use
  identifications based on the \emph{Chandra} X-ray sources alone. We
  identify M55 CX1 with the dwarf nova CV1 discovered by
  \citet{kpt+05}. From the position of the optical counterpart, using
  the astrometry presented above, we determine a \textit{Chandra}
  bore-sight correction of $\Delta\alpha=-0\farcs270\pm0\farcs044$ and
  $\Delta\delta=+0\farcs080\pm0\farcs045$. For NGC\,6366, we find that
  CXO\,J172729.9$-$050524 coincides with the bright star ($V=10.7$)
  BD$-04\degr4280$ (PPM\,706759; \citealt{rb88}). This star is
  saturated in our observations, but an accurate position is provided
  in the 2nd version of the USNO CCD Astrograph Catalog (UCAC2;
  \citealt{zuz+04}). The bore-sight correction is
  $\Delta\alpha=+0\farcs030\pm0\farcs091$ and
  $\Delta\delta=-0\farcs256\pm0\farcs082$. These offsets put the X-ray
  positions onto the International Celestial Reference System
  (ICRS). We note that the uncertainty in both corrections is
  dominated by the uncertainty in the X-ray position. Both
  corrections, however, are within the $0\farcs6$ (90\% confidence)
  accuracy in the pointing of \textit{Chandra} \citep{akc+00}.

\begin{figure}
    \resizebox{0.95\hsize}{!}{\includegraphics{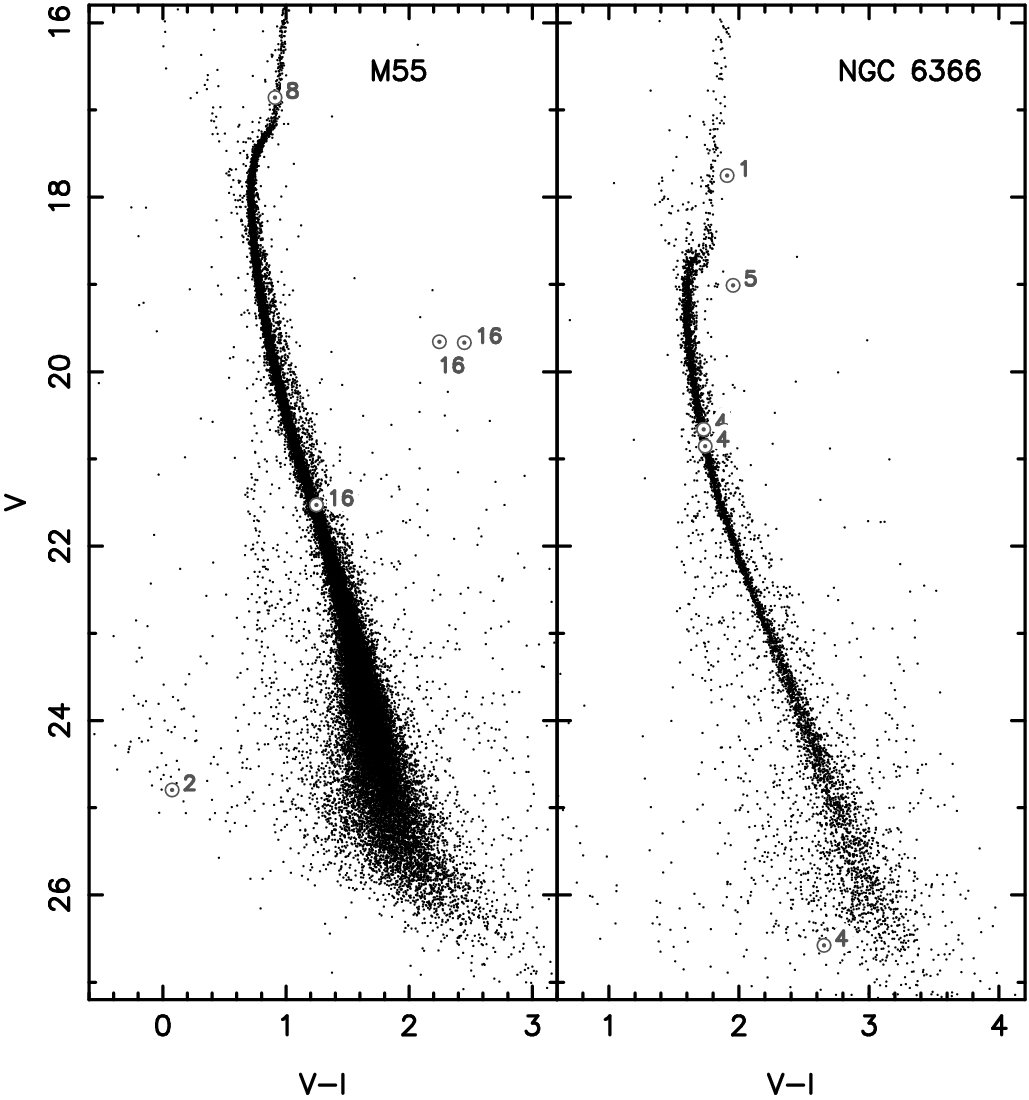}}
    \caption{Optical color-magnitude diagrams for M55 and NGC\,6366
      constructed from the \citet{asb+08} photometry. Candidate
      optical counterparts to the \textit{Chandra} X-ray sources are
      indicated and numbered.}
    \label{f5}
  \end{figure}

  With the optical and X-ray astrometry, the final uncertainty on the
  position of a X-ray source is the quadratic sum of the centroiding
  uncertainty of the X-ray source (those tabulated in Table\,\ref{t2}
  and \ref{t3}), the uncertainty in the X-ray bore-sight correction,
  and the uncertainty in the astrometry of the $R$-band image. The
  resulting $1\sigma$ uncertainties on the X-ray positions range
  between $0\farcs1$ for the brightest X-ray source to $0\farcs7$ for
  the fainter sources. In order to identify the optical counterparts
  to the X-ray sources, we treat every star inside the 99\% confidence
  error circle as a potential counterpart. These stars are indicated
  in the CMDs in Fig.~\ref{f4} and Fig.~\ref{f5} and finding charts
  are provided in Fig.~\ref{f6}.

  \section{Source classification}
  As in our previous papers \citep{plh+02,bph+04,kbp+06}, we first use
  our astrometry to identify stars within the, in this case,
  99\%\ confidence circles as possible optical counterparts to the
  X-ray sources. We then check whether these stars have unusual
  photometric properties as expected for candidate
  counterparts. Finally we combine the optical photometry of the
  candidate counterparts, the X-ray properties of the X-ray sources
  and the combined $f_\mathrm{X}/f_\mathrm{opt}$ to classify the
  probable counterparts. These steps are illustrated with
  Figures~\ref{f6}, \ref{f4}, \ref{f3} and \ref{f7}, respectively.
  For the latter figure, we estimate the visual magnitude as
  $V=(B+R)/2$ where no $V$-band magnitudes are available. The
  photometry and the offset between the \textit{Chandra} and optical
  positions are given for selected optical counterparts in
  Table\,\ref{t5}.

  \subsection{M55}
\begin{figure*}
    \centering
    \includegraphics[angle=270,width=2.8cm]{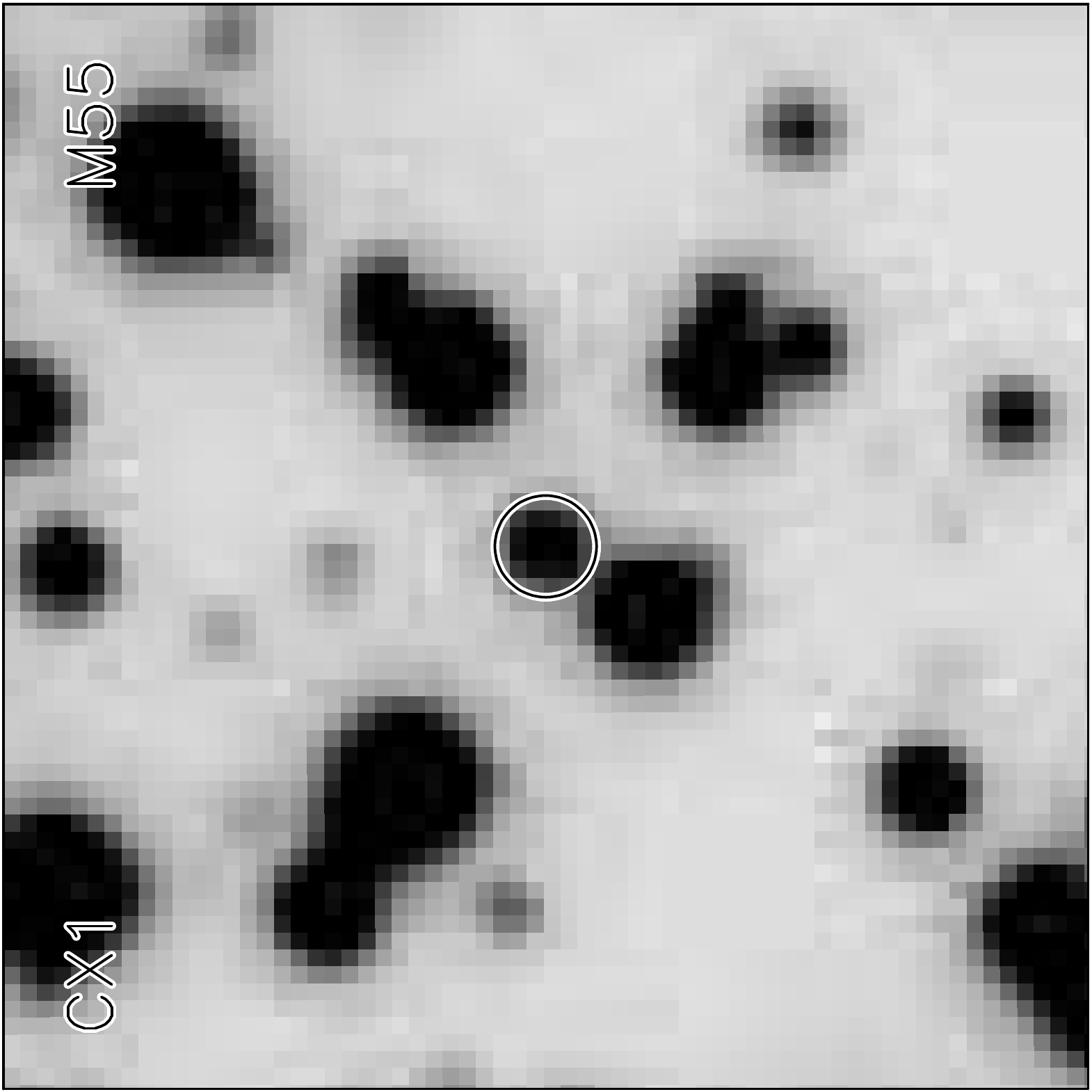}
    \includegraphics[angle=270,width=2.8cm]{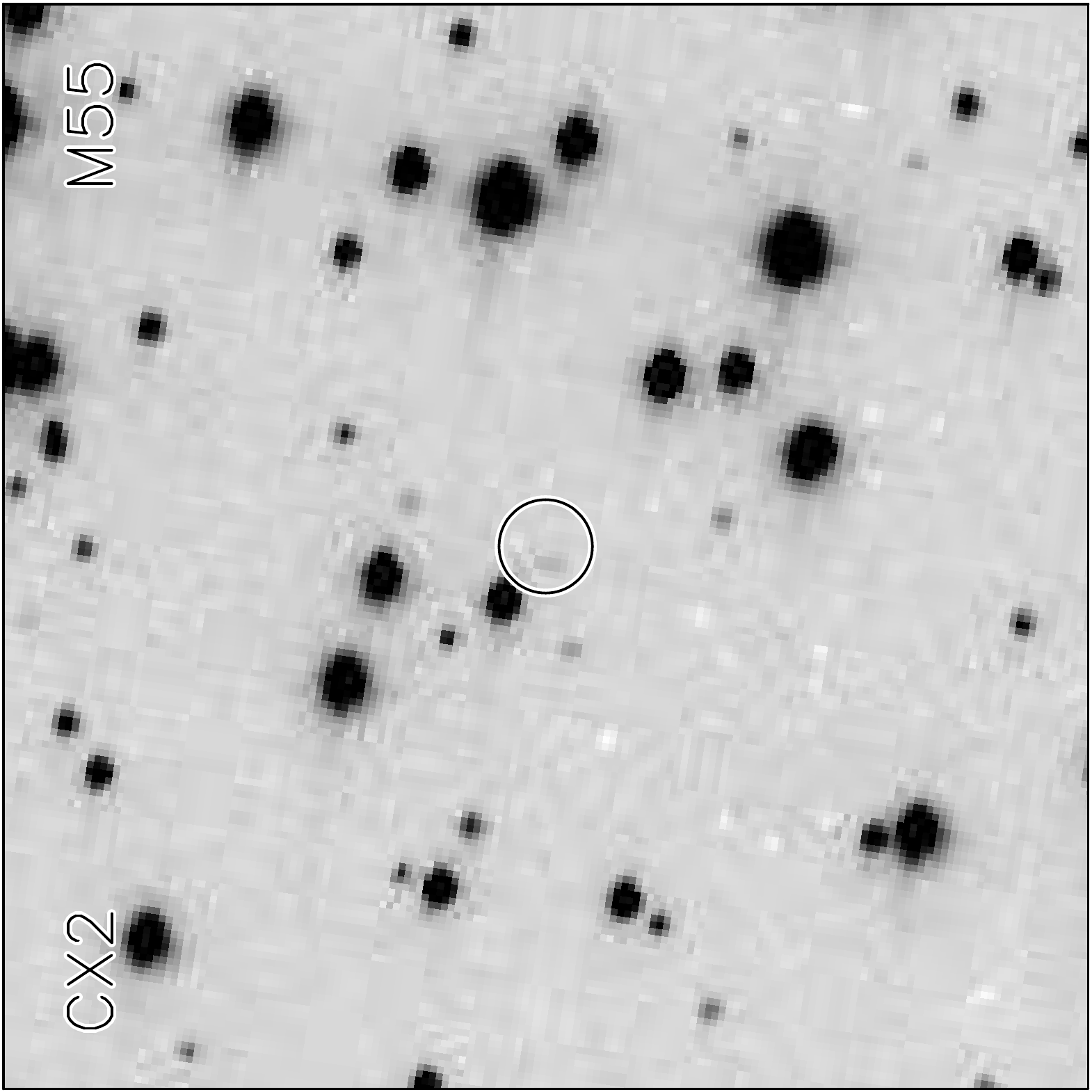}
    \includegraphics[angle=270,width=2.8cm]{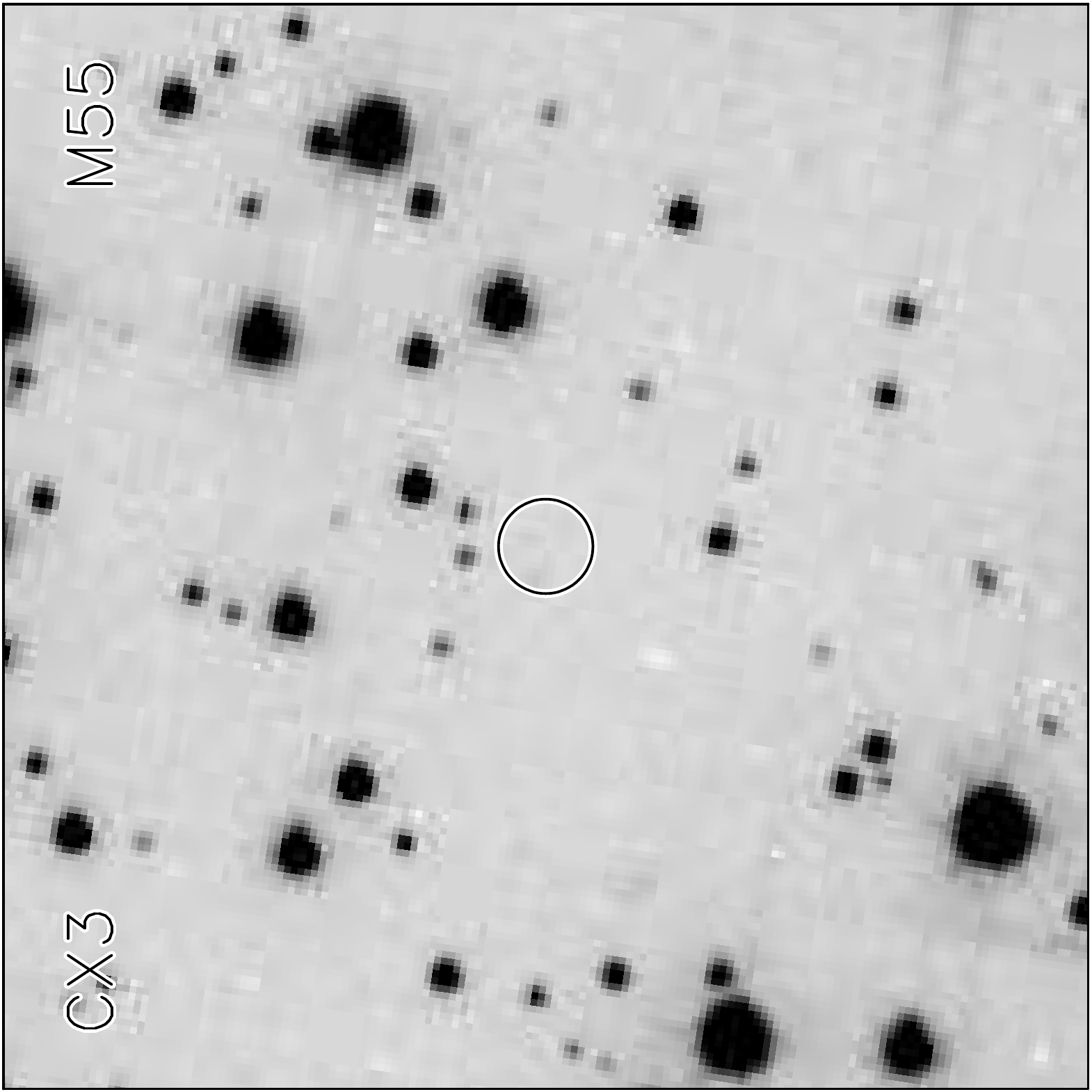}
    \includegraphics[angle=270,width=2.8cm]{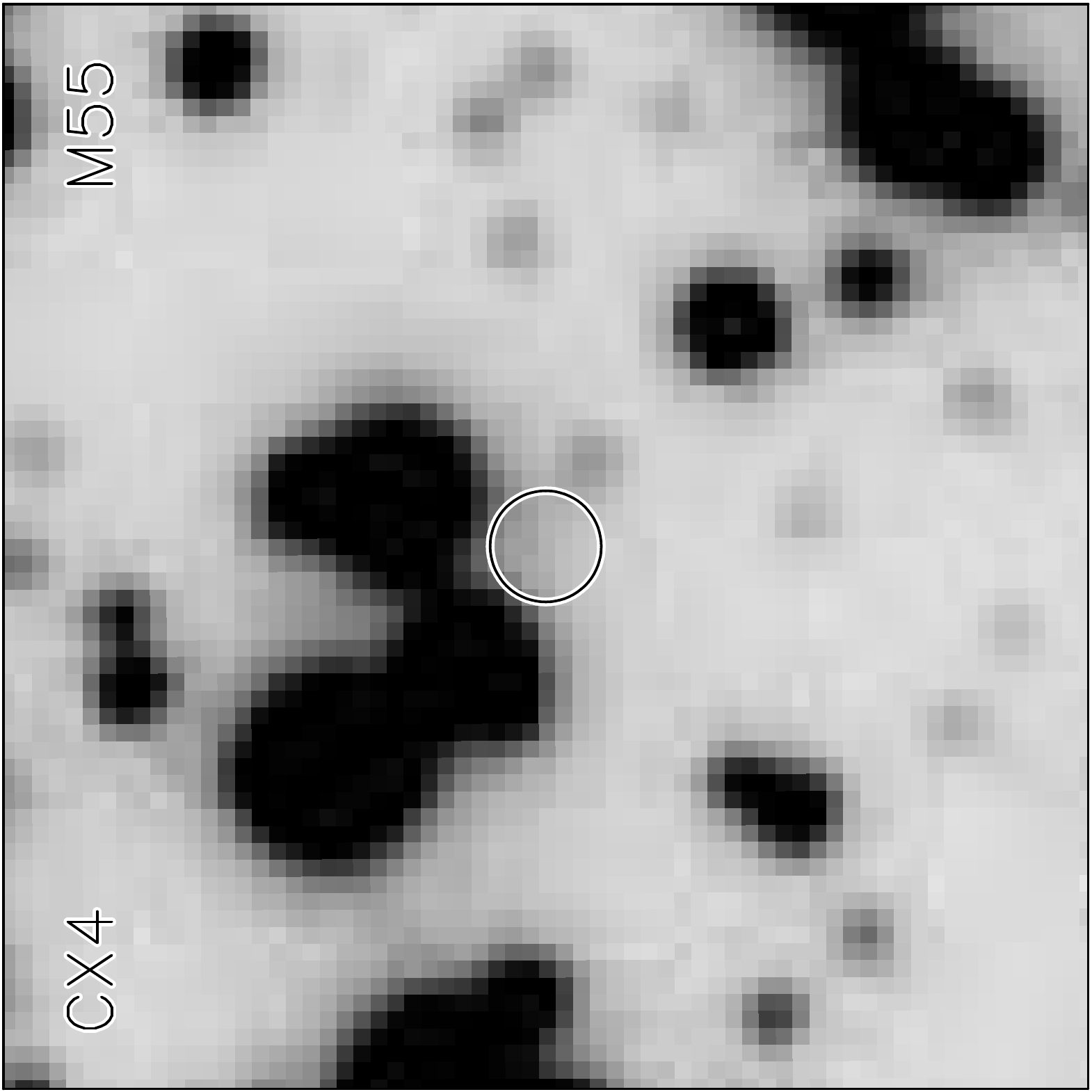}
    \includegraphics[angle=270,width=2.8cm]{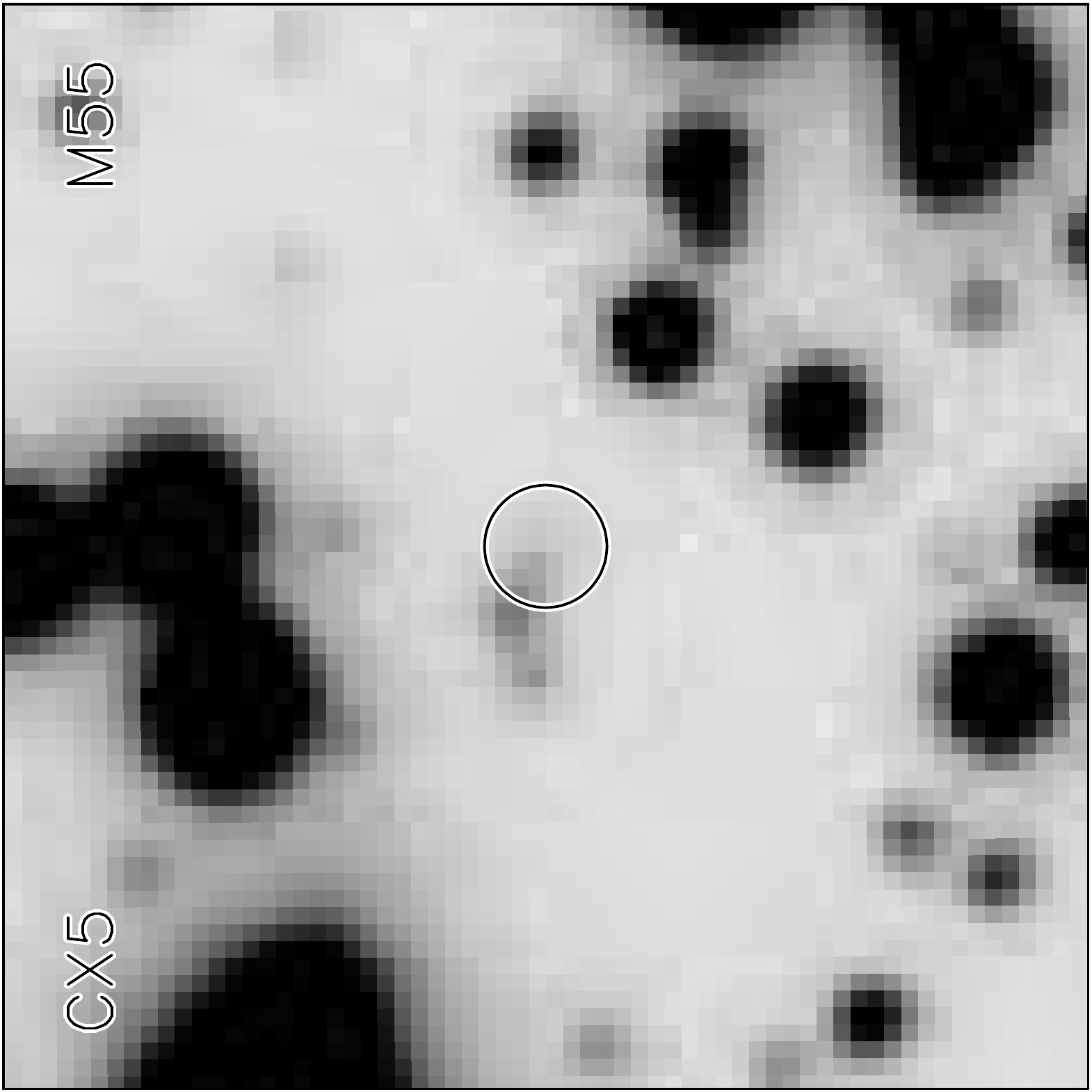}
    \includegraphics[angle=270,width=2.8cm]{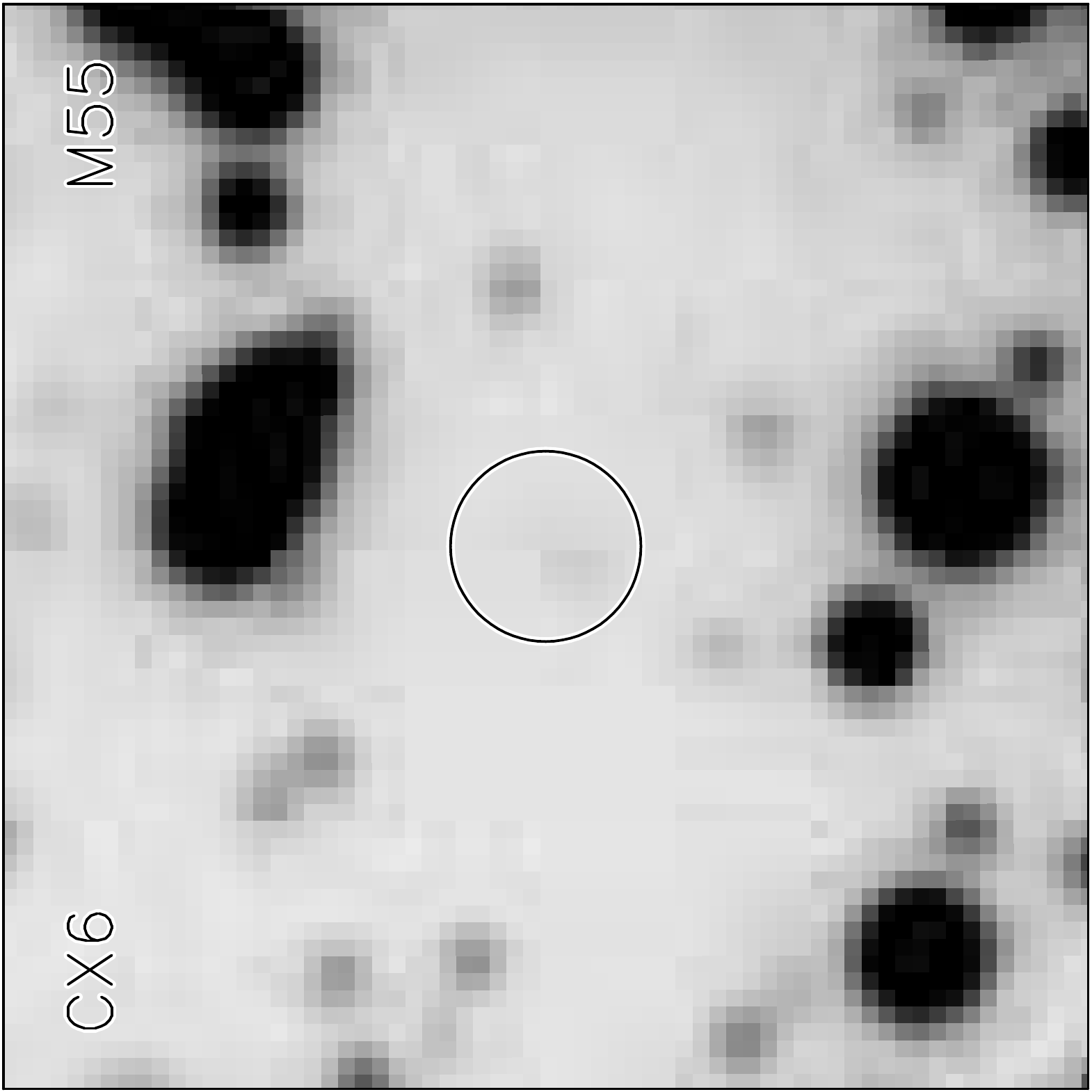}\\
    \includegraphics[angle=270,width=2.8cm]{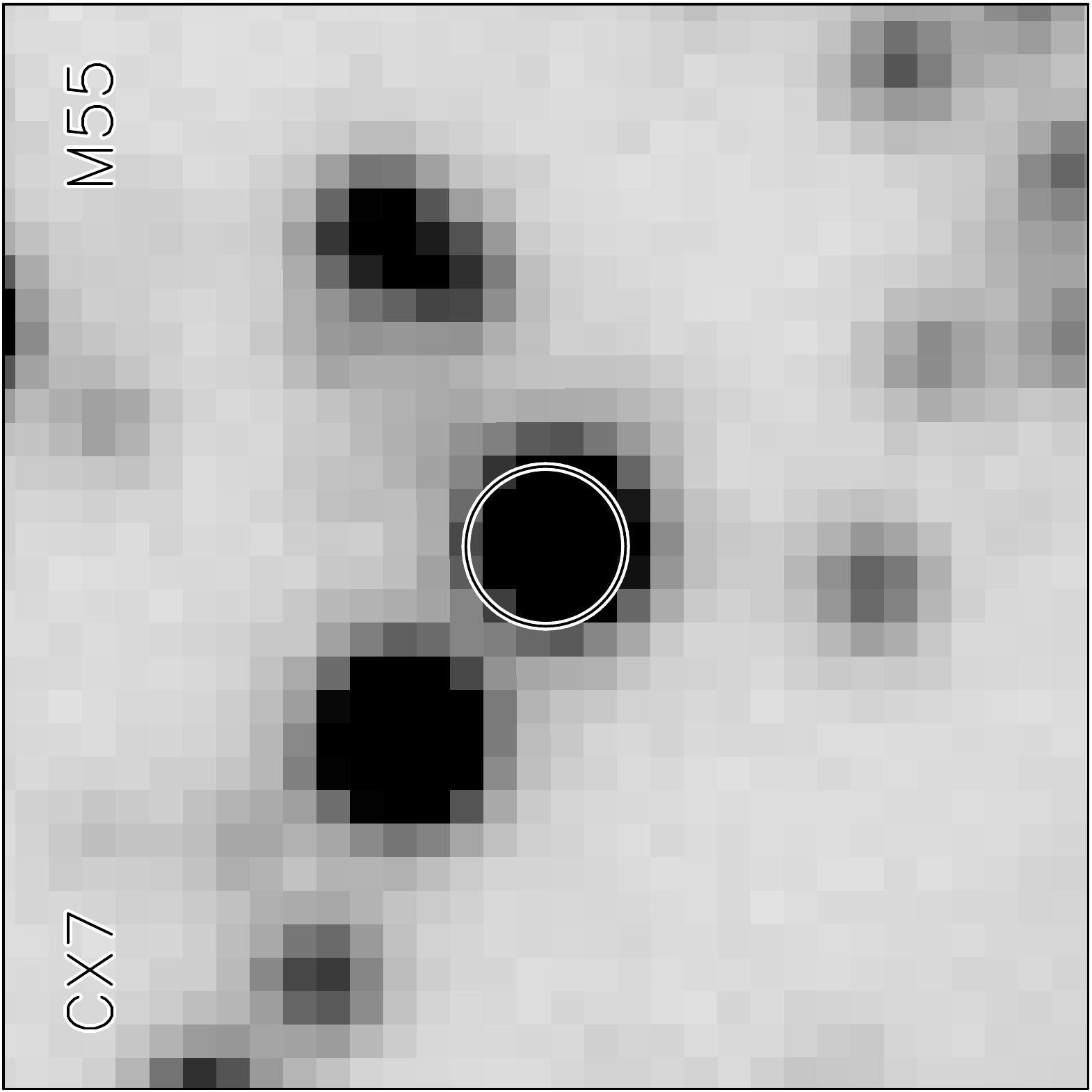}
    \includegraphics[angle=270,width=2.8cm]{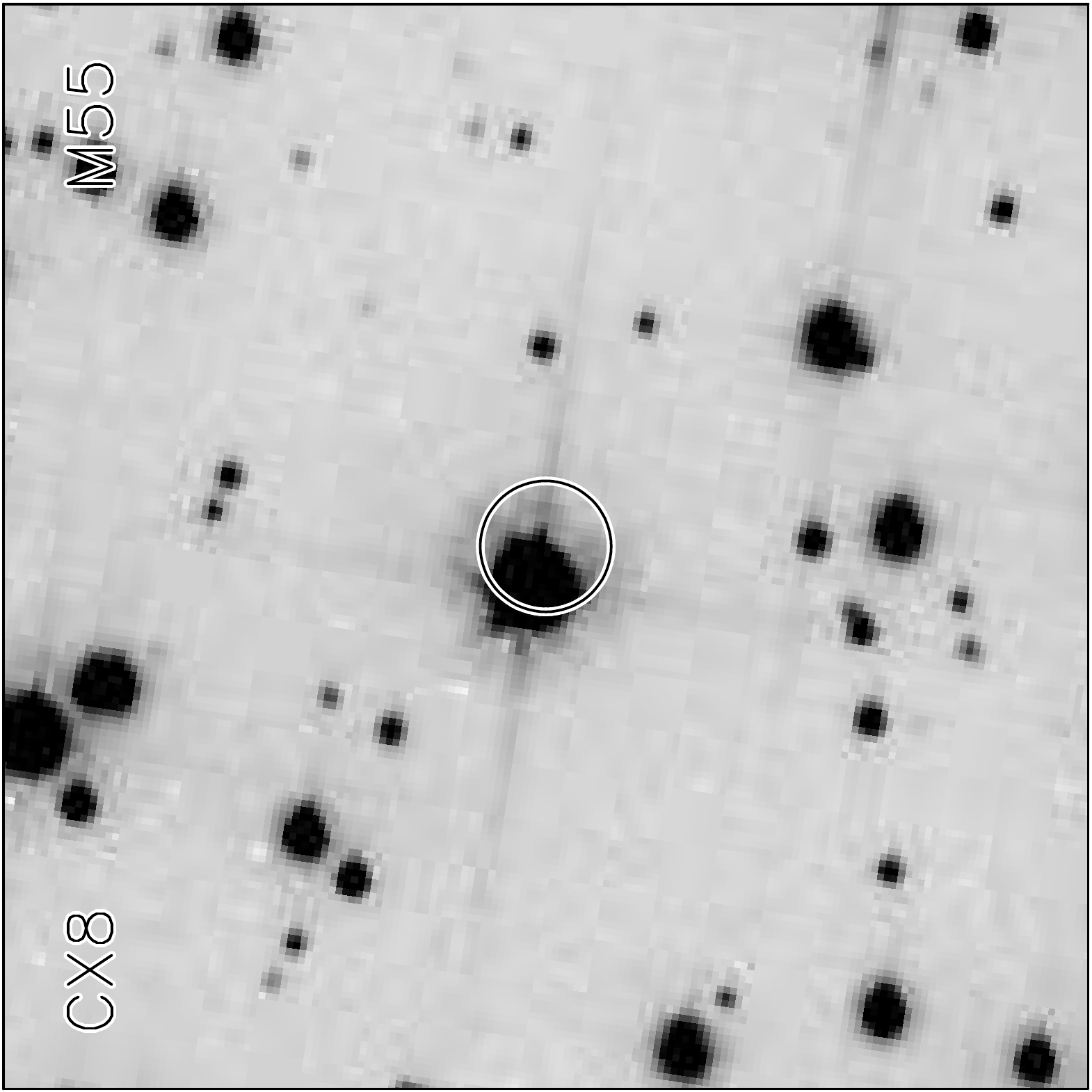}
    \includegraphics[angle=270,width=2.8cm]{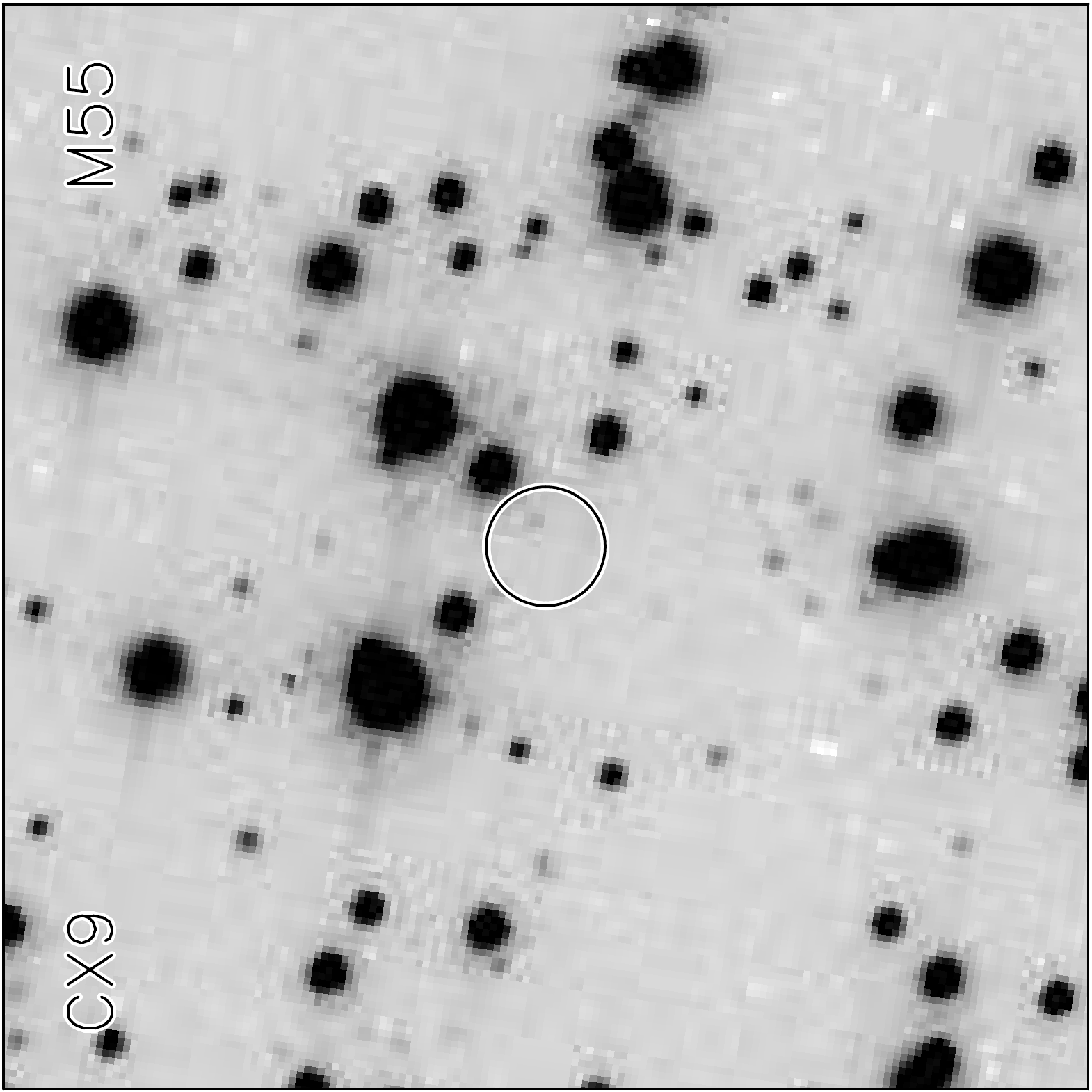}
    \includegraphics[angle=270,width=2.8cm]{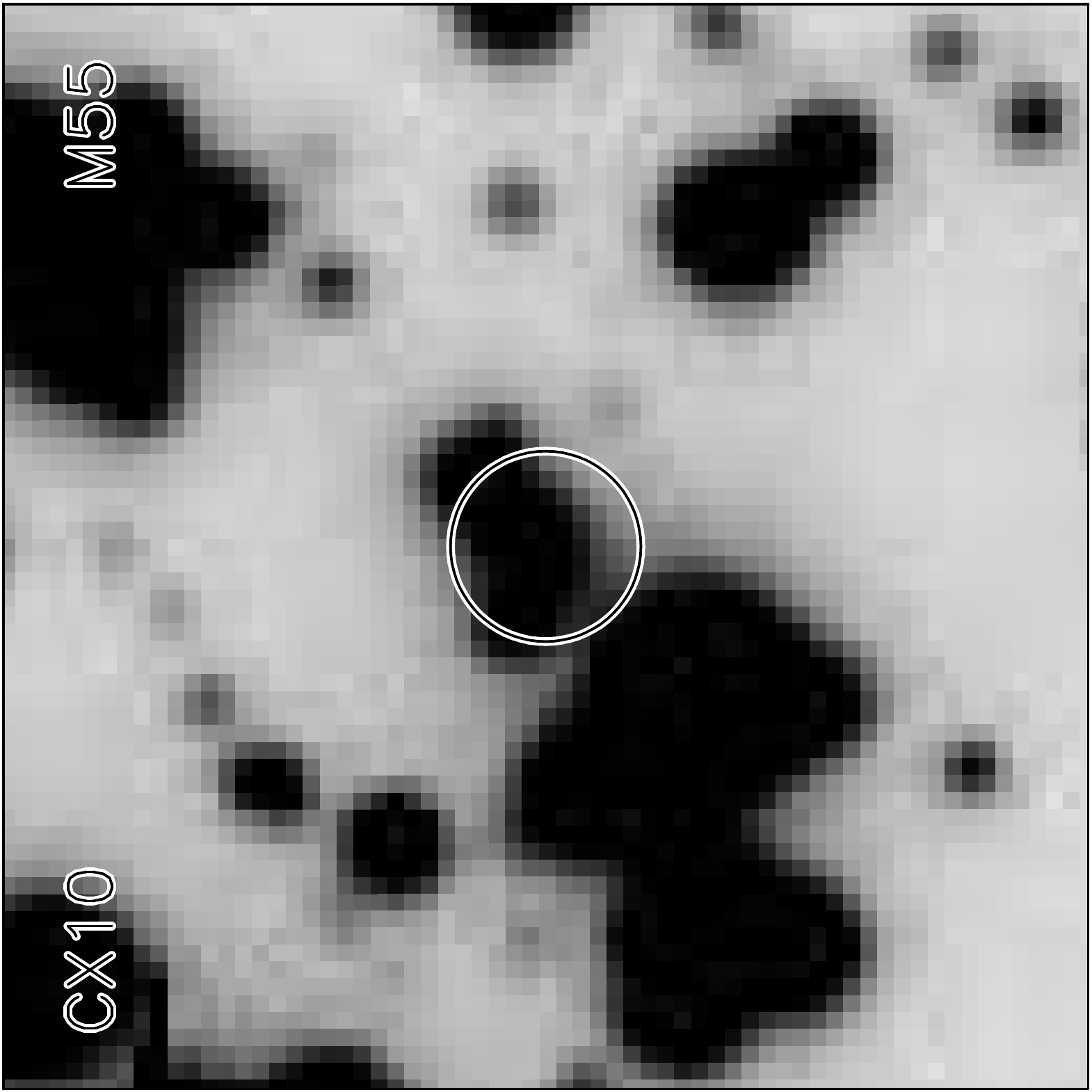}
    \includegraphics[angle=270,width=2.8cm]{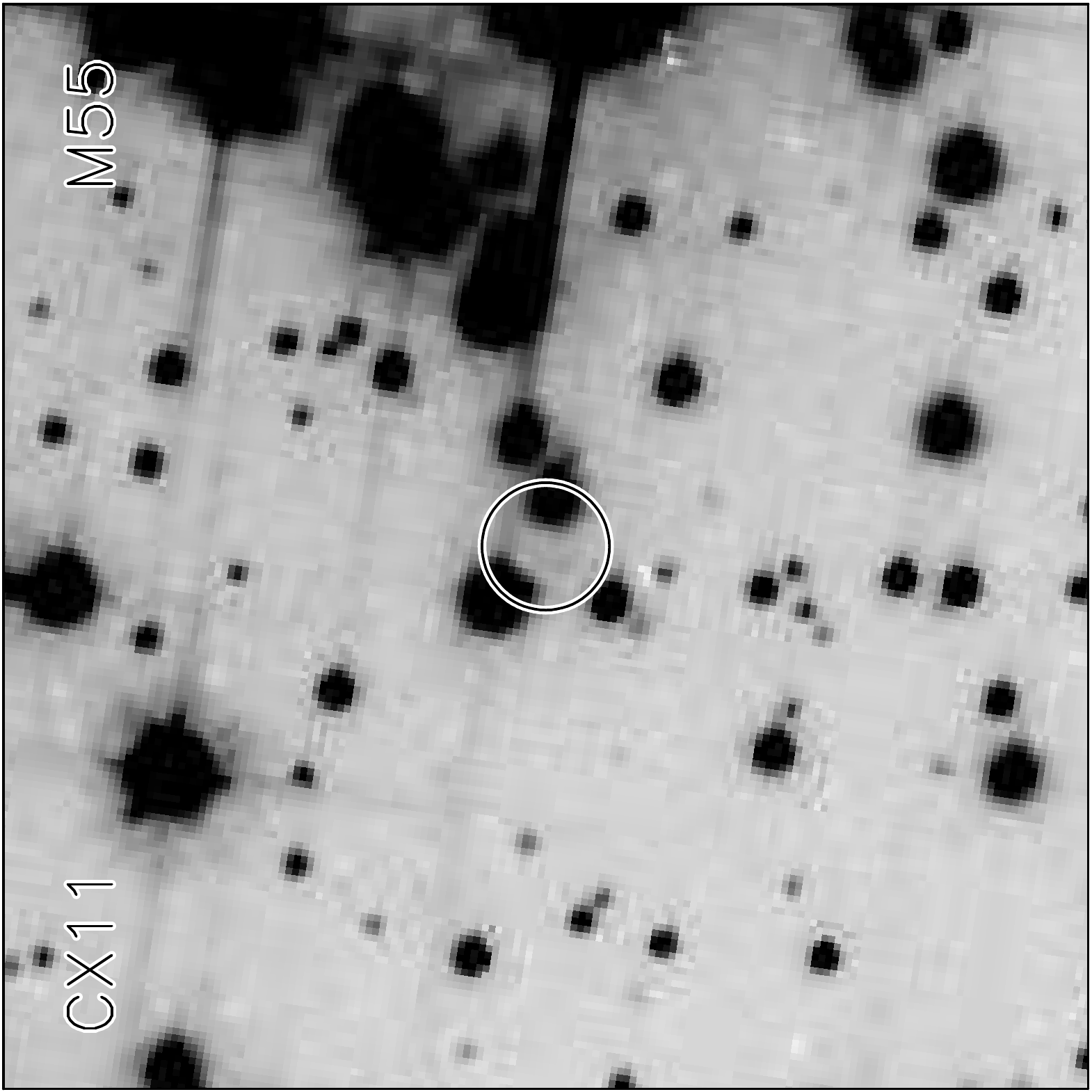}
    \includegraphics[angle=270,width=2.8cm]{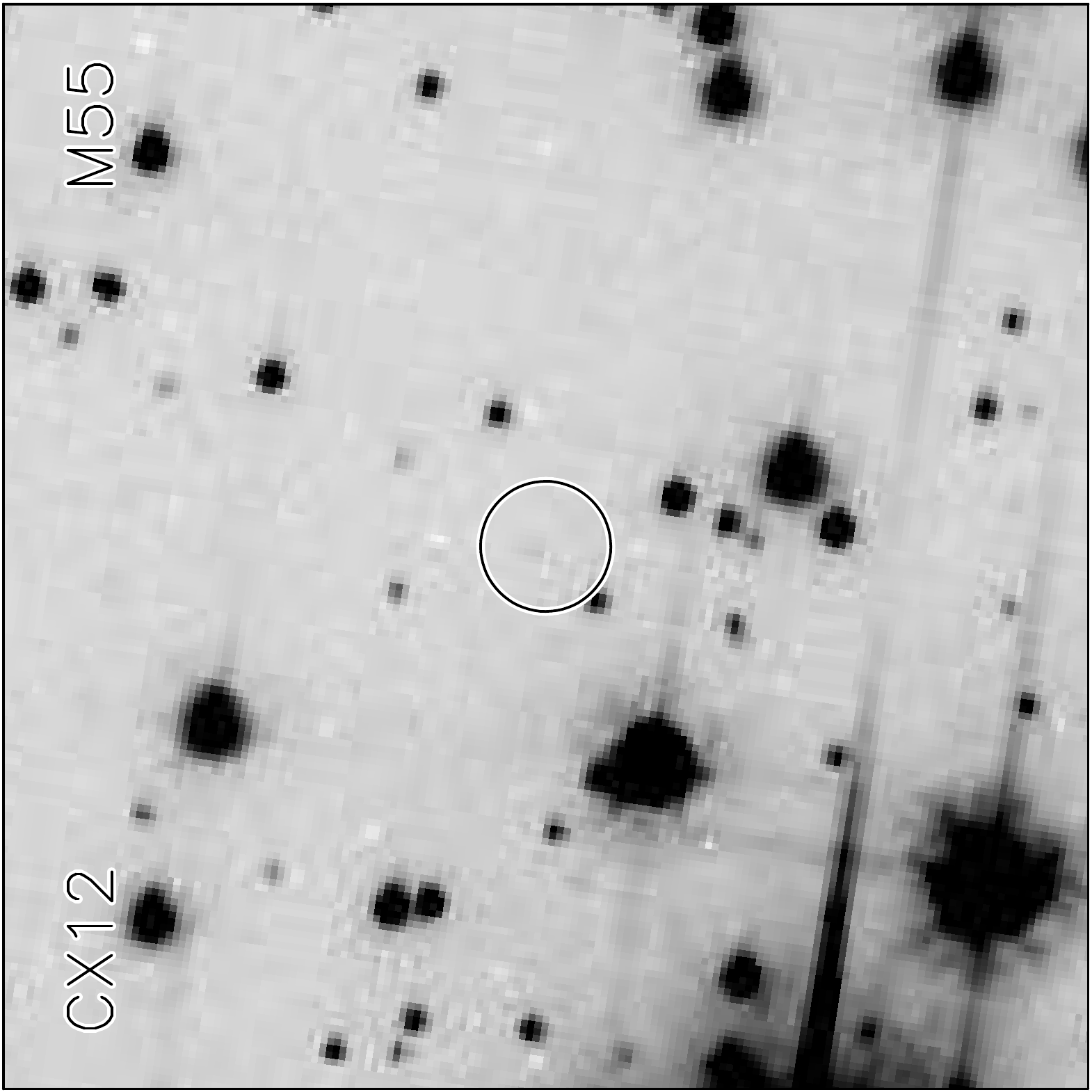}\\
    \includegraphics[angle=270,width=2.8cm]{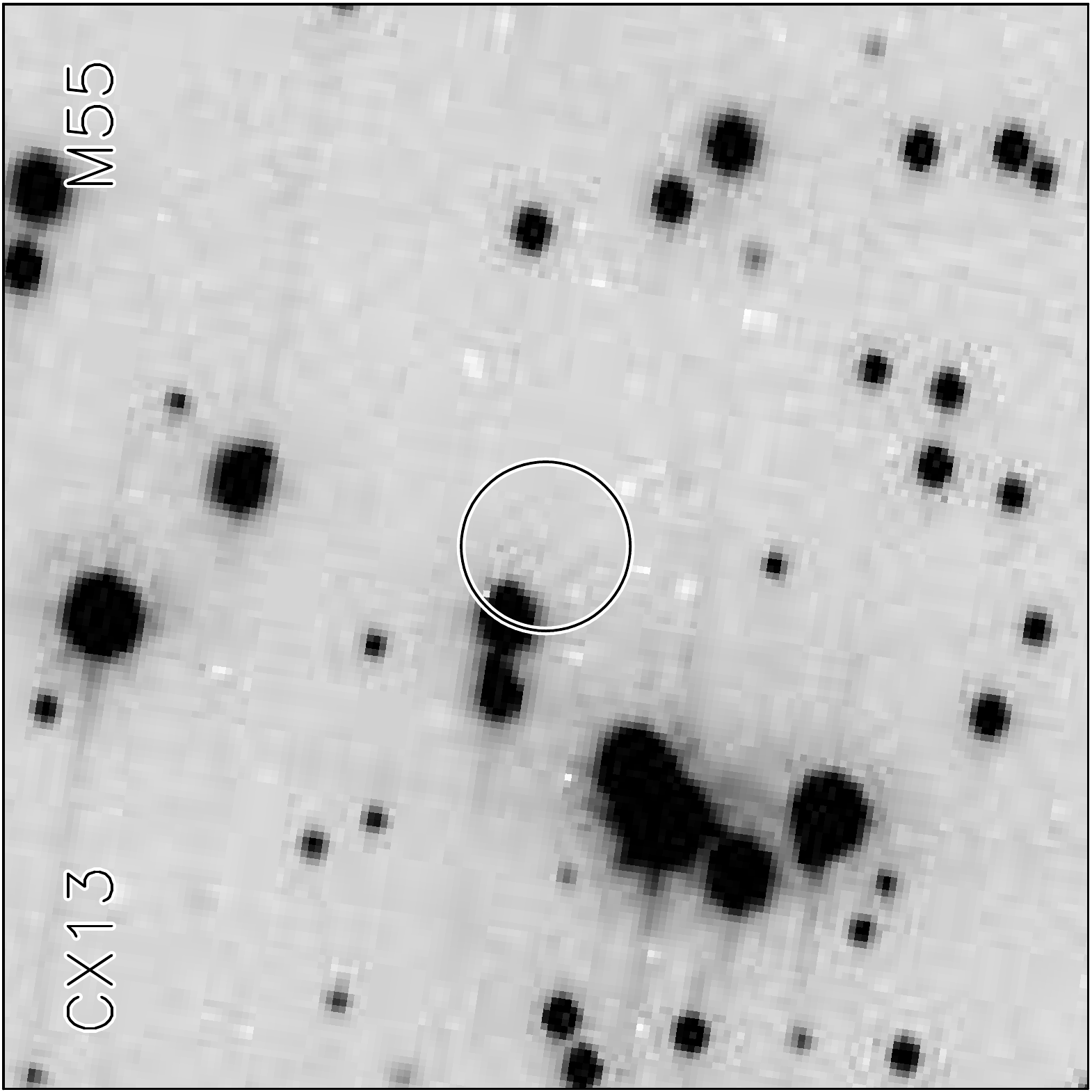}
    \includegraphics[angle=270,width=2.8cm]{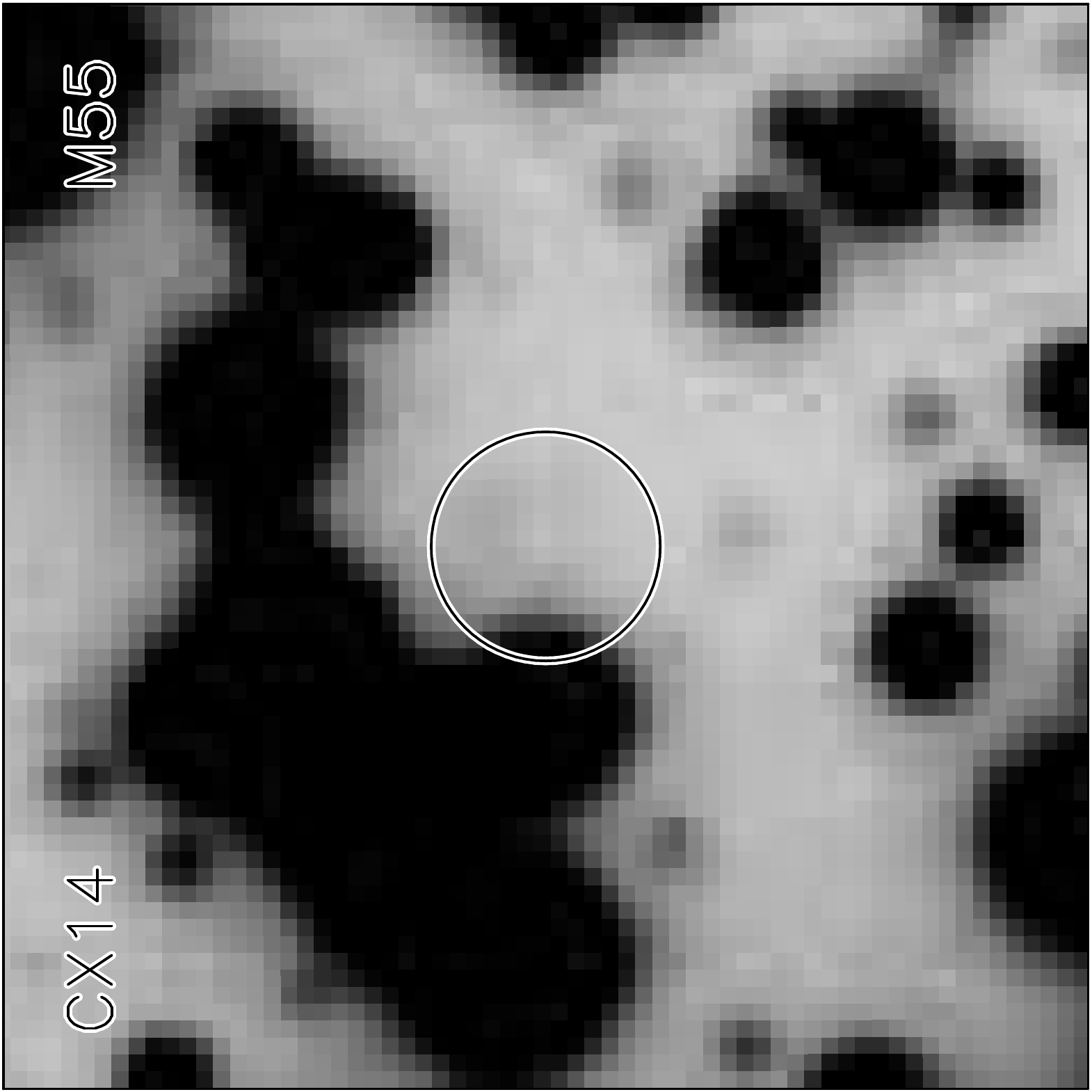}
    \includegraphics[angle=270,width=2.8cm]{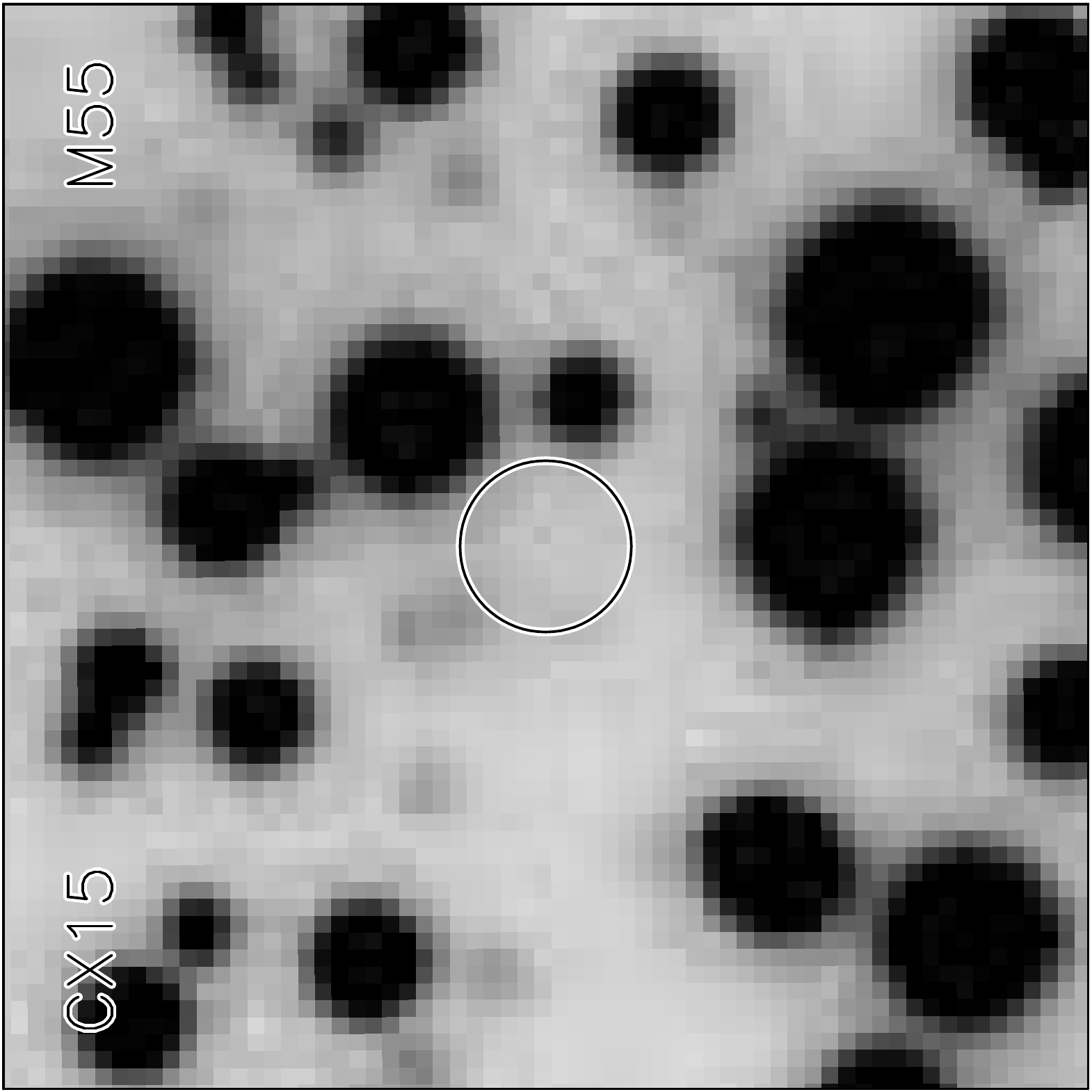}
    \includegraphics[angle=270,width=2.8cm]{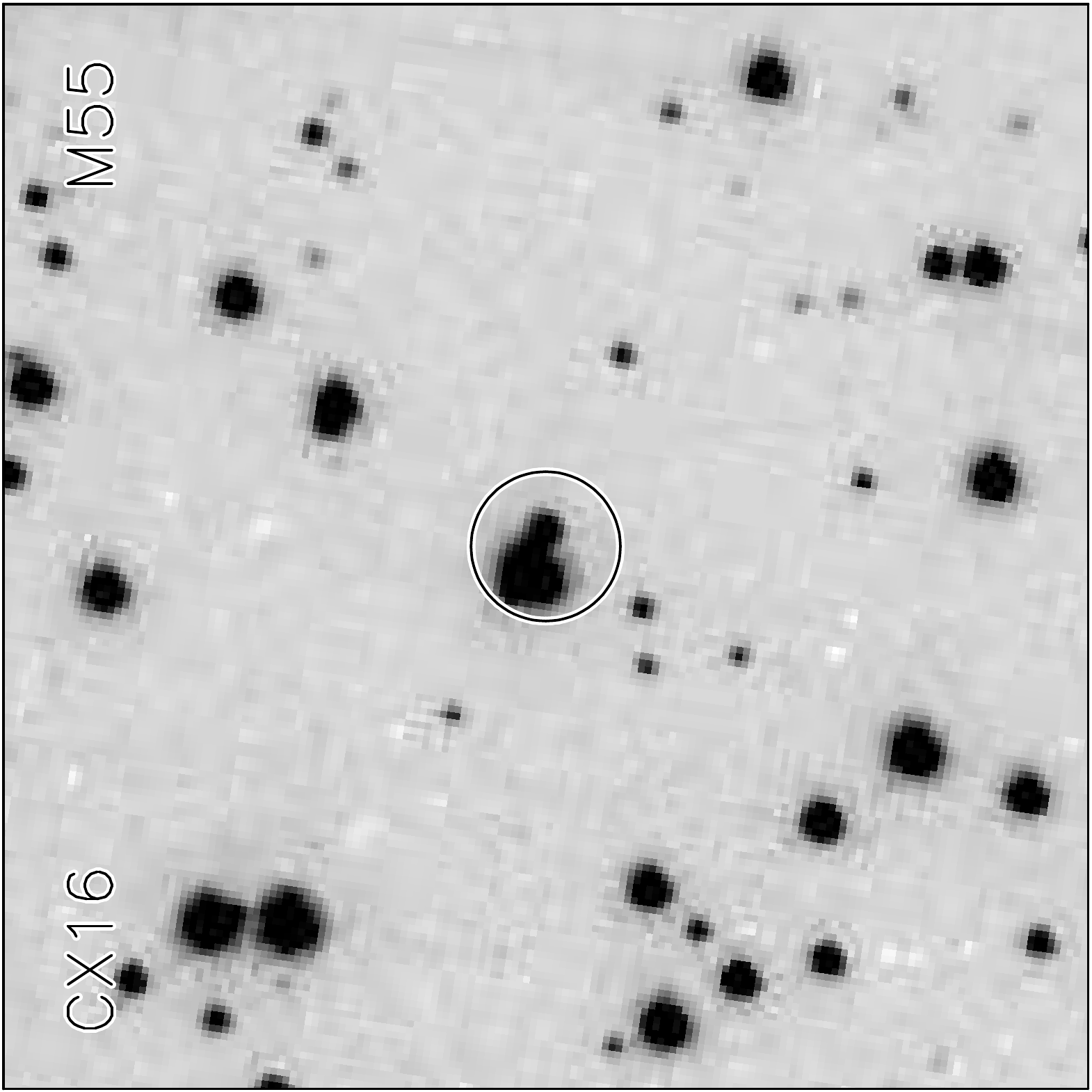}
    \includegraphics[angle=270,width=2.8cm]{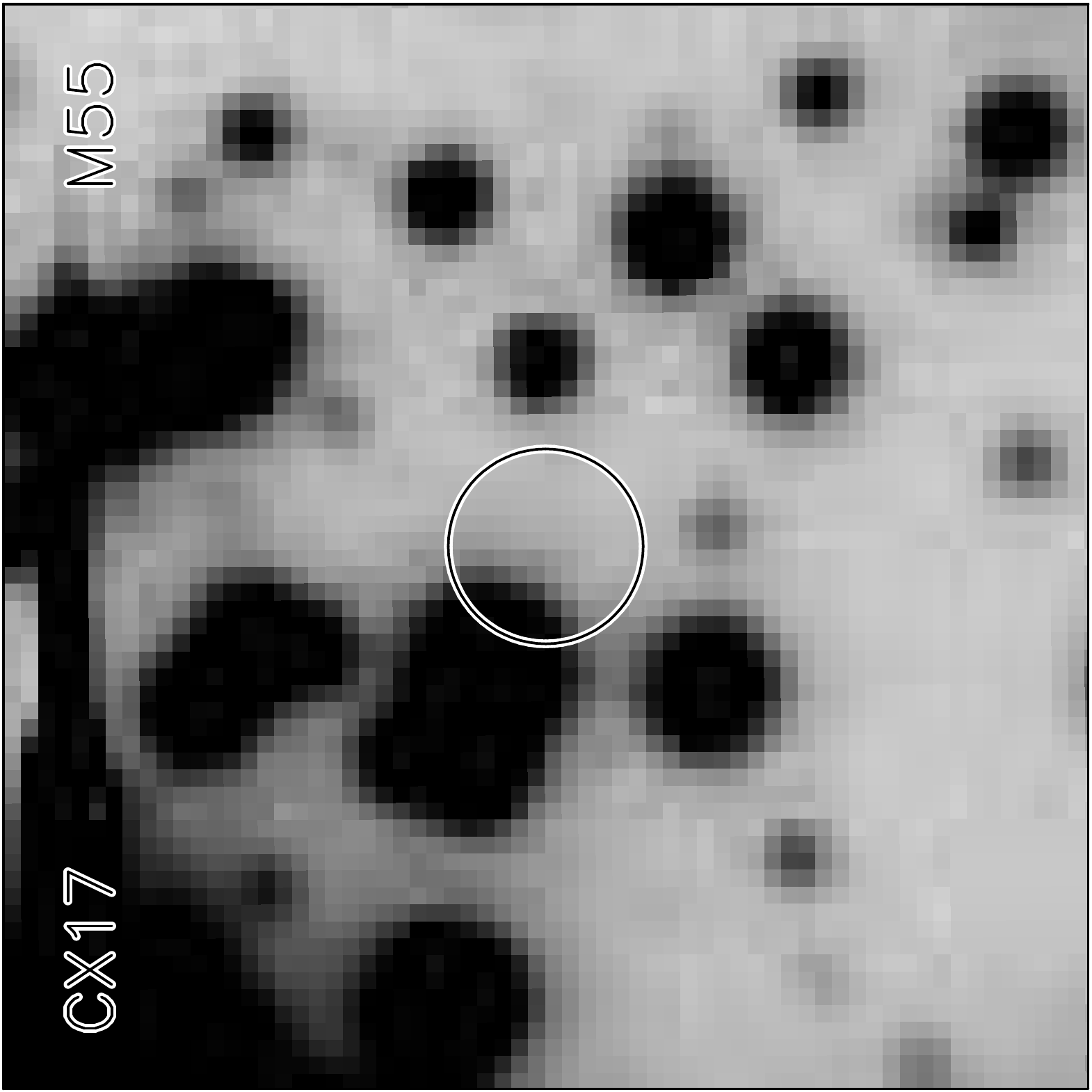}
    \includegraphics[angle=270,width=2.8cm]{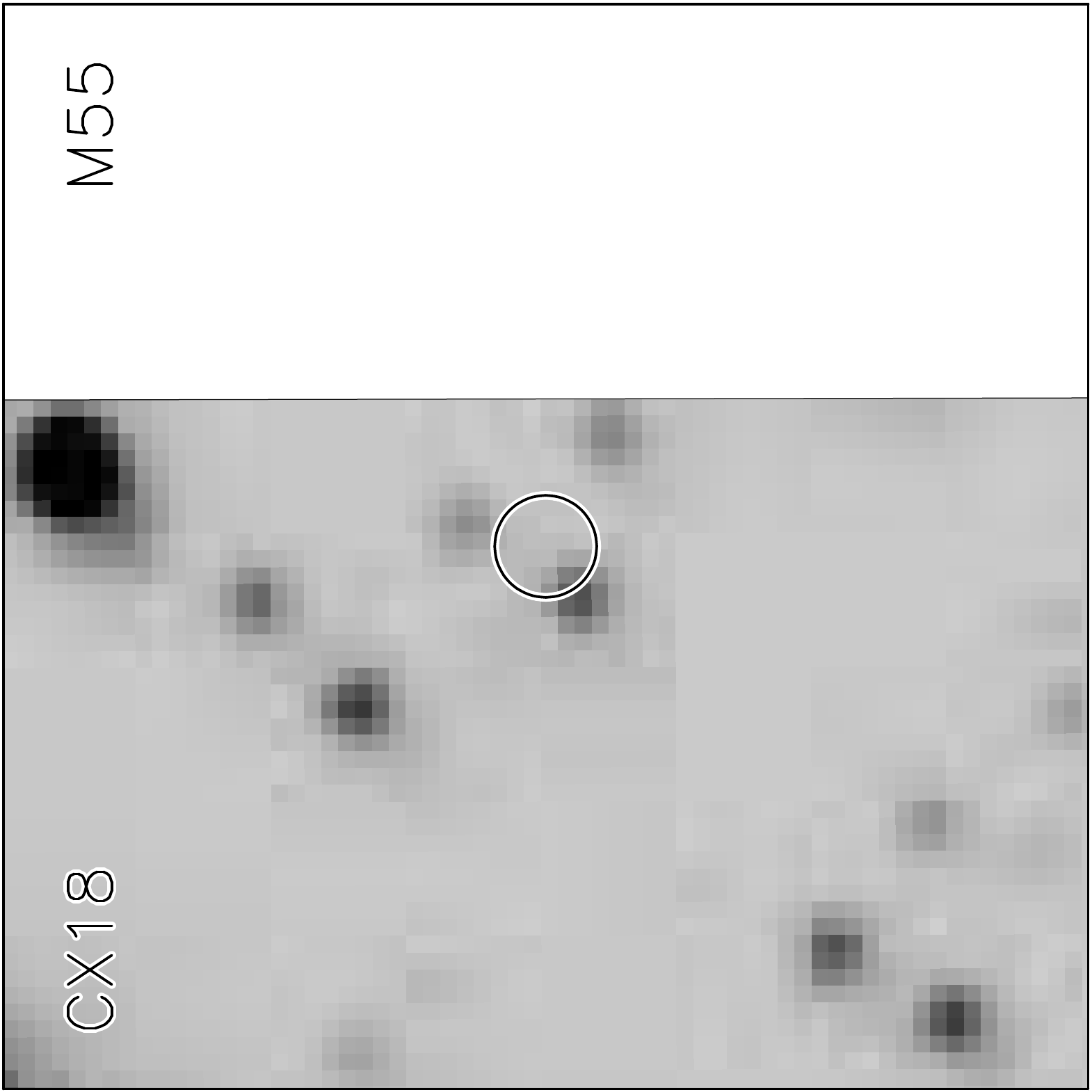}\\
    \includegraphics[angle=270,width=2.8cm]{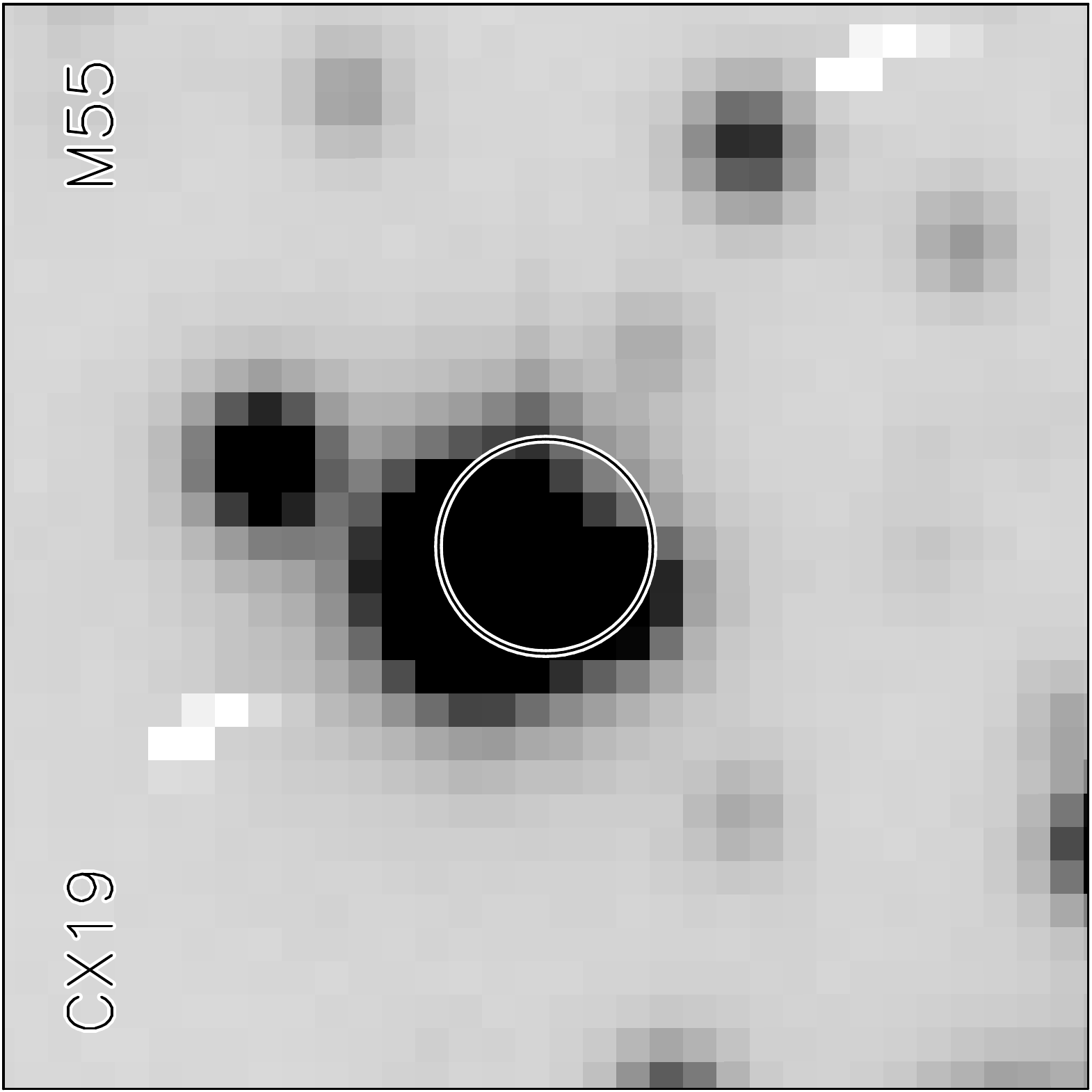}
    \includegraphics[angle=270,width=2.8cm]{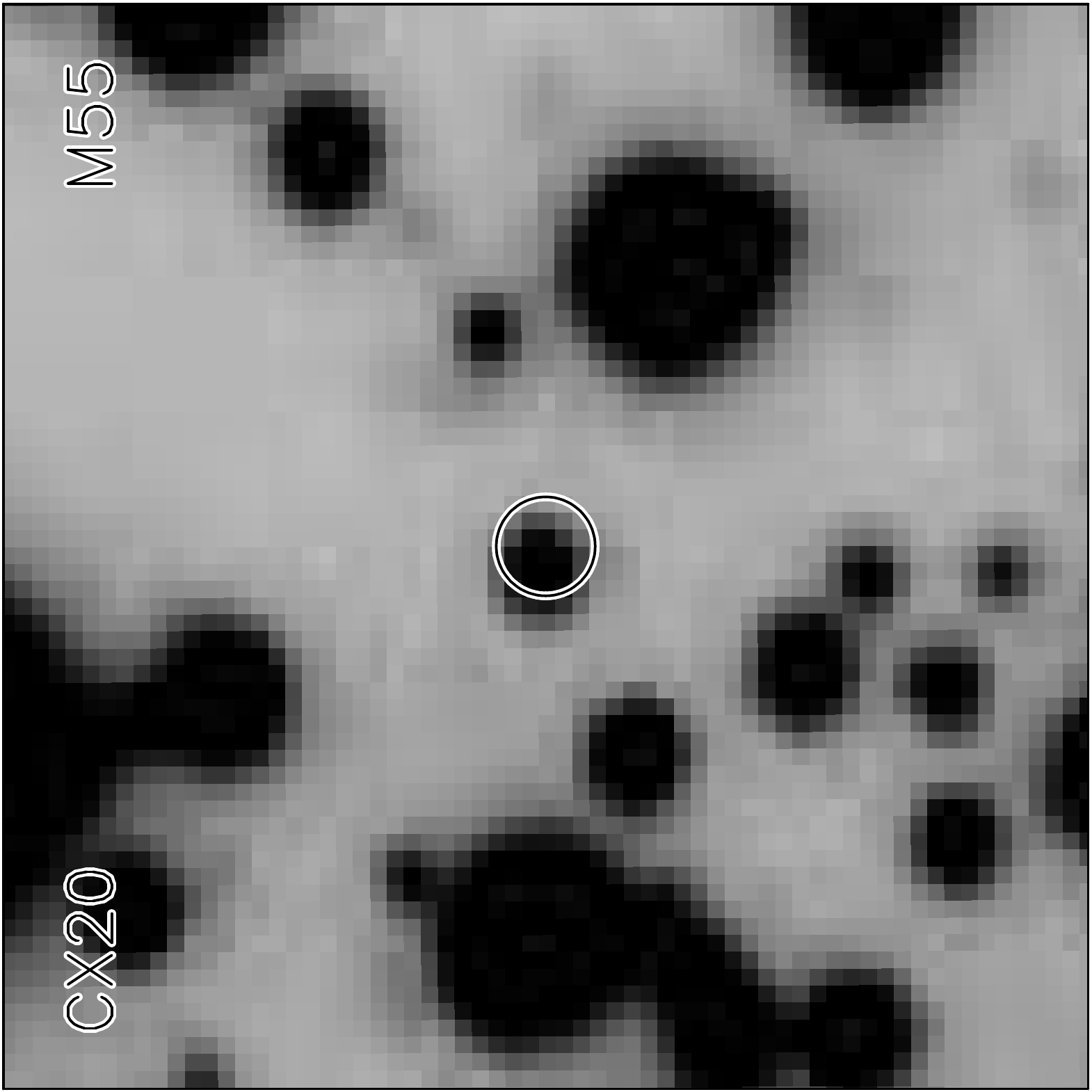}
    \includegraphics[angle=270,width=2.8cm]{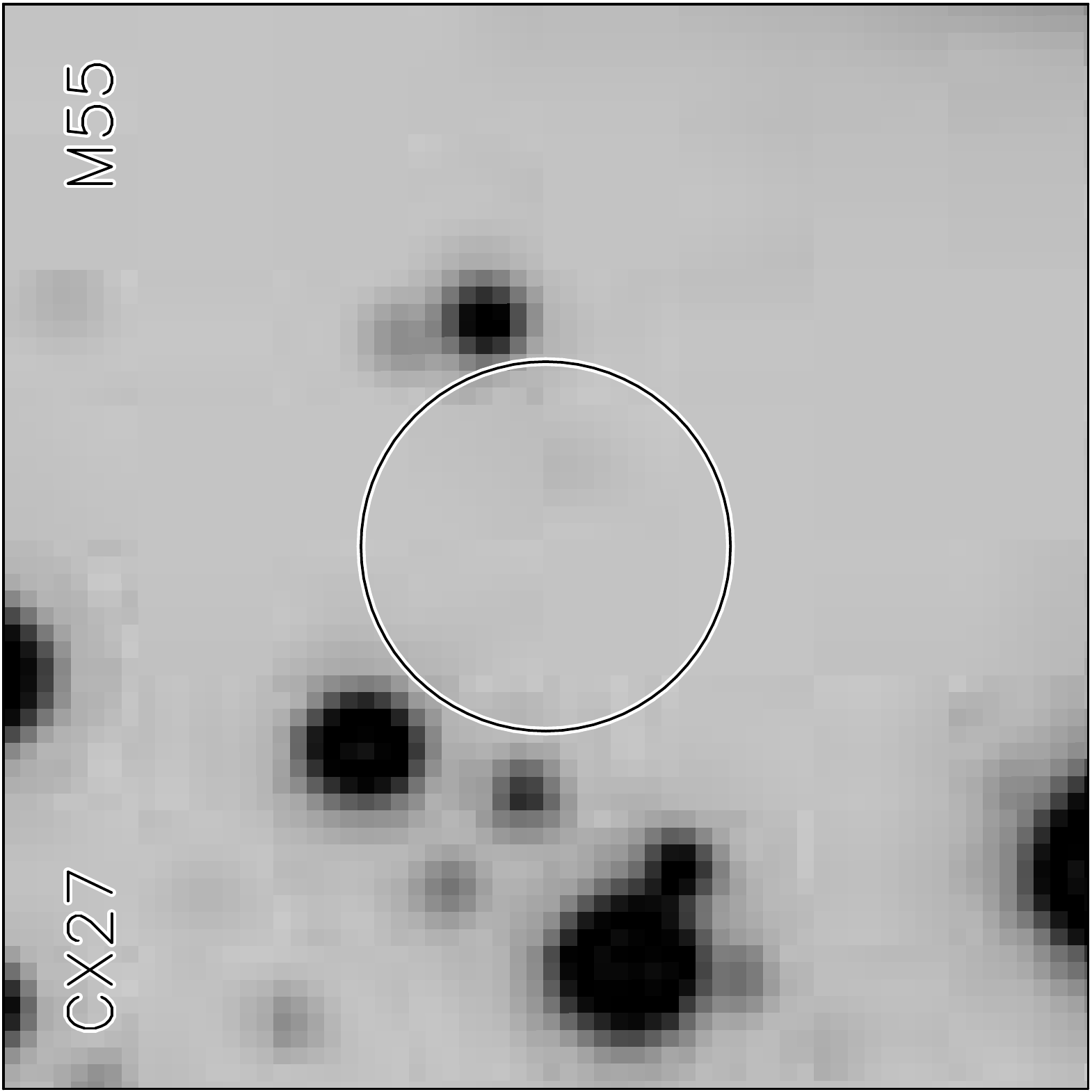}
    \includegraphics[angle=270,width=2.8cm]{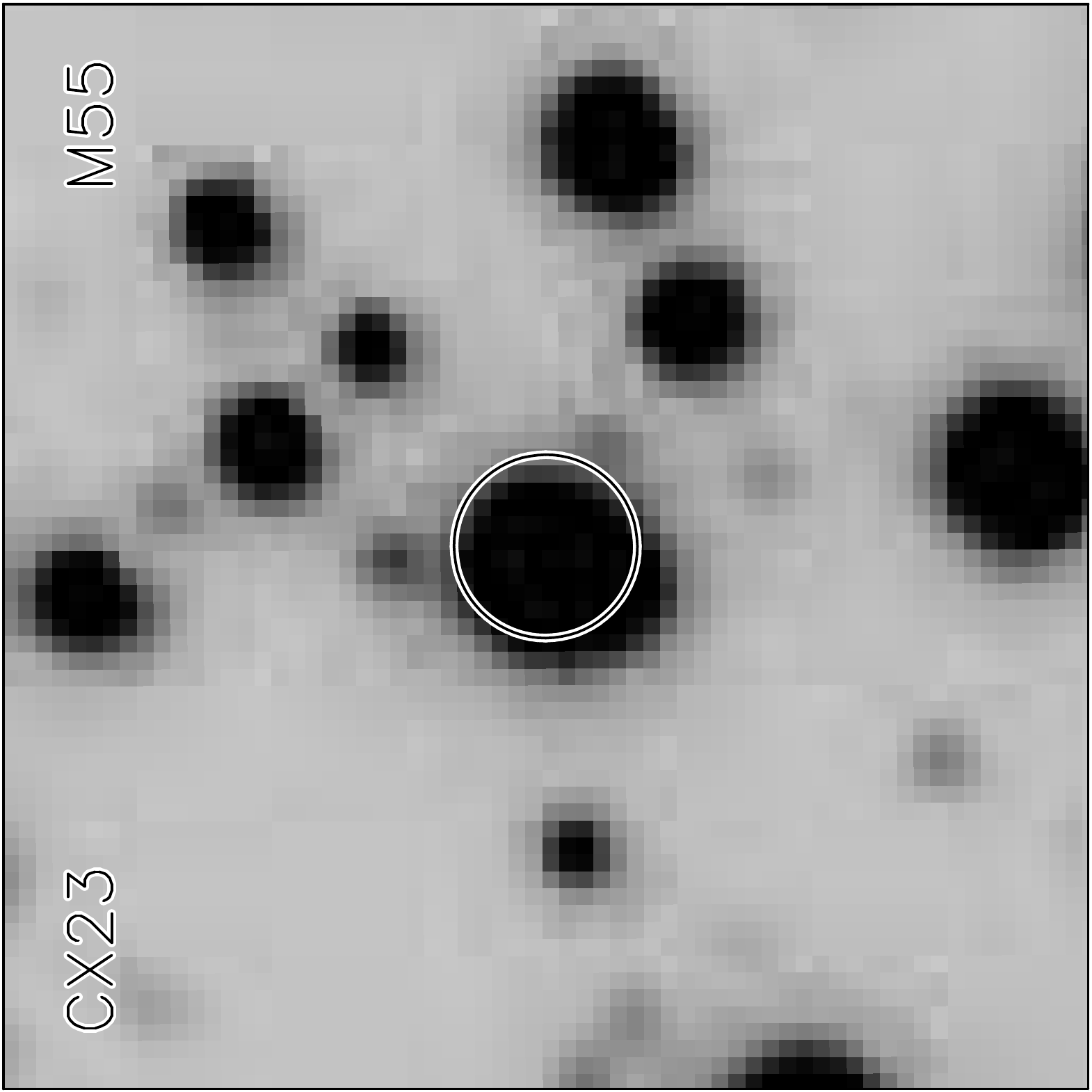}
    \includegraphics[angle=270,width=2.8cm]{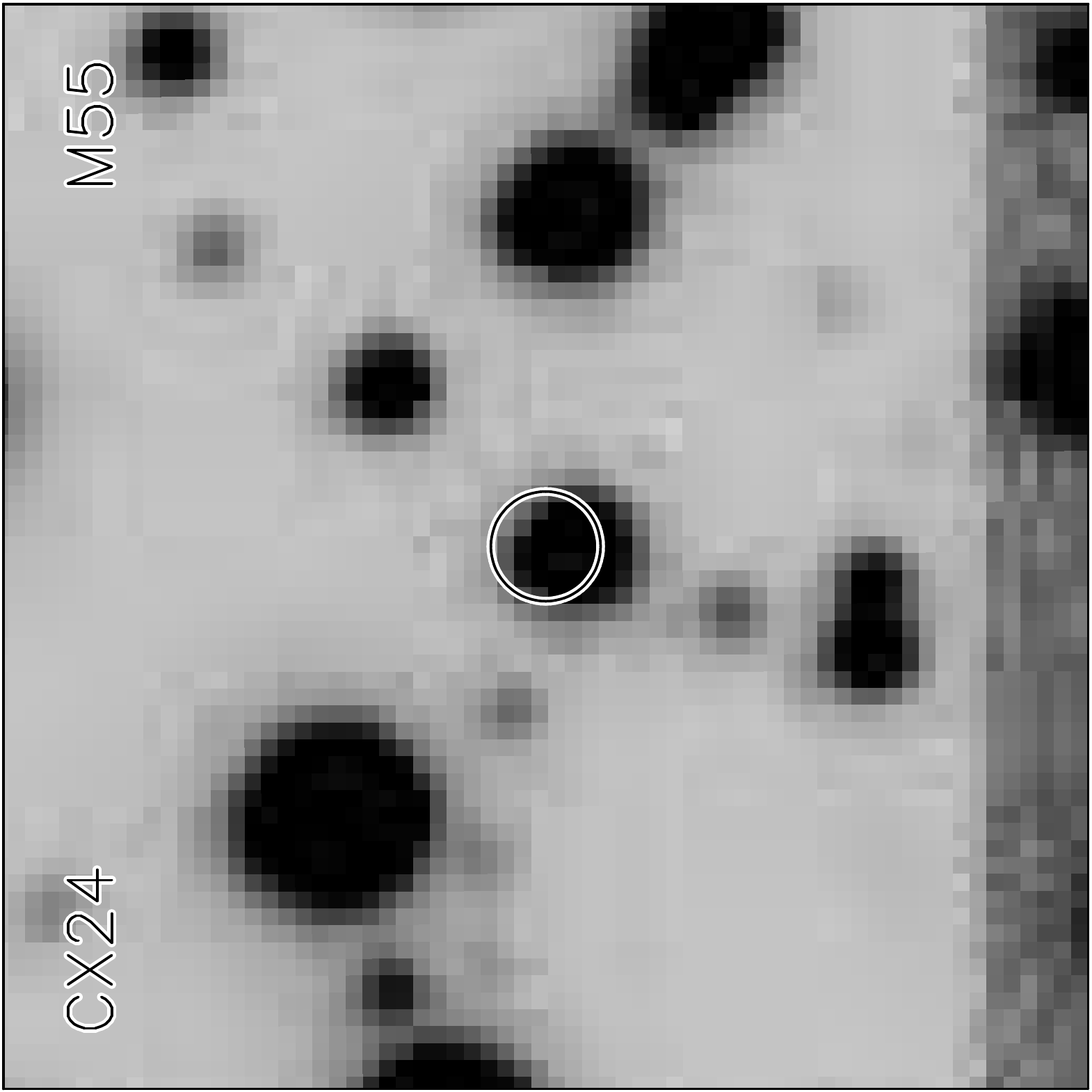}
    \includegraphics[angle=270,width=2.8cm]{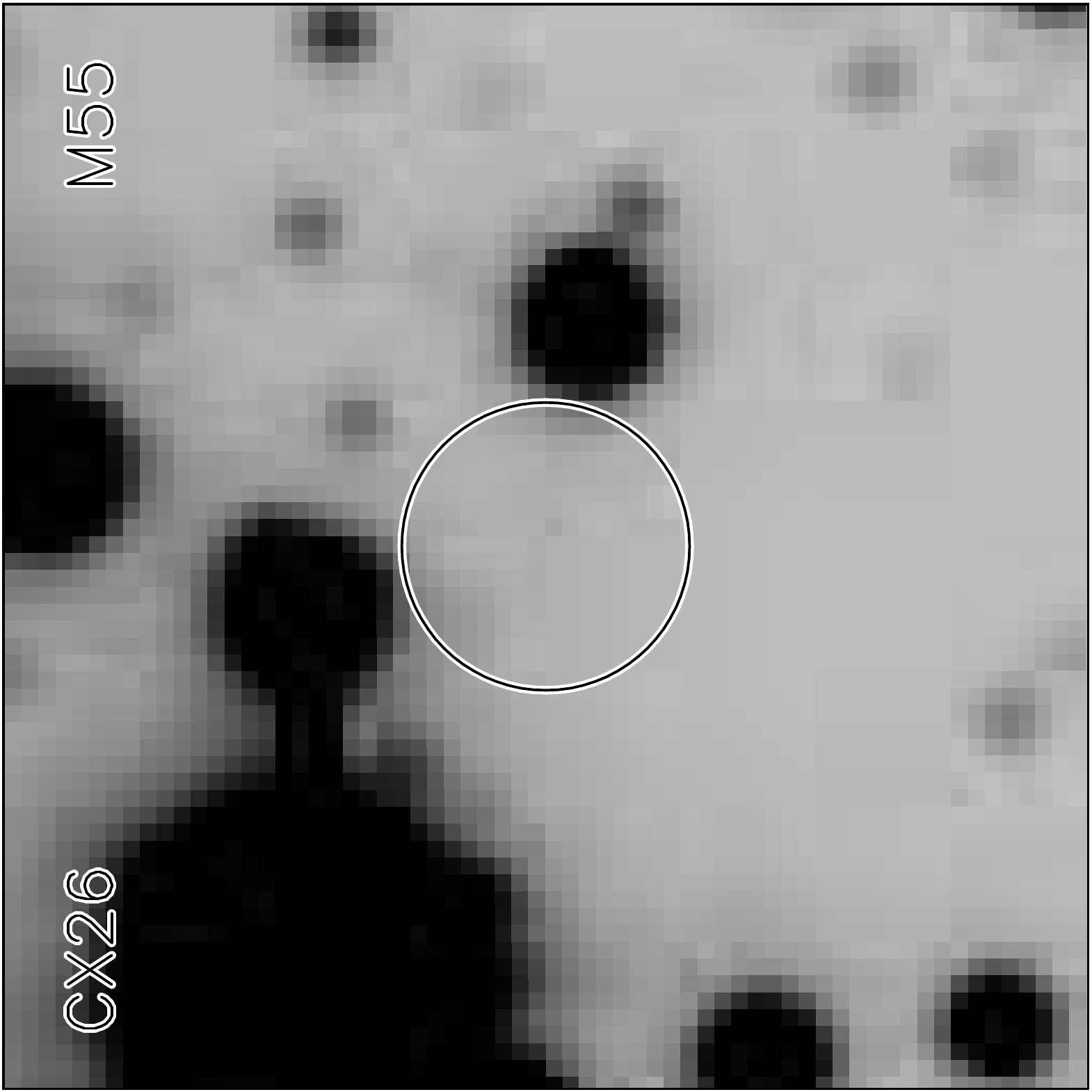}\\
    \includegraphics[angle=270,width=2.8cm]{9350f6ay.pdf}
    \includegraphics[angle=270,width=2.8cm]{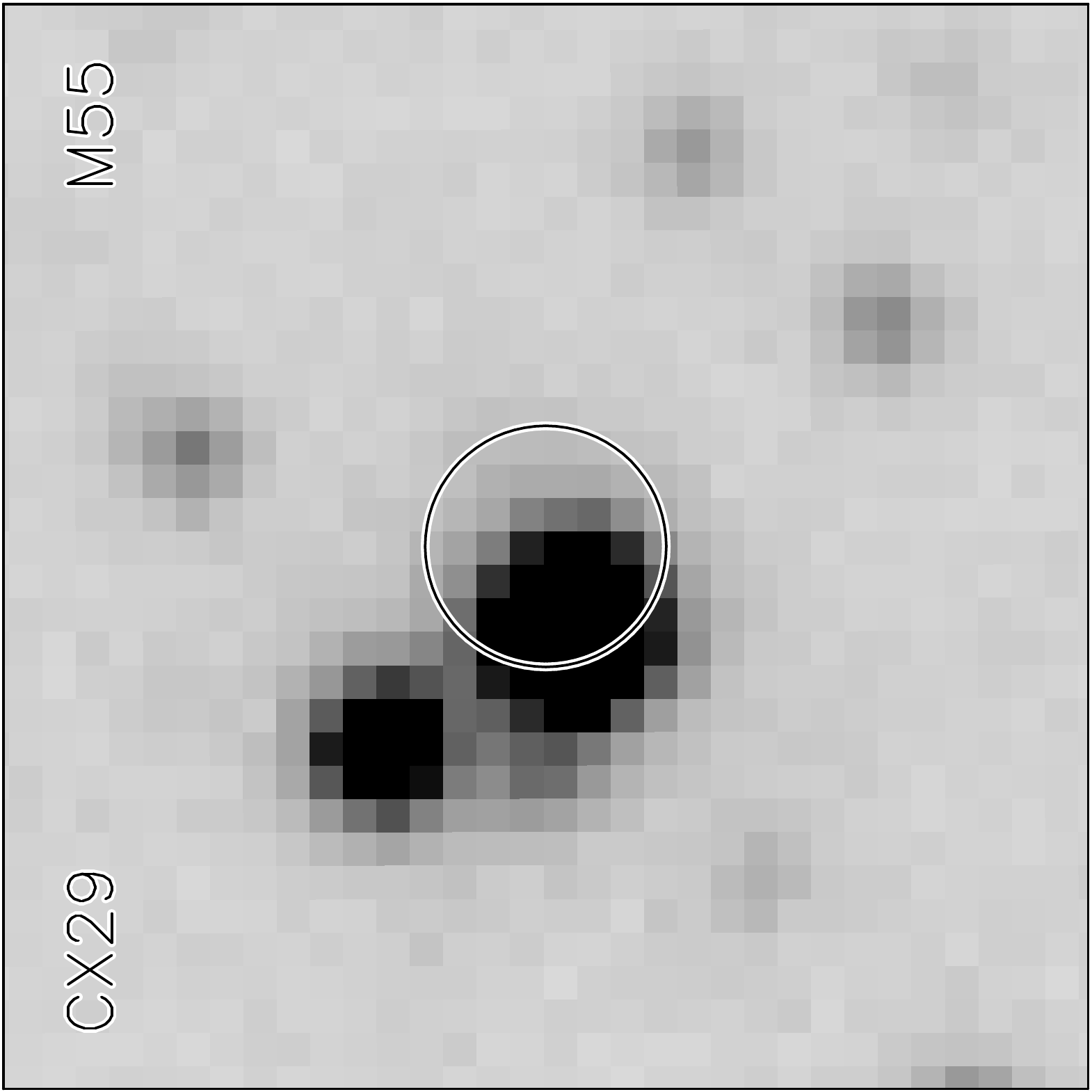}
    \phantom{\includegraphics[angle=270,width=2.8cm]{9350f6az.pdf}}
    \phantom{\includegraphics[angle=270,width=2.8cm]{9350f6az.pdf}}
    \includegraphics[angle=270,width=2.8cm]{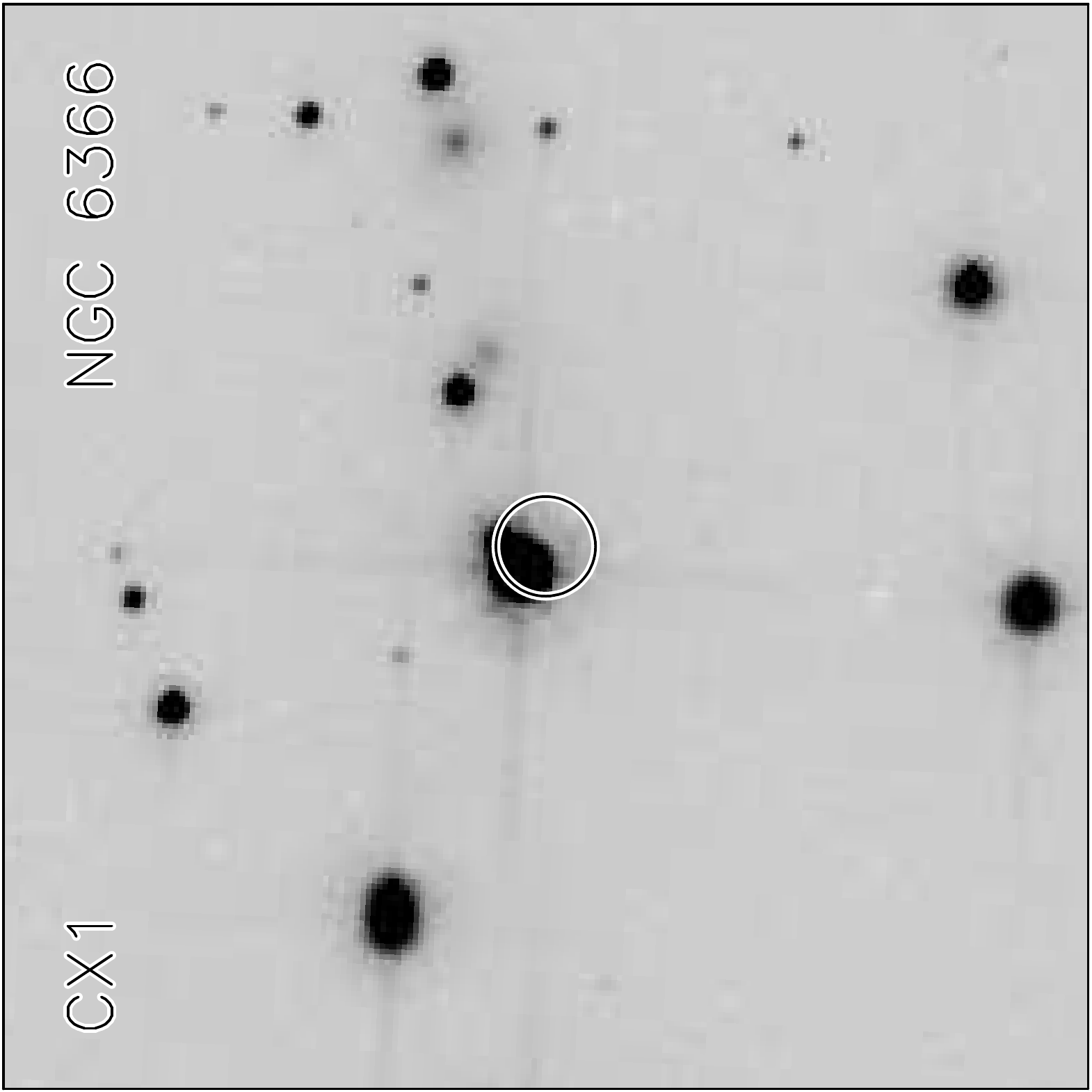}
    \includegraphics[angle=270,width=2.8cm]{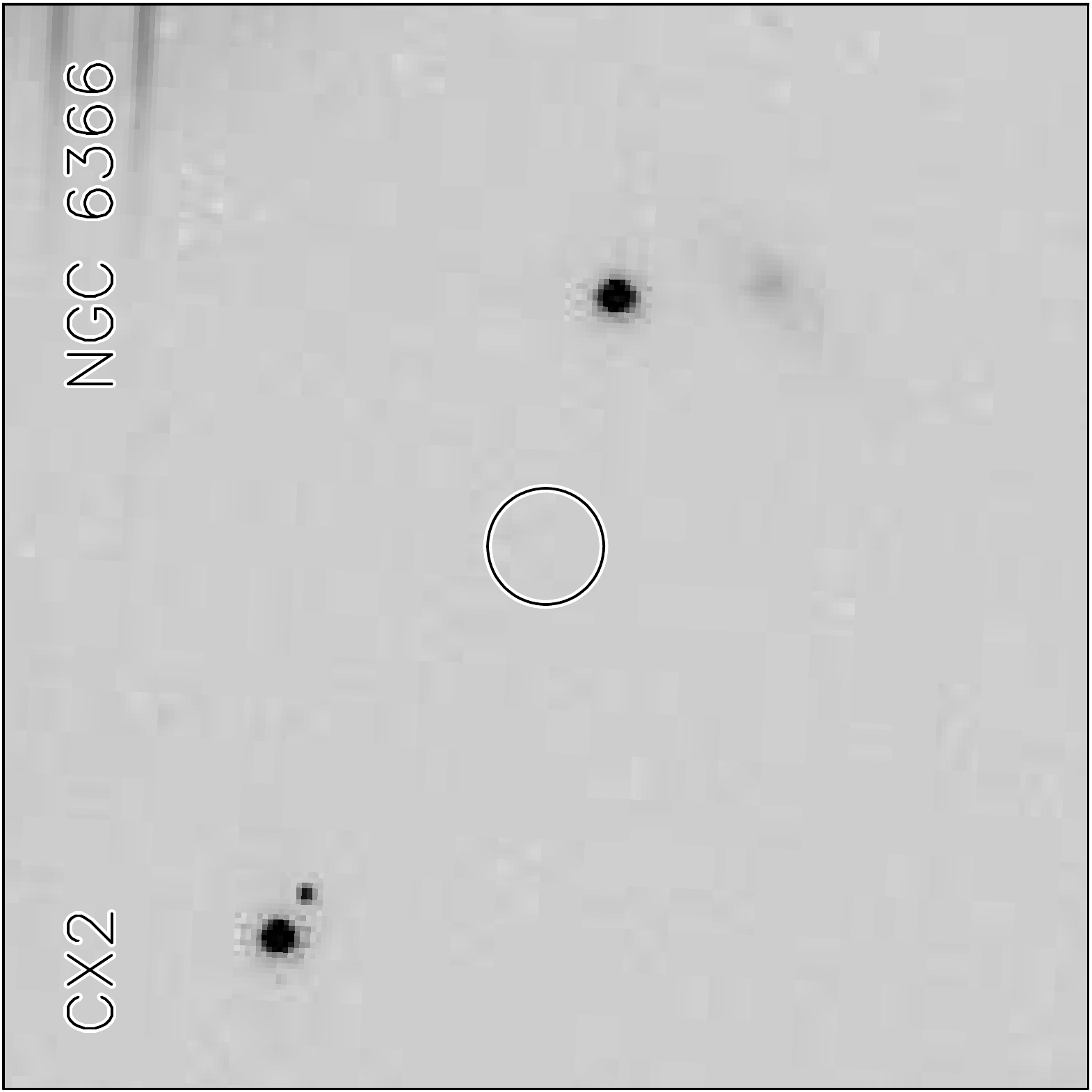}\\
    \includegraphics[angle=270,width=2.8cm]{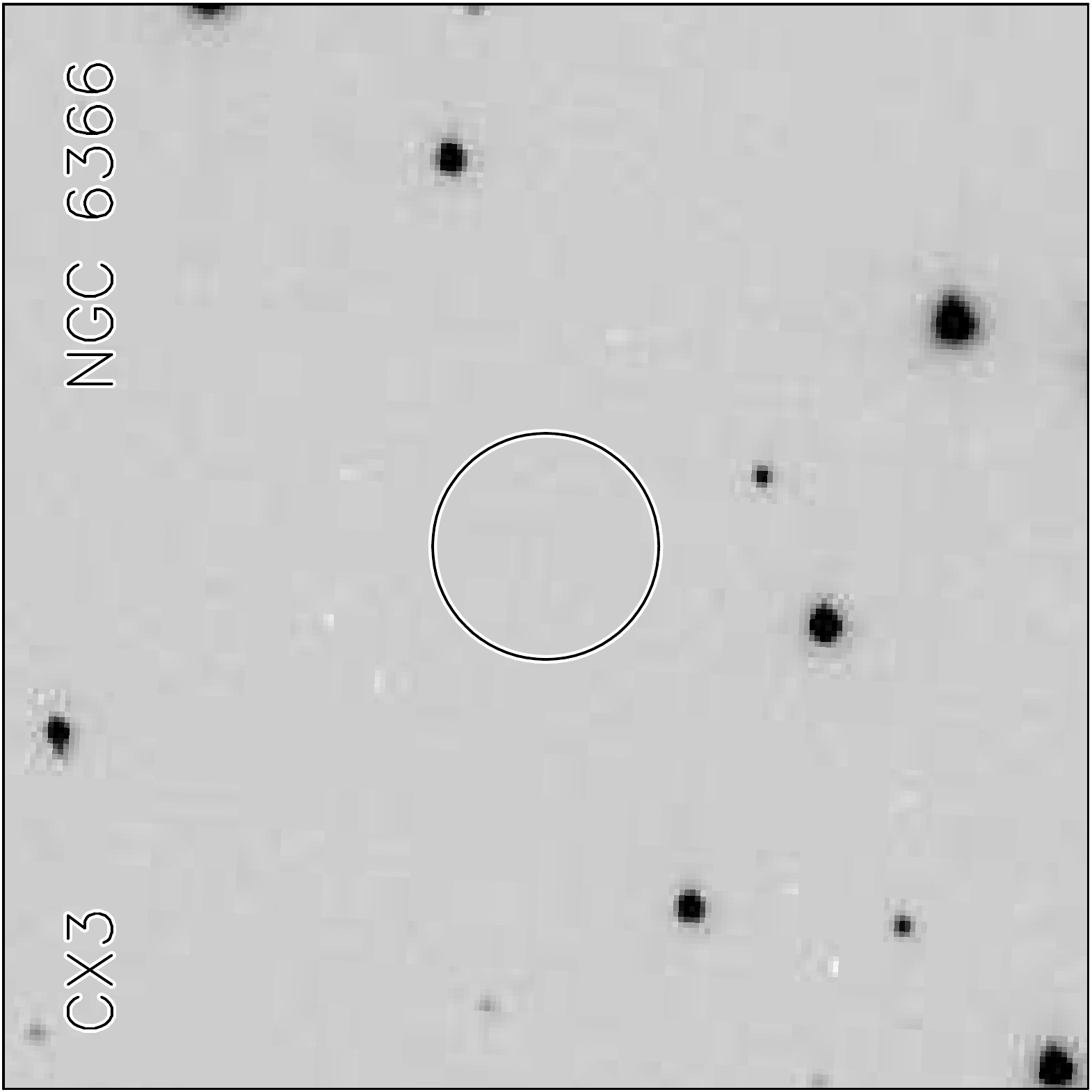}
    \includegraphics[angle=270,width=2.8cm]{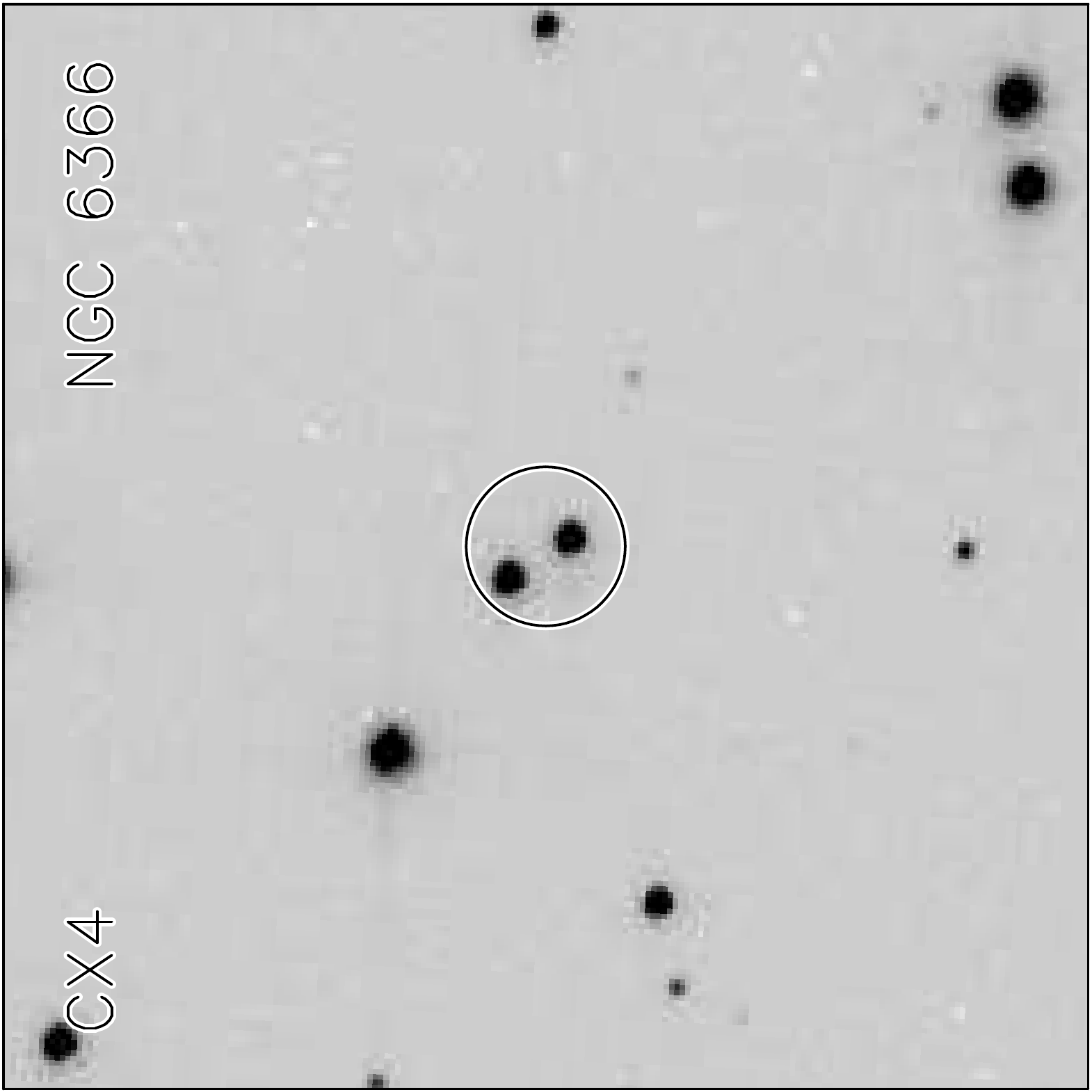}
    \includegraphics[angle=270,width=2.8cm]{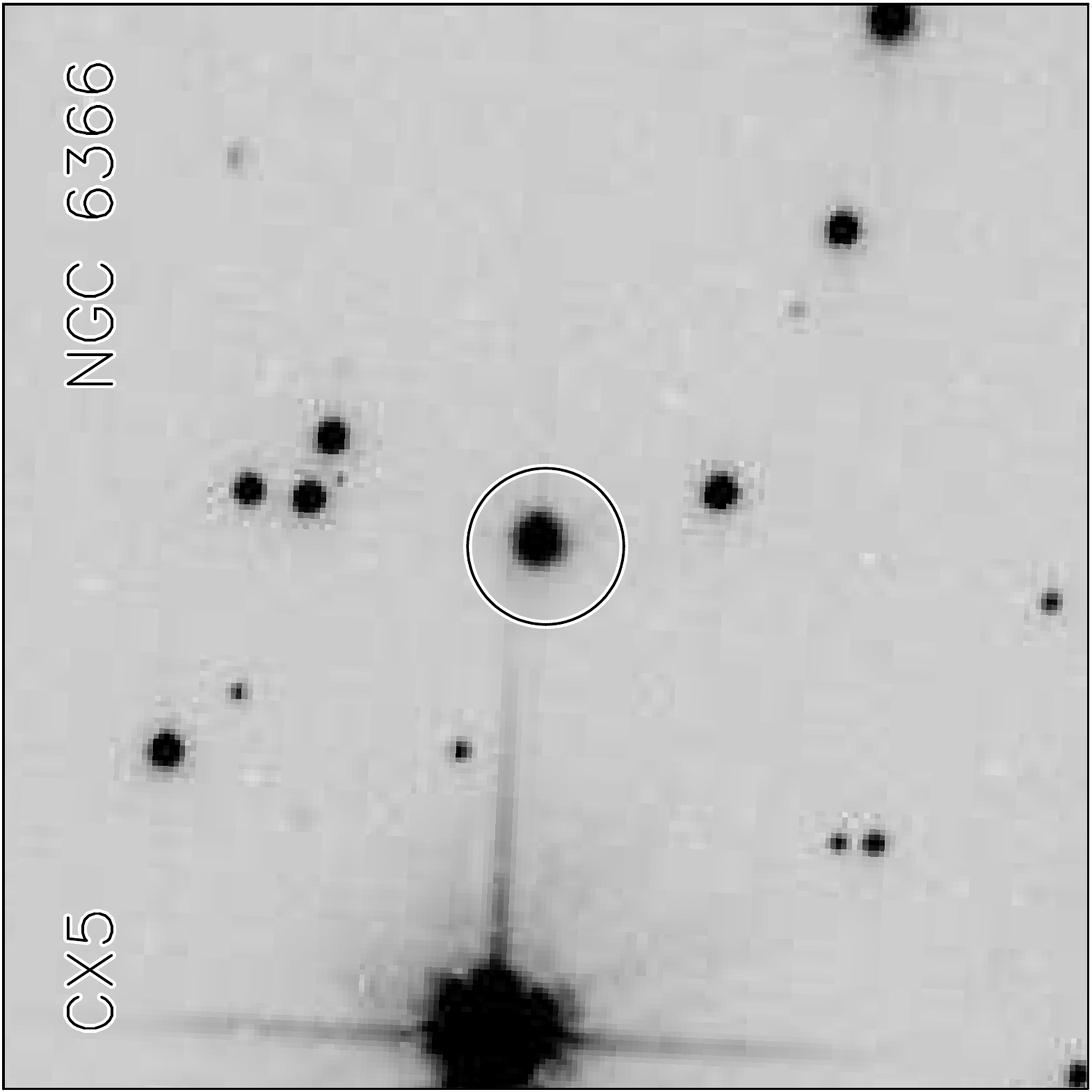}
    \includegraphics[angle=270,width=2.8cm]{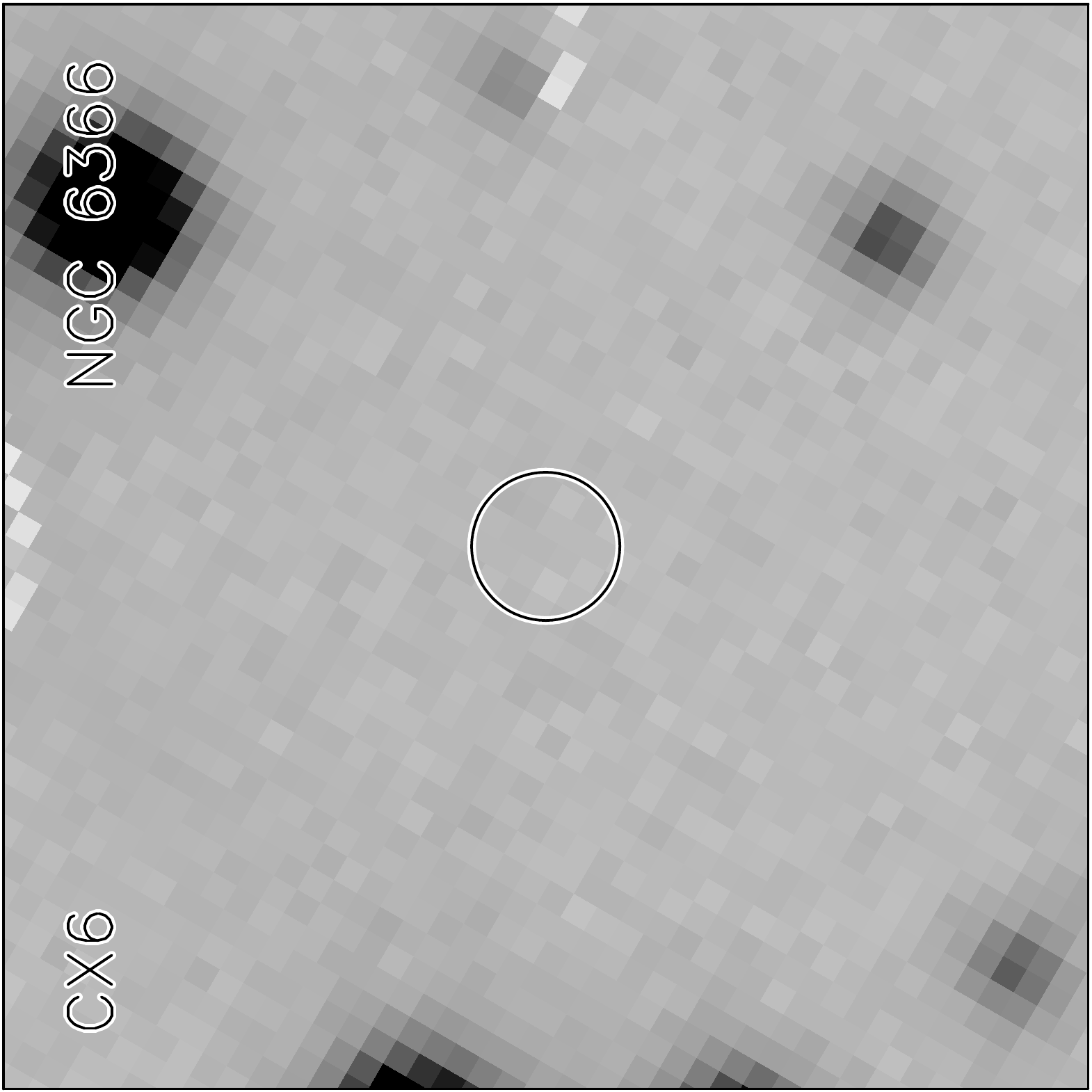}
    \includegraphics[angle=270,width=2.8cm]{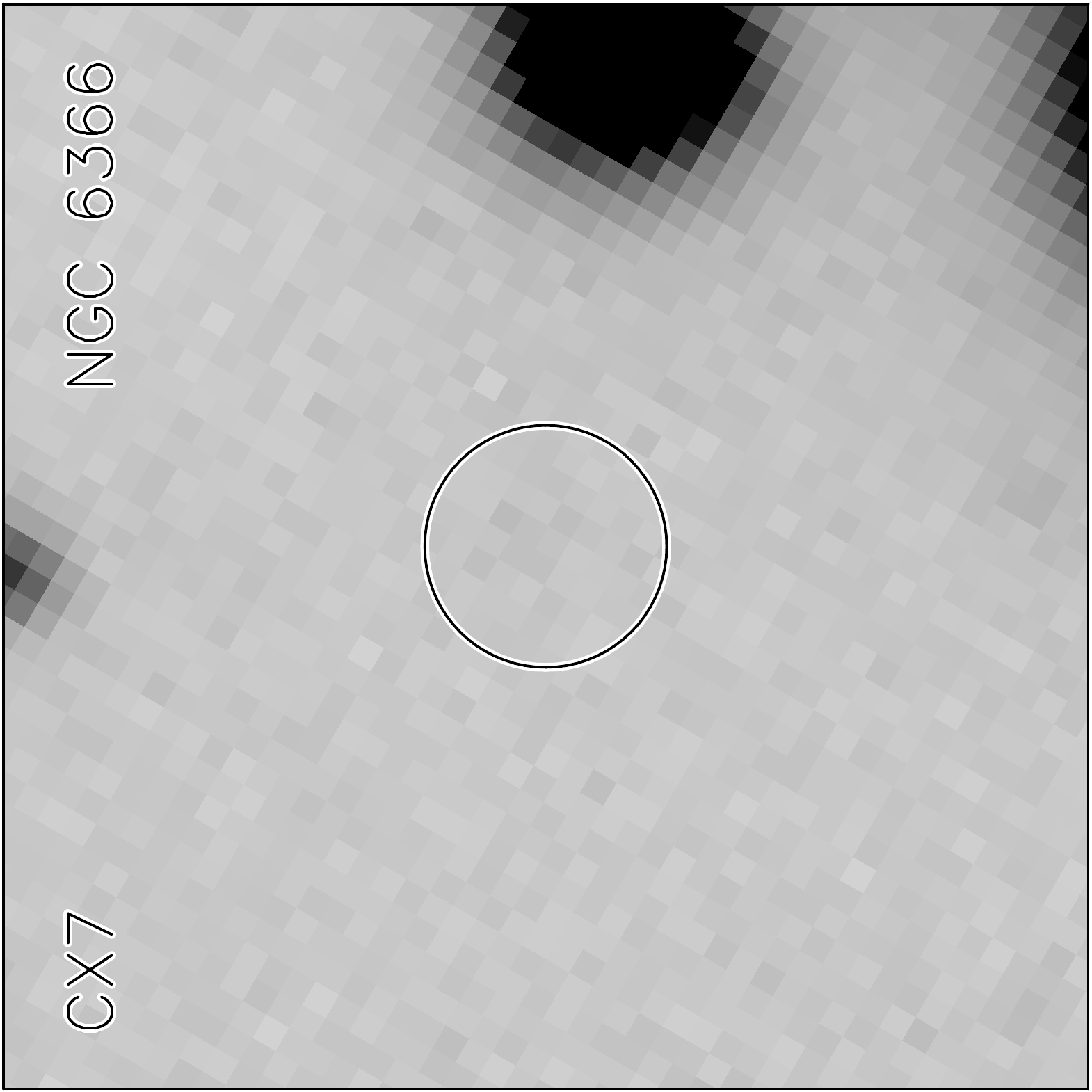}
    \includegraphics[angle=270,width=2.8cm]{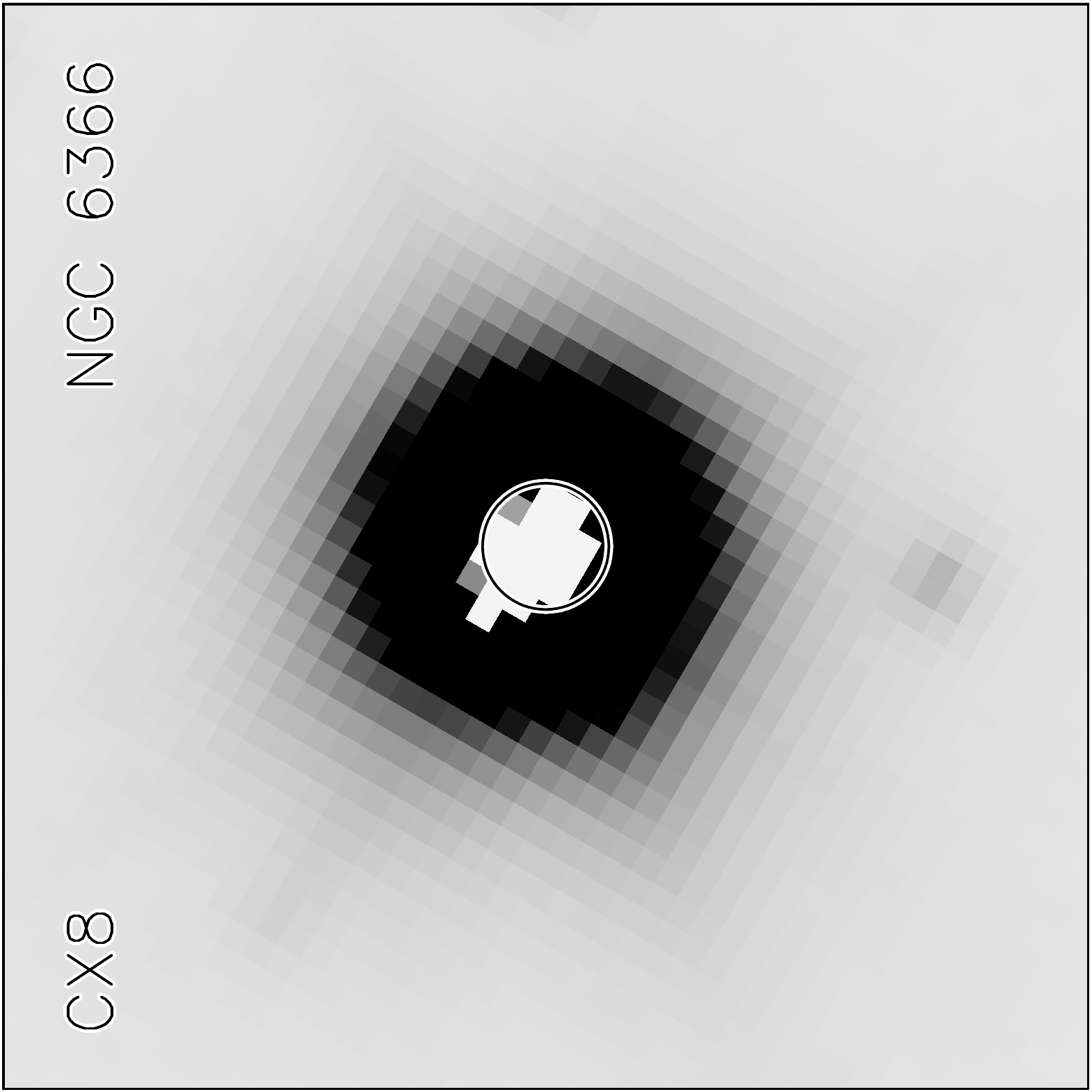}\\
    \includegraphics[angle=270,width=2.8cm]{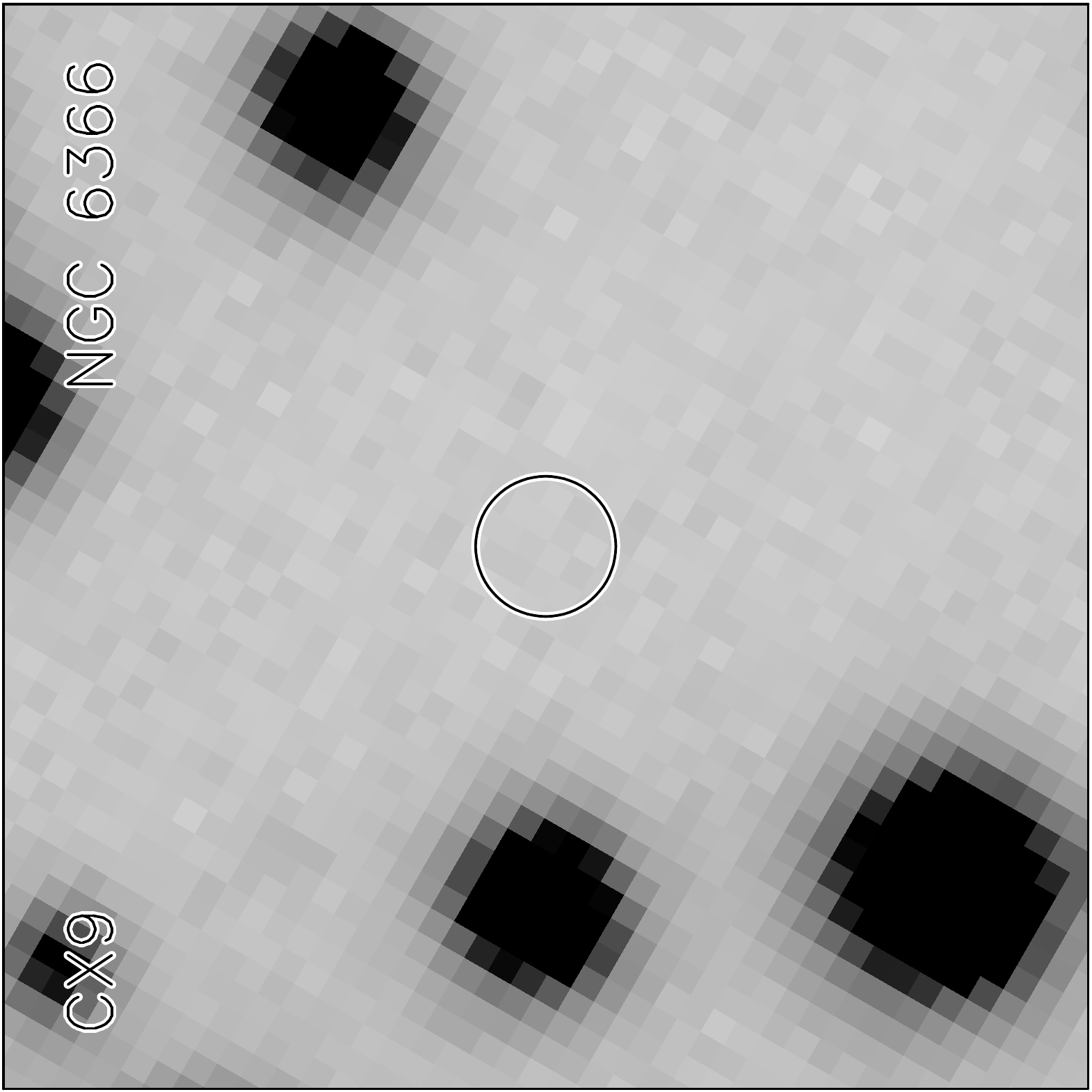}
    \includegraphics[angle=270,width=2.8cm]{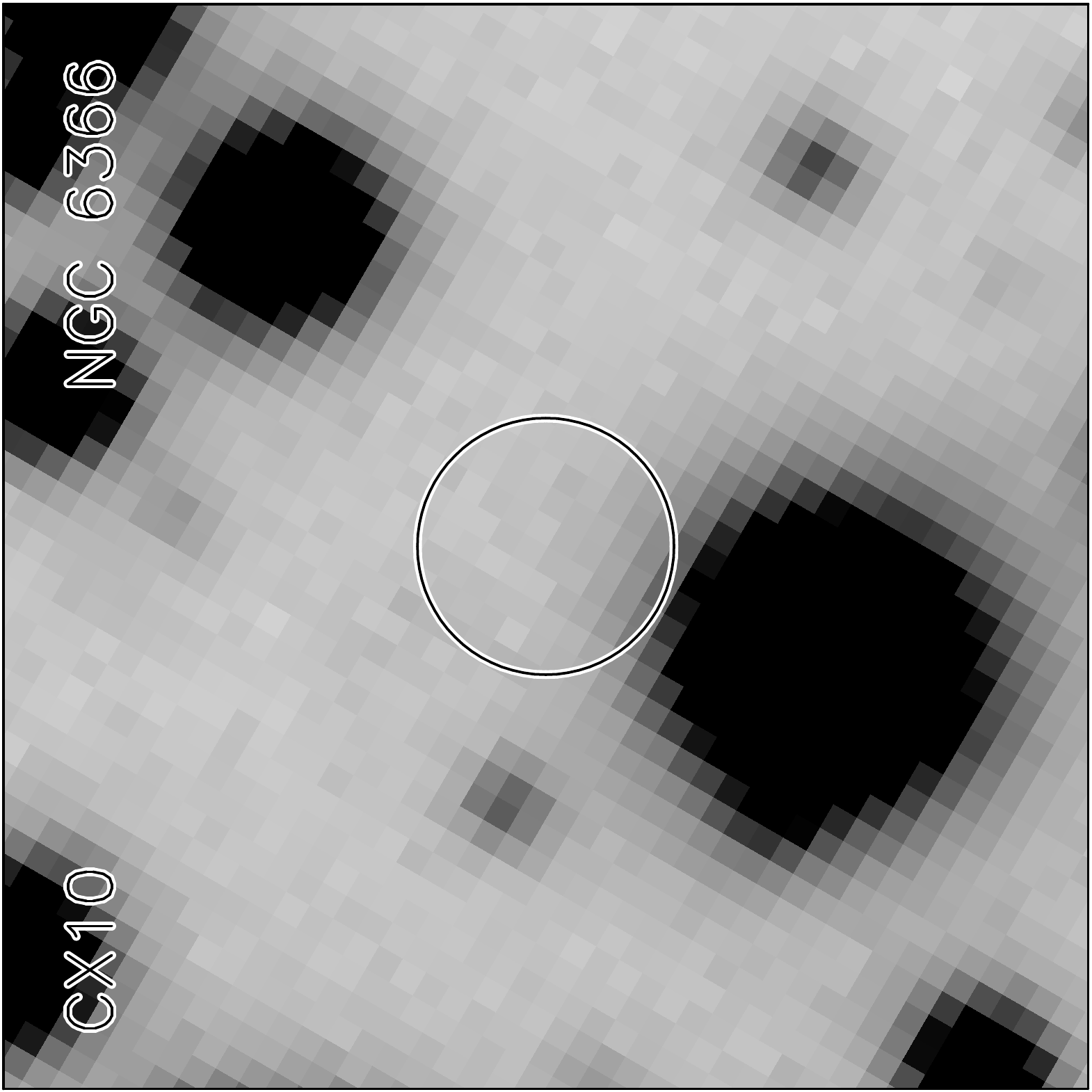}
    \includegraphics[angle=270,width=2.8cm]{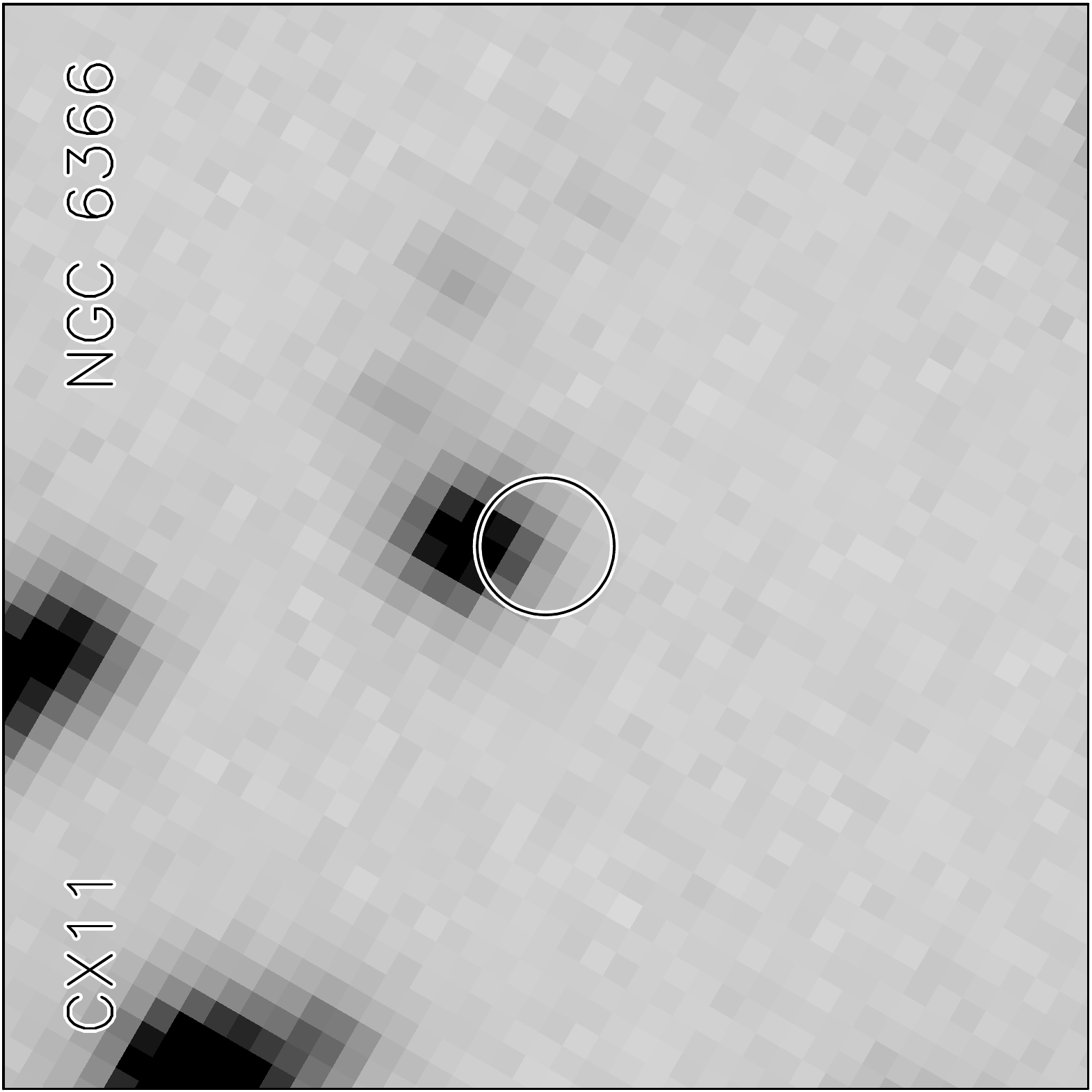}
    \includegraphics[angle=270,width=2.8cm]{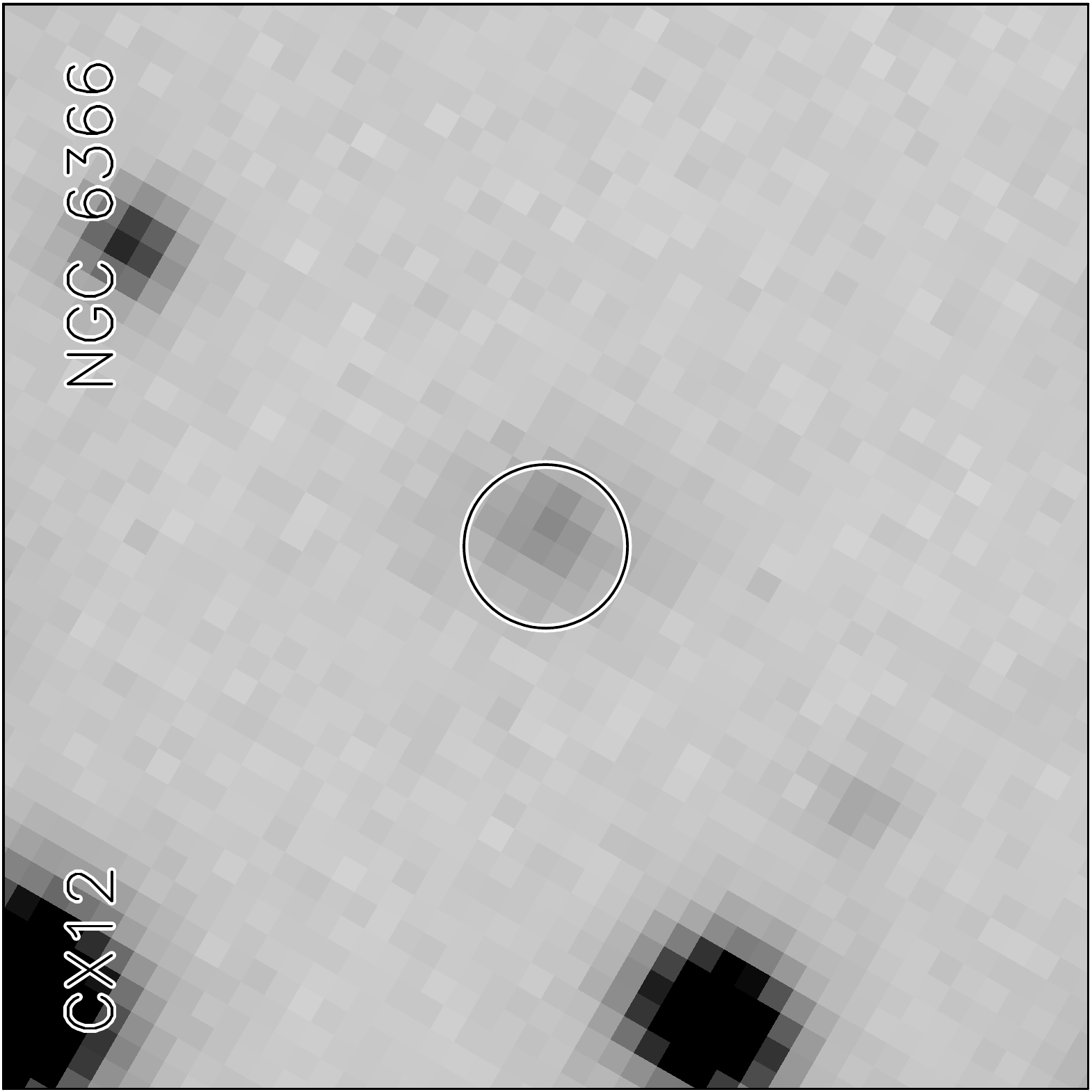}
    \includegraphics[angle=270,width=2.8cm]{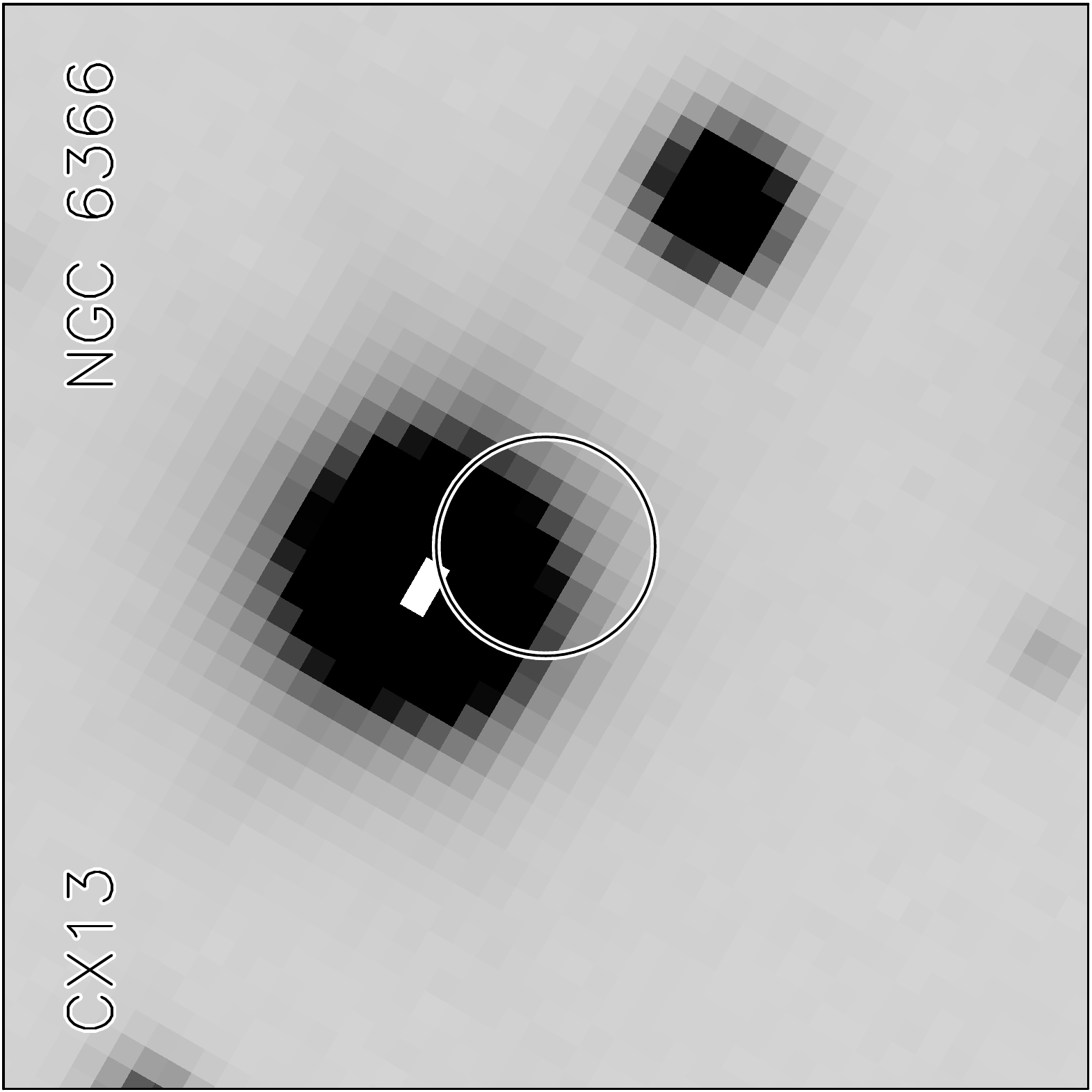}
    \includegraphics[angle=270,width=2.8cm]{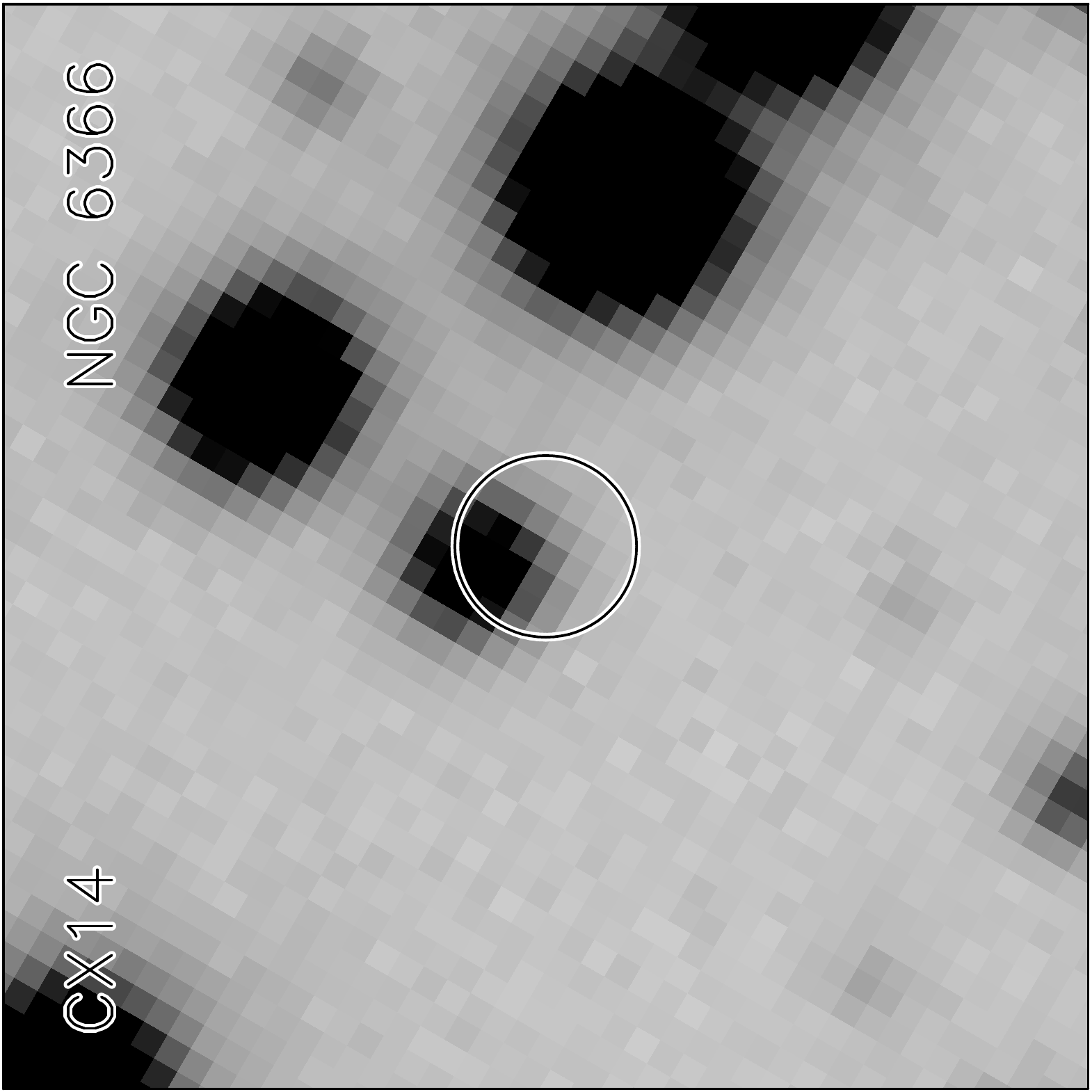}
    \caption{Finding charts for all \emph{Chandra} X-ray sources in
      M55 and NGC\,6366 which coincide with the FORS2 or ACS/WFC
      field-of-view (see Fig.~\ref{f1}). The first 26 charts, CX1 to
      CX29 are for sources in M55 and are $8\arcsec\times8\arcsec$ in
      size. Sources CX1 to CX17 are located in the cluster half-mass
      radius, the other sources are located outside of it. The next 14
      charts are for all X-ray sources in NGC\,6366 and these are
      $10\arcsec\times10\arcsec$ in size. The first 5 are for sources
      inside the half-mass radius, the other 9 outside the half-mass
      radius. The ACS/WFC images have a 5 times better pixel scale and
      show the drizzled \texttt{F606W} observations. The finding
      charts for the other sources were constructed from the $R$-band
      FORS2 observations. The uncertainties on the X-ray source
      position is indicated with a 99\% confidence error circle.  All
      images have North to the top and East to the left.}
    \label{f6}
  \end{figure*}

  Starting with M55, we note that the 99\%\ error circles in
  Fig.~\ref{f6} include relatively bright stars for CX1, CX7,
  CX8, CX10, and CX16.  The other circles contain no significant
  stars, or stars that we consider too far off-center to be probable
  counterparts.

\begin{figure}
    \resizebox{0.95\hsize}{!}{\includegraphics{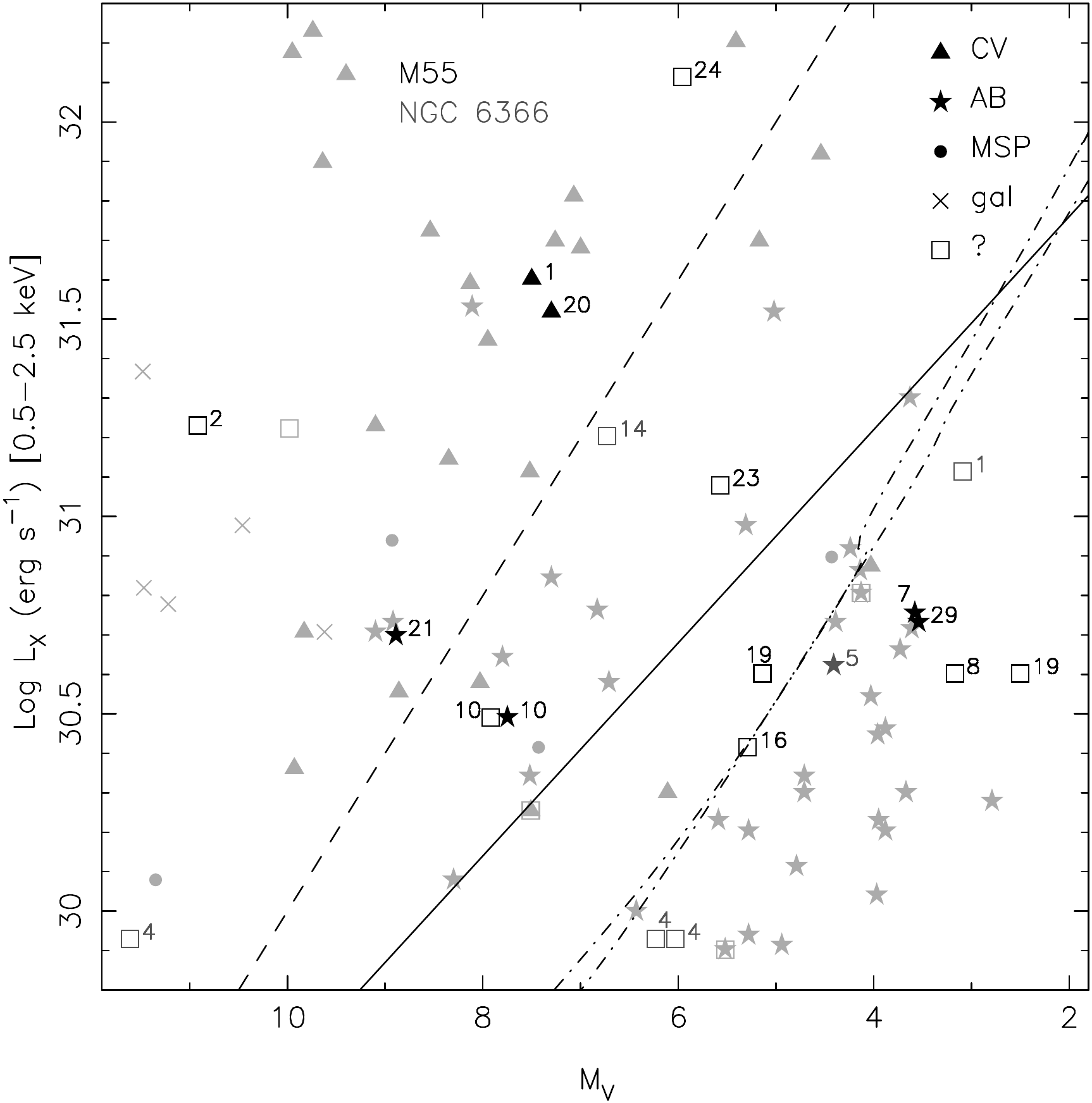}}
    \caption{The 0.5--2.5\,keV X-ray luminosity $L_\mathrm{X}$ against
      the absolute $V$-band magnitudes. The $V$-band magnitudes are
      estimated by $V=(B+R)/2$ to allow a comparison to X-ray sources
      with optical counterparts detected in other globular
      clusters. The X-ray sources are classified into the different
      classes as indicated. Besides the \emph{Chandra} X-ray sources
      discovered in M55 and NGC\,6366 (which are numbered) we also
      have included classified X-ray sources from 47\,Tuc, NGC\,6752,
      NGC\,288, NGC\,6397, $\omega$\,Cen, M22 and M4. The diagonal
      dashed line of constant X-ray to optical luminosity ratio ($\log
      L_\mathrm{X}=34.0-0.4 M_V$) roughly separates CVs from ABs
      (after \citealt{bph+04}). The solid diagonal line ($\log
      L_\mathrm{X}=32.3-0.27M_V$) is after \citet{vpb07}. The
      two dashed-dotted lines are computed from 11.2\,Gyr isochrones
      at $z=0.001$ and $z=0.019$ from \citet{gbbc00}, assuming
      $L_\mathrm{X}\simeq0.001L_\mathrm{bol}$.}
    \label{f7}
  \end{figure}

  CX1, as remarked above, is the dwarf nova discovered and
  studied by \citet{kpt+05}, whose optical position we find to be
  compatible with the \textit{Chandra} position, and which we used to
  correct the \textit{Chandra} coordinates. In Fig.~\ref{f4} we see
  that the optical counterpart of CX1 has strong H$\alpha$ emission
  as expected, but is less blue than expected for a dwarf nova, being at
  the blue edge of the main sequence. Its X-ray to optical flux ratio
  is as expected for a cataclysmic variable, well above the line $\log
  L_\mathrm{X}=34.0-0.4 M_V$ that roughly separates cataclysmic variables
  from the magnetically active binaries below it (Fig.~\ref{f7}).

  The error circle of CX2 shows a very faint blue object in the ACS/WFC
  observations. The source could be a cataclysmic variable, though it
  is below the detection limit in the FORS2 observations, so we do not
  know if it exhibits an excess of H$\alpha$ emission. Compared to
  classified cataclysmic variables in globular clusters, the optical
  counterpart is significantly fainter. The probable optical
  counterpart of CX7 is at a position slightly below and towards the
  red of the subgiant branch in the colour magnitude diagram
  Fig.~\ref{f4}, and thus a sub-subgiant in the terminology introduced
  by \citet{bvm98} in their study of M67. It has no significant
  H$\alpha$ emission. Its $f_\mathrm{X}/f_\mathrm{opt}$ indicates a
  magnetically active binary in agreement with the classification as a
  sub-subgiant. The possible optical counterpart of CX8 from its
  position in the colour magnitude diagram appears to be an ordinary
  subgiant; as this star is less well centered in the astrometric
  confidence circle, it is quite possible that the real counterpart of
  CX8 is a fainter object. CX10 has two possible counterparts, the
  brighter one is on the binary sequence in the colour magnitude
  diagram, but puzzlingly has strong H$\alpha$ absorption. Because the
  $R-\mathrm{H}\alpha$ is not affected by interstellar absorption,
  this offset from the main sequence cannot be explained as due to the
  star being a fore- or background object. This suggests it is an
  active binary.  The \emph{HST} observations show that the single
  star found in the error circle of CX16 in the FORS2 observations is
  in fact a blend of three stars. Two of these stars are rather red
  and too far from the main sequence of M55 to be a member. Their red
  colours suggest they are background objects. The third star lies on
  the cluster main-sequence in $V-I$, and could indicate that the star
  is an active binary. Because the star is blended in the FORS2
  observations, it is unclear to which object the H$\alpha$ absorption
  can be attributed.

  For the sources outside the half-mass radius of M55, CX19 has two
  candidate counterparts, one on the subgiant branch, the other on the
  main-sequence. Neither counterpart displays a significant emission
  or absorption of H$\alpha$ flux. CX20 on the other hand has very
  strong H$\alpha$ emission and is very blue in $B-R$. With a X-ray
  luminosity of $L_\mathrm{X}=3.3\times10^{31}$\,erg\,s$^{-1}$ it is
  very similar to CX1 and possibly a cataclysmic variable. The
  counterpart to CX21 is on the main-sequence and could be an active
  binary. However, the $f_\mathrm{X}/f_\mathrm{opt}$ ratio of this
  source places it in the region that is primarily populated by CVs,
  the X-ray source could be a cataclysmic variable. We will return to
  this source in Section\,\ref{s5.2}. CX23 shows similarities with
  CX16 in the sense that it is very red and has a large excess of
  H$\alpha$ emission. It is likely a background
  object. The optical counterpart to CX24 is slightly blue but shows
  no H$\alpha$ emission or absorption compared to that of the cluster
  stars. It is the brightest X-ray source in our sample, and it has an
  $f_\mathrm{X}/f_\mathrm{opt}$ ratio that would be indicative of a
  cataclysmic variable. However, the absence of H$\alpha$ suggests it
  belongs to the Galactic field. The optical counterpart to CX29 may
  be yet another sub-subgiant; it is located below the subgiant branch
  and has a X-ray luminosity comparable to that of CX7 in M55 and CX5
  in NGC\,6366. We do note that the source lies at the edge of the
  error circle.

  \subsection{NGC\,6366}
  Turning now to NGC6366, we consider four possibilities for the
  extended emission of source CX1b; a planetary nebula, a supernova
  remnant, a group or cluster of galaxies and a nova
  remnant. Planetary nebulae and supernovae tend to have strong
  H$\alpha$ emission. The apparent absence of such emission in our
  H$\alpha$ images of CX1 (Fig.\,\ref{f2}) argues against these two
  possibilities. A second argument against a planetary nebula is that
  the X-ray spectra of planetary nebula observed with \textit{Chandra}
  are rather soft, limited to energies below 1.5\,keV (e.g.\,
  BD$+30\degr3639$, \citealt{ksvd00}; NGC\,7027, \citealt{kvs01};
  NGC\,6543, \citealt{cgg+01}), whereas 50\% of the photons of CX1b is
  above 2\,keV (Fig.\,\ref{f2}). A second argument against a supernova
  remnant is that it is an unlikely object to be found in a globular
  cluster, or as a foreground object at a Galactic latitude of
  16\degr.

  The clusters of galaxies studied by \citet{ktp+04} and
  \citet{pjkt05} have typical X-ray diameters of order 1\,Mpc and
  X-ray luminosities in excess of $10^{43}$\,erg\,s$^{-1}$. Both the
  flux and diameter of CX1b imply a distance of more than a Gpc if
  this source is a typical cluster. The intra-cluster gas in such
  bright clusters strongly dominates the total flux, and so the
  superposition of CX1a and CX1b would be accidental if CX1b is
  a galaxy cluster. Groups of galaxies have a smaller extent and
  lower X-ray luminosity, against which an individual galaxy can
  stand out (e.g.\ \citealt{nmf07}). However, even our deepest $B$ and
  $R$-band images show no obvious galaxies near the position of CX1,
  and this excludes a group or cluster of galaxies out to several
  hundred Mpc.

  Finally, a nova outburst may give rise to extended X-ray emission,
  as illustrated by the case of GK\,Per (Nova Persei 1901) which was
  detected as an extended source with \textit{Chandra}, surrounding
  the point source of the cataclysmic variable GK\,Per itself
  \citep{bal05}. Both in size and in luminosity CX1 is a plausible
  old nova at the distance of NGC\,6366, but the spectrum of the
  extended source CX1b is rather harder than the soft extended
  source surrounding GK\,Per. The amount of matter in a nova remnant
  is much less than in a planetary nebula, perhaps compatible with the
  absence of extended H$\alpha$ emission; indeed, \citet{sod95} argue
  that the faint emission around GK\,Per in the H$\alpha$ filter is
  due largely to \ion{N}{ii} emission.

  For the moment, we consider an old nova to be the more likely
  explanation for the X-ray source CX1 in NGC\,6366. Novae are known
  in globular clusters; T\,Sco (1860) in M80, V1148\,Sgr (1943) in
  NGC\,6553 and Nova 1938 in M14 \citep{saw38,may49,hw64}. The nova
  hypothesis explains the position of CX1 near the cluster center,
  and the superposition of CX1a on CX1b. It remains to be shown,
  however, that the X-ray spectrum of an old nova can be as hard as
  that of CX1b.

  The two bright possible optical counterparts of CX4 are both
  ordinary main sequence stars according to the colour magnitude
  diagram; the $f_\mathrm{X}/f_\mathrm{opt}$ suggests a magnetically
  active binary, in which case the companion does not contribute
  significantly to the optical fluxes. However, CX4 is a hard X-ray
  source which may be hard to reconcile with a magnetically active
  binary. It may be that the source is not related to the globular
  cluster and is assocated with the faint optical counterpart seen in
  the ACS/WFC observations. The optical counterpart of CX5 is yet
  another sub-subgiant as CX7 in M55.  For NGC\,6366 CX8 coincides
  with the bright star BD\,$-04\degr4280$. With $V=10.7$, it is
  saturated in our FORS2 observations (see Fig.\,\ref{f6}) and hence
  not present in Fig.\,\ref{f4} and Table\,\ref{t5}. At this
  brightness it is at least 8\,mag brighter than the cluster turn-off
  and hence highly unlikely to be a member of NGC\,6366. Finally CX14
  has a candidate counterpart that is blue in $B-R$ with respect to
  the cluster main-sequence but has no H$\alpha$ emission or
  absorption. The absence of H$\alpha$ emission suggests it belongs to
  the Galactic field.

  \begin{table*}
    \centering
    \caption[]{Candidate optical counterparts to the \emph{Chandra}
      X-ray sources. The positional offsets are given relative to
      the \emph{Chandra} values in Tables\,\ref{t2} and
      \ref{t3}. The X-ray luminosity is in the 0.5--2.5\,keV energy
      range. The first optical counterpart to CX16 is a blend of the
      three other stars.}\label{t5}
    \begin{tabular}{lrrrrrrrc}
      \hline \hline
      CX & \multicolumn{1}{c}{$\Delta\alpha$} &
      \multicolumn{1}{c}{$\Delta\delta$} & \multicolumn{1}{c}{$R$}
      & \multicolumn{1}{c}{$B-R$} &
      \multicolumn{1}{c}{$R-\mathrm{H}\alpha$} & \multicolumn{1}{c}{$V$} & \multicolumn{1}{c}{$V-I$} & $L_\mathrm{X}$ \\
      & & & & & & & & (erg\,s$^{-1}$\,cm$^{-2}$) \\
      \hline
      \multicolumn{7}{l}{M55} \\
       1 &   $0\farcs00$ & $-0\farcs00$ & $20.69(1)$ &  $1.36(2)$ &  $0.58(2)$ &            &           & $4.0\times10^{31}$\\
       2 &   $0\farcs18$ & $-0\farcs03$ &            &            &            & $24.80(2)$ & $0.1(3)$  & $1.7\times10^{31}$\\
       7 &  $-0\farcs06$ & $-0\farcs11$ & $16.75(1)$ &  $1.40(1)$ &  $0.03(1)$ &            &           & $5.7\times10^{30}$\\
       8 &   $0\farcs28$ &  $0\farcs12$ & $16.45(1)$ &  $1.19(1)$ & $-0.00(1)$ & $16.86(1)$ & $0.91(1)$ & $4.0\times10^{30}$\\
      10 &  $-0\farcs04$ &  $0\farcs03$ & $20.51(4)$ &  $2.23(6)$ & $-0.36(8)$ &            &           & $3.1\times10^{30}$\\
      10 &   $0\farcs53$ &  $0\farcs28$ & $20.90(2)$ &  $1.79(3)$ &  $0.01(3)$ &            &           & $3.1\times10^{30}$\\
      16 &   $0\farcs14$ &  $0\farcs09$ & $17.82(1)$ &  $2.68(3)$ &  $0.31(1)$ &            &           & $2.6\times10^{30}$\\
      16 &   $0\farcs24$ &  $0\farcs04$ &            &            &            & $19.66(2)$ & $2.25(2)$ & $2.6\times10^{30}$\\
      16 &   $0\farcs24$ &  $0\farcs15$ &            &            &            & $19.67(2)$ & $2.50(2)$ & $2.6\times10^{30}$\\
      16 &  $-0\farcs13$ & $-0\farcs00$ &            &            &            & $21.53(1)$ & $1.25(1)$ & $2.6\times10^{30}$\\[0.5em]

      19 &   $0\farcs25$ &  $0\farcs47$ & $15.78(1)$ &  $1.18(1)$ &  $0.07(1)$ &            &           & $4.0\times10^{30}$\\
      19 &   $0\farcs35$ & $-0\farcs52$ & $18.45(1)$ &  $1.01(2)$ & $-0.09(2)$ &            &           & $4.0\times10^{30}$\\
      20 &   $0\farcs20$ &  $0\farcs09$ & $21.25(6)$ & $-0.16(6)$ &  $0.51(7)$ &            &           & $3.3\times10^{31}$\\
      21 &  $-0\farcs09$ & $-0\farcs48$ & $21.70(1)$ &  $2.12(3)$ & $-0.00(2)$ &            &           & $5.0\times10^{30}$\\
      23 &   $0\farcs13$ &  $0\farcs06$ & $18.13(2)$ &  $2.63(3)$ &  $0.50(4)$ &            &           & $1.2\times10^{31}$\\
      24 &   $0\farcs04$ & $-0\farcs26$ & $19.41(1)$ &  $0.73(1)$ &  $0.03(1)$ &            &           & $1.3\times10^{32}$\\
      29 &   $0\farcs61$ & $-0\farcs19$ & $16.68(1)$ &  $1.46(1)$ &  $0.04(1)$ &            &           & $5.4\times10^{30}$\\
      \hline
      \multicolumn{7}{l}{NGC\,6366} \\
       1 &   $0\farcs21$ &  $0\farcs24$ &  $16.73(1)$ & $2.66(1)$ &  $0.03(1)$ & $17.75(1)$ & $1.91(1)$ & $1.3\times10^{31}$\\
       4 &   $0\farcs29$ &  $0\farcs34$ &  $19.84(1)$ & $2.32(2)$ & $-0.03(2)$ & $20.66(1)$ & $1.73(1)$ & $8.5\times10^{29}$\\
       4 &  $-0\farcs08$ & $-0\farcs23$ &  $20.02(1)$ & $2.36(2)$ &  $0.00(2)$ & $20.85(1)$ & $1.74(1)$ & $8.5\times10^{29}$\\
       4 &   $0\farcs18$ & $-0\farcs27$ &             &           &            & $26.6(3)$  & $2.6(3)$  & $8.5\times10^{29}$\\
       5 &  $-0\farcs05$ &  $0\farcs07$ &  $18.11(1)$ & $2.56(2)$ &  $0.06(1)$ & $19.01(1)$ & $1.96(1)$ & $4.2\times10^{30}$\\[0.5em]
      14 &   $0\farcs22$ &  $0\farcs68$ &  $20.78(1)$ & $1.83(1)$ &  $0.04(2)$ &            &           & $1.6\times10^{31}$\\
      \hline
    \end{tabular}
  \end{table*}

  \section{Discussion}
  \subsection{Comparison with earlier observations} Both M55 and
  NGC\,6366 have been observed with \emph{ROSAT}, and \emph{XMM} also
  observed M55.

  \subsubsection{M55}\label{s511}
\begin{figure}
     \resizebox{0.95\hsize}{!}{\includegraphics[angle=270]{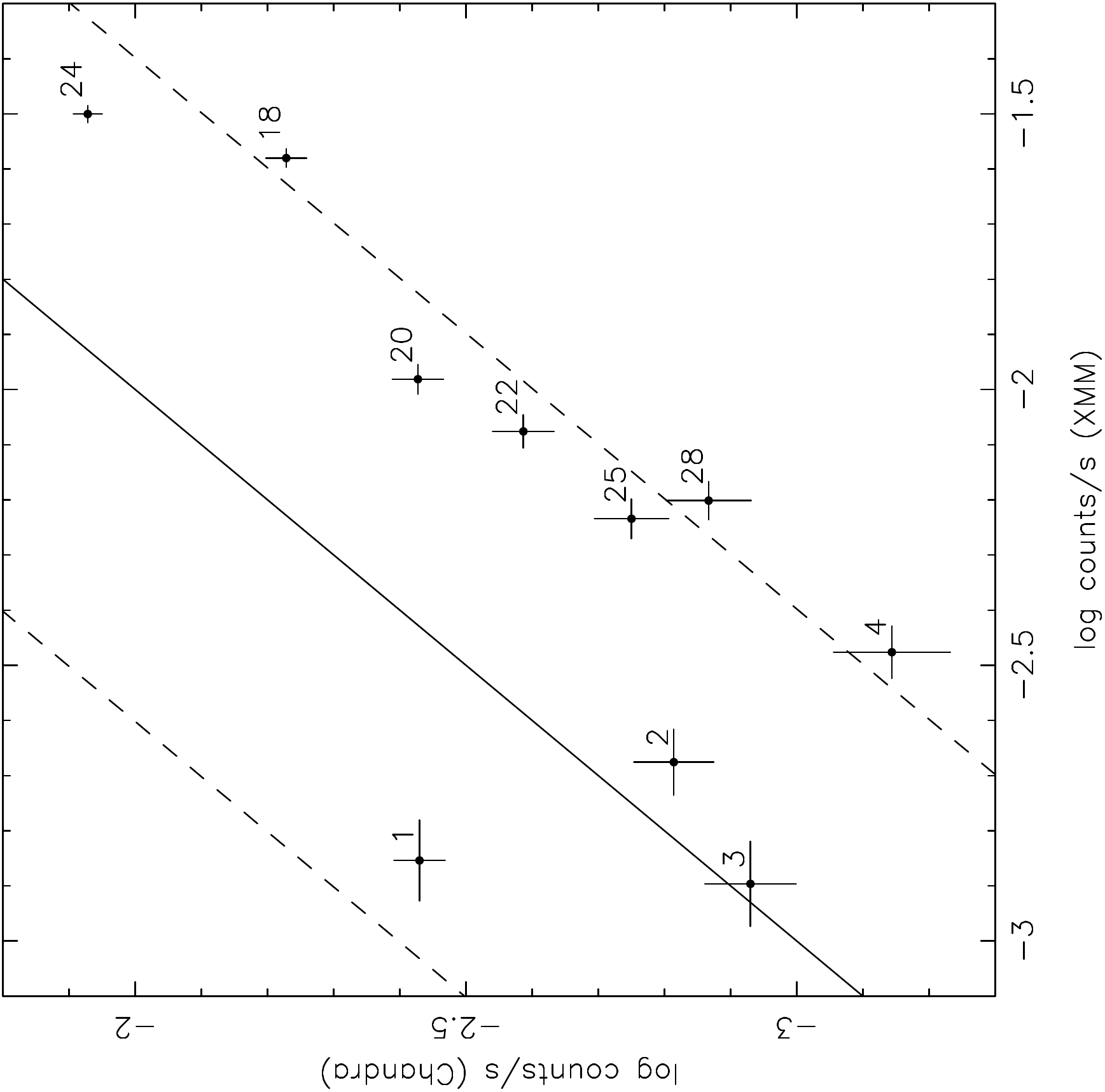}}
     \caption{Detected countrate in the 0.5--6.0\,keV energy band
     ($X_\mathrm{soft}+X_\mathrm{hard}$, Table\,\ref{t2}) of the
     \emph{Chandra} ACIS-S observation as a function of the \emph{XMM}
     EPIC (PN and both MOS detectors) countrate in the 0.5--10\,keV
     energy band (Table\,2 of \citealt{wwb06}). The solid and dashed
     lines show equal countrates and 4 times higher/lower countrates,
     respectively. For most sources, the \emph{XMM} countrate is about
     4 times higher than the \emph{Chandra} countrate, as expected for
     a constant source.}
     \label{f8} 
  \end{figure}

  In Fig.\,\ref{f8} we compare the countrates of sources in M55
  observed with \emph{Chandra} (from Table\,\ref{t2}) with those
  observed with \emph{XMM} EPIC (from Table\,2 of
  \citealt{wwb06}). All \emph{XMM} sources within the field-of-view of
  \emph{Chandra} have been detected; the \emph{XMM} sources are the
  sources with the highest countrates in Chandra, plus CX29 (XMM45).
  Comparison of the \emph{Chandra} and \emph{XMM} source positions
  shows that the error estimates by Webb et al.\ for the \emph{XMM}
  source positions are overestimated by about a factor 2.  As
  expected, the \emph{XMM} countrate for most sources is $\sim4$ times
  higher than the \emph{Chandra} countrate: for example, a 2\,keV
  bremsstrahlung spectrum absorbed by the column towards M55, with
  0.001 cts\,s$^{-1}$ in \emph{Chandra} ACIS-S will produce about
  0.0026\,cts\,s$^{-1}$ on the PN and 0.00075\,cts\,s$^{-1}$ on each
  of the two MOS detectors, for a total of 0.0041 cts\,s$^{-1}$ in
  \emph{XMM}\footnote{We estimate this with the PIMMS tool at {\tt
      http://heasarc.gsfc.nasa.gov/Tools/w3pimms.html}; we have
    multiplied the countrates within a $15\arcsec$ extraction radius
    as given by PIMMS with 1.47 to estimate the total countrates}.

  Remarkably, all three sources with the lowest \emph{XMM} count rates
  were brighter during the \emph{Chandra} observation. One of
  these is the cataclysmic variable CV1, our source CX1. \citet{wwb06}
  state that this source was three times fainter, in the 0.5--10\,keV
  band, during the \emph{XMM} observation than during earlier
  \emph{ROSAT} observations.  Since the \emph{ROSAT} measurements are
  limited to the 0.2--2.5\,keV band, we prefer to make the comparison
  of the luminosities in this energy range.  The hardness ratio of CX1
  as measured with \emph{Chandra} (see Fig.\,\ref{f3}) is that of a
  bremsstrahlung spectrum with temperature
  $kT_\mathrm{b}\simeq2$\,keV. Assuming this spectrum, the distance
  and absorption of M55 as above, and the countrates given in Table\,3
  of \citet{ver01}; cf.\ \citet{jvh96} and in Table\,2 of
  \citet{wwb06}, we obtain 0.5--2.5\,keV luminosities, in units of
  $10^{31}$\,erg\,s$^{-1}$, of $6.0\pm1.5$ for both the \emph{ROSAT}
  PSPC observation (March 1991) and the \emph{ROSAT} HRI observation
  (September 1997), and of $0.44\pm0.09$ for the \emph{XMM}
  observation (October 2001); our \emph{Chandra} observation (May
  2004, Table\,1) gives $4.1\pm0.4$.  Thus, the X-ray flux was an
  order of magnitude lower during the \emph{XMM} observation than
  during the \emph{ROSAT} and \emph{Chandra} observations.
  \citet{wwb06} assume that the X-ray flux is higher during an
  optical/ultraviolet dwarf nova outburst, and conclude that CV1 was
  in outburst during both \emph{ROSAT} observations.  However, most
  dwarf novae have {\em lower} X-ray fluxes during outburst than in
  quiescence (\citealt{wvb+96,vwm99} and the discussion of Fig.\,2 in
  \citealt{vbrp97}). We conclude that CV1 was \emph{in outburst}
  during the \emph{XMM} observations, and in quiescence during the
  \emph{ROSAT} and \emph{Chandra} observations, in good agreement with
  the observations by \citet{kpt+05} that CV1 spends most of the time
  in quiescence.

  We detect the optical counterpart to CV1 at $R=20.7$ and $B-R=1.34$,
  where it lies only about 0.3\,mag blue-wards of the cluster
  main-sequence. With $R-\mathrm{H}\alpha=0.58$, it exhibits a large
  excess of H$\alpha$ emission, reminiscent of cataclysmic variables.
  At $B=22.1$, our photometry of CV1 puts it close to the quiescent
  level of about $B=22.5$ seen in the light-curve by
  \citet{kpt+05}. Furthermore, the proximity in color of the source to
  the cluster main-sequence was also seen by \citet{kpt+05} during
  quiescence, where the source actually coincides with the
  main-sequence in $B-V$ and $V-I$ (see Fig.\,4 of Kaluzny et al.).
  Hence, we conclude that CX1/CV1 was also in quiescence during our
  FORS2 observations.

  Fig.\,\ref{f8} shows that CX20/XMM14 had \emph{Chandra} and
  \emph{XMM} countrates as expected for a constant luminosity between
  the two observations.  The source was not detected with
  \emph{ROSAT}; in re-analysing the \emph{ROSAT} PSPC observation,
  using the method outlined in \citet{ver01}, we obtain a $2\sigma$
  upper limit to the countrate of 0.0006 cts\,s$^{-1}$ (channels
  50--240), which corresponds roughly to a flux at 0.5--2.5\,keV of
  $8.9\times10^{-15}$\,erg\,cm$^{-2}$\,s$^{-1}$. Analysis of the HRI
  data gives an upper limit of 10 source counts, or
  $8\times10^{-15}$\,erg\,cm$^{-2}$\,s$^{-1}$.  Within the error,
  these limits are compatible with the flux found with \emph{Chandra}
  (Table\,\ref{t2}) and with \emph{XMM}. \citet{wwb06} argue that this
  source has been variable by at least a factor 22 between the
  \emph{ROSAT} and \emph{XMM} observations. We find no evidence for
  such variation.

  \citet{kpt+05} discuss an interesting blue variable star, which they
  name M55-B1. From its variability between observing seasons, in the
  absence of short-term ($\sim$ hr) variability, they argue that it is
  a background quasar rather than a cataclysmic variable. This source
  is not detected on the \emph{Chandra} ACIS-S3 chip, but on an
  adjacent chip.  It is clearly detected in X-rays, with an unabsorbed
  flux in the 0.5--8.0\,keV band of
  $5.1\times10^{-15}$\,erg\,cm$^{-2}$\,s$^{-1}$ for absorption fixed
  at that of M55.  A free power law fit gives a higher absorption
  ($N_\mathrm{H}=7.4\times10^{21}$\,cm$^{-2}$) and higher unabsorbed
  flux of $8.8\times10^{-15}$ erg\,s$^{-1}$.

  \subsubsection{NGC\,6366}
  Only the brightest \textit{Chandra} source was detected with
  \textit{ROSAT}, at a 0.5--2.5\,keV luminosity \citep{ver01}
  compatible with that during the \textit{Chandra} observation.
  \textit{ROSAT} observed CX1a and CX1b as one source.  The observed
  spectrum of the point source is relatively soft (see Fig.\,\ref{f2},
  and Sect.\,\ref{s2.3}), which with the high absorption towards
  NGC\,6366 implies a very soft intrinsic spectrum. Whereas many
  cataclysmic variables have hard spectra, a very soft spectrum is
  still compatible with identification as an old nova.

%  Only the brightest \emph{Chandra} source was detected by
%  \emph{ROSAT}. The X-ray luminosity given by \citet{ver01} is 6.5
%  times higher than the luminosity derived from \emph{Chandra}
%  (Table\,1). The steady decline of surface brightness from center to
%  limb (Fig.\,3) indicates that the X-ray source is intrinsically
%  extended, rather than a conglomerate of a few point sources. The
%  observed X-ray spectrum is soft, considering the high absorption
%  towards NGC\,6366 (Fig.\,4), implying a very soft intrinsic
%  spectrum.  A soft extended (foreground or) cluster source could be a
%  planetary nebula; $20\arcsec$ corresponds to about 0.35\,pc at the
%  distance of NGC\,6366.  One does not expect strong variability of
%  such an extended source, and the optical images, including the
%  H$\alpha$ image, show no evidence for a nebula (Fig.\,6). If we
%  multiply the escape velocity of a white dwarf, 5000 km\,s$^{-1}$
%  say, with a life time of 10\,yr for the X-ray emission, we obtain an
%  estimated size of a nova shell of 0.05\,pc, far too small for the
%  observed X-ray extent.  A soft extended background source could be a
%  cluster of galaxies; but no strong variability would be expected in
%  this case either.

  \subsection{Ratio of optical to X-ray luminosity}\label{s5.2}
  Our classification of the X-ray sources for which we have an optical
  identification is based in part on the X-ray to optical luminosity
  ratio. \citet{bph+04} showed that a line of constant
  \textit{Chandra} X-ray to optical luminosity roughly separates the
  cataclysmic variables from the active binaries in 47\,Tuc, NGC\,6752
  and M4 (data from \citealt{ghem01,eghg03a,plh+02,bph+04}). This
  separatrix corresponds to the dashed line in Fig.\,\ref{f7},
  and is given by
  \begin{equation} 
    \log L_\mathrm{X} \mbox{(erg\,s$^{-1}$) [0.5--2.5\,keV]}=34.0 -0.4M_V
    \label{sepbas}
  \end{equation} 

  Remarkably, and perhaps disconcertingly, this line is rather higher
  than the upper bound to the X-ray luminosities of stars and RS~CVn
  binaries near the Sun. An upper bound is given in \citet{vpb07} as
  \begin{equation}
    \log L_\mathrm{X} \mbox{(erg\,s$^{-1}$) [0.5--2.5\,keV]}=32.3
    -0.27M_V
    \label{sepver}
  \end{equation} 
  This line is an absolute upper bound in that it lies above all
  \textit{ROSAT} measurements for nearby stars (selected from
  \citealt{hssv99}) and for RS\,CVn systems \citep{dlfs93}, where we
  multiply the 0.1--2.4\,keV luminosities given by H\"unsch et
  al.\ and Dempsey et al.\ with 0.4 to obtain the 0.5--2.5\,keV
  luminosities.  It is shown as a solid line in Fig.\,\ref{f7}. It has
  been remarked already on the basis of \textit{Einstein} 0.2--4\,keV
  measurements that the X-ray luminosity of main-sequence stars
  increases with increasing stellar rotation velocity, until it
  saturates at

  \begin{equation}
    L_\mathrm{X}\simeq0.001 L_\mathrm{bol}
    \label{satur}
  \end{equation}
  where $L_\mathrm{bol}$ is the bolometric luminosity of the star
  (e.g.\ \citealt{vw87}, esp.\ their Fig.\,6).

  We show in Fig.\,\ref{f7} the line of $0.001L_\mathrm{bol}$ for
  the 11.2\,Gyr isochrones calculated by \citet{gbbc00} for
  metal-poor stars ($z=0.001$), appropriate for globular clusters, and
  for stars of solar abundance ($z=0.019$), appropriate for stars in
  the solar neighbourhood.  These lines differ little, and give a more
  conservative upper bound to the maximum 0.5--2.5\,keV luminosity of
  stars near the Sun.

  The region between the separatrix Eq.\,\ref{sepbas} on the one hand
  and the upper bounds given by Eq.\,\ref{sepver} or Eq.\,\ref{satur}
  contains a dozen optical counterparts of X-ray sources in globular
  clusters that have been classified as magnetically active binaries.
  If this classification is correct, we must accept that such binaries
  can have rather higher X-ray luminosities in globular clusters than
  in the solar neighbourhood. Since the bolometric luminosity as a
  function of $M_V$ is similar for low and solar
  metallicities, this would imply that the saturation of X-rays occurs
  at higher $L_\mathrm{X}/L_\mathrm{bol}$ in globular clusters.
  Alternatively, we may wish to reconsider the classification of these
  sources. If we want to classify them as cataclysmic variables, we
  must accept that globular clusters host cataclysmic variables that
  are not blue with respect to the main sequence, even in the $B$-band
  filter. These systems clearly warrant further study. We note that
  CV1 in M55 does lie on the main-sequence in $B-V$ in quiescence
  (\citealt{kpt+05}, see Sect.\,\ref{s511}).

  \begin{table}
    \caption{Values for central density $\rho_0$, core radius $r_\mathrm{c}$, 
      distance $d$, and absolute visual magnitude $M_V$, taken from
      Harris (1996, version Feb 2003), and for M4 from 
      \citet{rfi+97}. Collision number $\Gamma$ and mass within
      the half-mass radius $M_\mathrm{h}$ are scaled on the values for M4.
      \label{t6}}
    \begin{tabular}{lrrlrrr}
      \hline \hline
      cluster & log $\rho_0$ & $r_\mathrm{c}$ & $d$ & $M_V$ & $\Gamma$ & $M_\mathrm{h}$ \\
      &  ($L_\odot$~pc$^{-3}$)  & ($\arcsec$)  & (kpc) & & \\
      \hline
      M4 &  4.01 &   49.8 &   1.73 & $-$6.9 & $\equiv$1.00 & $\equiv$1.00 \\
      NGC\,6397 &  5.68 &    3.0 &   2.3 & $-$6.6 &   2.05 &    0.78 \\
      47\,Tuc  &  4.81 &   24.0 &   4.5 & $-$9.4 &  24.91 &   10.19 \\
      NGC\,288  &  1.80 &   85.2 &   8.8 & $-$6.7 &   0.04 &    0.86 \\
      M55 &  2.15 &  169.8 &   5.3 & $-$7.6 &   0.18 &    1.82 \\
      NGC\,6366 &  2.42 &  109.8 &   3.6 & $-$5.8 &   0.09 &    0.35 \\
      \hline
    \end{tabular}
  \end{table}

  \subsection{Origin of the X-ray sources}
  To compare the numbers of X-ray sources in globular clusters,
  \citet{pla+03} chose a luminosity limit of
  $L_\mathrm{X}>4\times10^{30}$\,erg\,s$^{-1}$ [0.5--6.0\,keV], which
  is detectable in almost all the clusters they investigated.  We have
  detected 16 sources above this limit within the half-mass radius of
  M55, with an expected number of 8-9 unrelated sources; and 5 within
  the half-mass radius of NGC\,6366, with an expected number of 4
  unrelated sources. As already argued above, from these numbers alone
  we cannot exclude that all detected sources are unrelated.  The
  situation changes if we take into account our optical
  identifications (Section\,4), which indicate that at least three
  sources (CX1, CX7, CX10) are highly probable members of M55, and
  that two sources (CX4, CX5) are highly probable members of
  NGC\,6366. We have also argued that the point source plus extended
  source is an old nova in NGC\,6366.

  In Table\,\ref{t6} we compare the collision numbers
  $\Gamma\equiv{\rho_0}^{1.5}{r_\mathrm{c}}^2$ and the masses within
  the half-mass radii $M_\mathrm{h}$ (as measured with the absolute
  magnitude assuming a fixed mass-to-light ratio) for several
  well-observed clusters; where we scale both $\Gamma$ and
  $M_\mathrm{h}$ on the globular cluster M4.

  \citet{pla+03} report $41\pm2$ sources above the luminosity limit in
  the core of 47\,Tuc.  If the number of X-ray sources would scale
  with the collision number, we would predict 0.3 sources in M55, and
  0.15 in NGC\,6366. The number of detected very probable member X-ray
  sources is significantly higher in both clusters. If the number of
  X-ray sources would scale with the mass, we would predict about 7
  sources in M55, and 3 to 4 in NGC\,6366. The number of detected
  member X-ray sources is remarkably close to this.

  To study the origin of X-ray sources in globular clusters in further
  detail, \citet{ph06} investigate the hypothesis that the number of
  sources scales with the collision number and mass as
  \begin{equation}
    N = a\Gamma + b M\label{e:numbers}
  \end{equation}
  and show that the dependence on mass is significantly present in the
  observations. To overcome the small-number statistics of clusters
  with small source numbers, they combine the numbers of such clusters
  before fitting them to Eq.\,\ref{e:numbers}.

  In a recent paper, \citet{vpb07} suggest a more general way of
  fitting the observed numbers to Eq.\,\ref{e:numbers}.  Briefly, they
  determine for each cluster the most probable combination of numbers
  $N_\mathrm{c}$ of member and $N_\mathrm{b}$ of unrelated sources
  that gives the observed number of sources within the half-mass
  radius $N_\mathrm{h}$, based on the expected number of cluster
  sources $\mu_\mathrm{c}$ according to the model of
  Eq.\,\ref{e:numbers}, and on the expected number of unrelated
  sources $\mu_\mathrm{b}$. In doing so they assume that the
  probability of finding $N$ sources when $\mu$ are expected is given
  by the Poisson function
  \begin{equation}
    P(N,\mu) = {\mu^N\over N!}e^{-\mu}
  \end{equation}
  The fitting procedure consists of varying $a$ and $b$ to maximize
  \begin{equation}
    P = \prod_j
    [P(N_\mathrm{c},\mu_\mathrm{c})P(N_\mathrm{b},\mu_\mathrm{b})]_j\label{totprob}
  \end{equation}
  where $j$ indexes the clusters. 

\begin{figure}
     \resizebox{0.95\hsize}{!}{\includegraphics[angle=270]{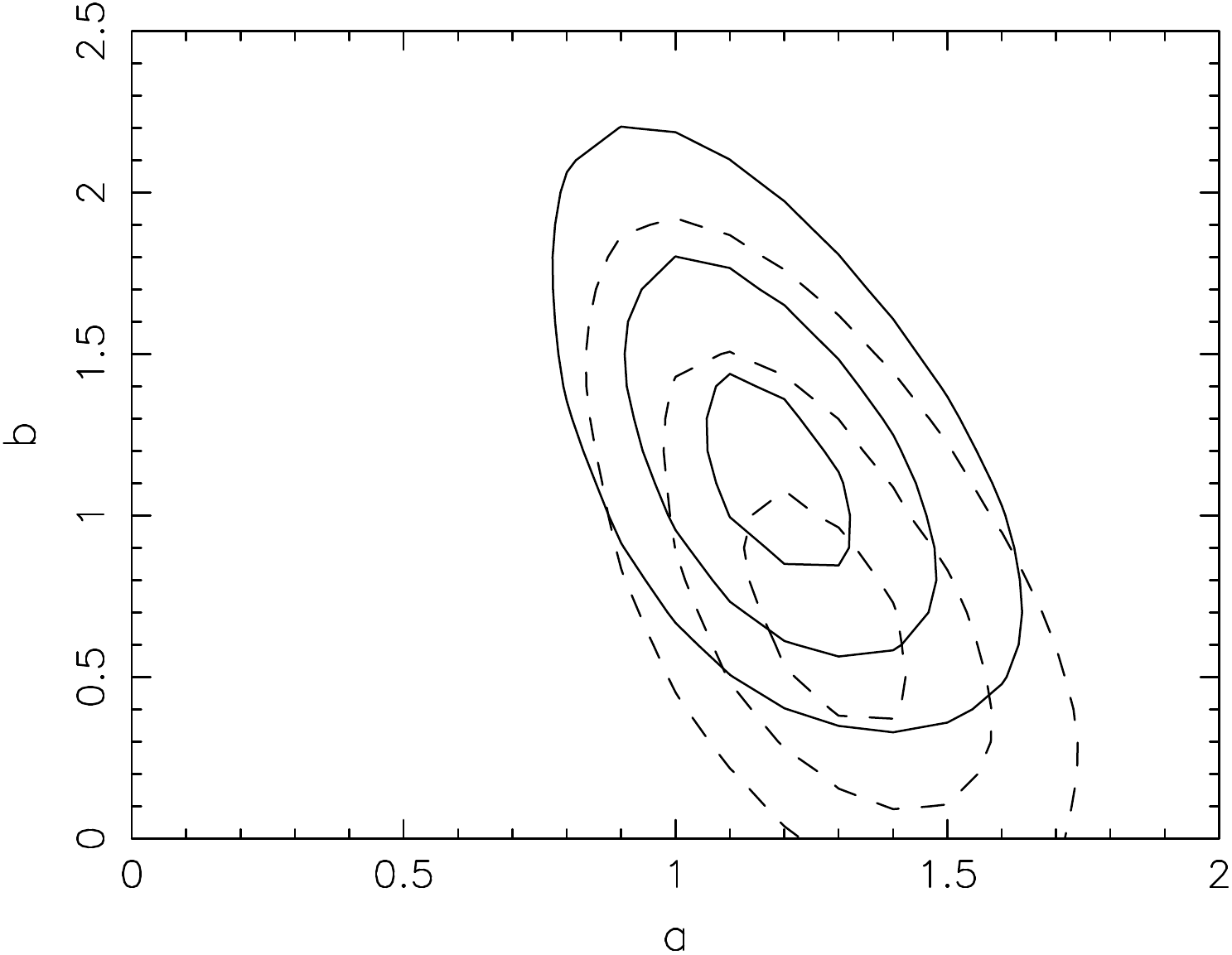}}
     \caption{Probability contours around the best solutions for a fit
       of the observed number of sources within the half-mass radii of
       12 globular clusters to Eq.\,\ref{e:numbers} where we allow
       zero cluster members in some clusters (dashed contours), and
       where we impose lower limits to the number of cluster members
       of 2 for NGC\,288, 3 for M55 and 2 for NGC\,6366 (solid
       contours). We assume that $-2(P-P_\mathrm{bf})$, with the
       probability $P$ given by Eq.\,\ref{totprob} and $P_\mathrm{bf}$
       indicating the best fit probability, follows a $\chi^2$
       distribution, and show $\Delta\chi^2=1$, 4 and 9.  The
       imposition of minimum numbers of cluster member X-ray sources
       strongly increases the significance of the mass dependence in
       Eq.\,\ref{e:numbers}.}\label{fig:probs}
  \end{figure}

  The values for $N_\mathrm{h}$ and $\mu_\mathrm{c}$ that we use are
  taken from \citet{pla+03}. We ignore the clusters for which only a
  lower limit to $N_\mathrm{h}$ is known. We add three clusters:
  NGC\,288, M55 and NGC\,6366; the values we use are 11/8.0, 16/8.3
  and 5/4.0 for $N_\mathrm{h}/\mu_\mathrm{c}$ for these three
  clusters, respectively.  The best solution is found for
  $a=1.3\Gamma_\mathrm{M4}$ and $b=0.6M_\mathrm{M4}$. The contours of
  the acceptable solutions are given in Fig.\,\ref{fig:probs}, and
  it is seen that $b=0$ is acceptable at about the $2\sigma$
  level. Thus, from the observed numbers alone, we have no firm
  evidence for a dependence on mass.

  We repeated the fitting procedure, but excluding solutions for which
  $N_\mathrm{c}$ is less than the minimum number of highly probable
  cluster members.  For these fits we assume that 2 sources are
  certain members of NGC\,288, 3 of M55 and 2 of NGC\,6366. As
  expected, these constraints force the solutions to a stronger
  dependence on mass: the best solution is found for
  $a=1.2\Gamma_\mathrm{M4}$ and $b=1.1M_\mathrm{M4}$. As seen in
  Fig.\,\ref{fig:probs}, the dependence on mass is now significant.

  The best values of $a$ and $b$ are relatively stable against small
  variations in the numbers $\mu_\mathrm{c}$ and $N_\mathrm{h}$ for
  individual clusters, as comparison with the results of \citet{vpb07}
  confirm. The probability of the best solution does vary with small
  variations in $\mu_\mathrm{c}$ and $N_\mathrm{h}$ for individual
  clusters.  For example, increasing the minimum of cluster member
  X-ray sources in NGC\,6366 from 2 to 3 does not change the best
  values for $a$ and $b$, but decreases the probability of the best
  solution for NGC\,6366, for which the best model predicts
  $\mu_\mathrm{c}=0.4$, and thus also slightly decreases the overall
  probability.

  The cluster which is responsible for the lowest probability is
  NGC\,6397, which houses 12 sources, of which our best model assigns
  $N_\mathrm{c}=11$ to be cluster members, even though it expects only
  $\mu_\mathrm{c}=3.6$. The remarkably large number of sources in this
  cluster has been remarked upon before, and various explanations have
  been offered. In our model, however, whereas still the worst fit
  cluster, it is not altogether unacceptable. The main difference
  between this result and the strong deviation for NGC\,6397 as found
  by e.g. \citet{pla+03} appears to be due to a different definition
  of $\Gamma$. \citet{pla+03} integrate a King model to estimate
  $\Gamma$, whereas we compute it as
  $\Gamma={\rho_0}^{1.5}{r_\mathrm{c}}^2$.  Our probabilities do not
  take into account the uncertainties in $\Gamma$ and $M_\mathrm{h}$,
  nor the observed number $N_\mathrm{h}$ which through its dependence
  of the flux limit is affected by the assumed distance to the
  cluster. Clearly these uncertainties warrant further investigation.

  An important result illustrated by Fig.\,\ref{fig:probs} is that the
  masses of the clusters may {\em not} be taken as proxies for their
  collision numbers. Even though there is some correlation between
  mass and collision number, the spread in collision number at given
  mass, and in mass at given collision number, is very large. Indeed
  our fits show that the number of X-ray sources within a cluster is
  very badly predicted by the mass alone, as illustrated by the highly
  significant offset of the best solution from the $a=0$ line.

  \begin{acknowledgements}
    We thank Jacco Vink for drawing our attention to the extended
    emission of old novae. We thank Ata Sarajedini and Jay Anderson
    for providing us with the data for Fig.\,\ref{f5} in advance of
    publication.  W.\,H.\,G.\,L. is grateful for generous support from
    NASA. C.\,G.\,B.\, acknowledges support from NWO.
  \end{acknowledgements}

%  \bibliographystyle{aa} 
%\bibliography{references}

\end{document}